\documentclass[twocolumn,prd,floatfix,preprintnumbers,a4paper,nofootinbib,superscriptaddress]{revtex4-1}

\usepackage{epsfig}
\usepackage{graphics}
\usepackage{graphicx}
\usepackage{amsmath,amssymb}
\usepackage{amsfonts}
\usepackage{bm}
\usepackage[usenames,dvipsnames]{color}
\usepackage{amssymb}
\usepackage{psfrag}
\usepackage{times}
\usepackage[varg]{txfonts}
\usepackage{xspace}
\usepackage{placeins} 
\usepackage{capt-of} 
\usepackage{enumerate}

\usepackage[colorlinks, pdfborder={0 0 0}]{hyperref}
\definecolor{LinkColor}{rgb}{0.75, 0, 0}
\definecolor{CiteColor}{rgb}{0, 0.5, 0.5}
\definecolor{UrlColor}{rgb}{0, 0, 0.75}
\hypersetup{linkcolor=LinkColor}
\hypersetup{citecolor=CiteColor}
\hypersetup{urlcolor=UrlColor}
\let\oldtheequation\theequation
\makeatletter
\def\tagform@#1{\maketag@@@{\ignorespaces#1\unskip\@@italiccorr}}
\renewcommand{\theequation}{(\oldtheequation)}
\makeatother

\usepackage[utf8]{inputenc}
\usepackage{ulem}
\makeatletter
\g@addto@macro\normalsize{%
  \addtolength\abovedisplayskip{0.5\baselineskip}
  \addtolength\belowdisplayskip{0.5\baselineskip}
  \addtolength\abovedisplayshortskip{0.5\baselineskip}
  \addtolength\belowdisplayshortskip{0.5\baselineskip}
}
\makeatother

\newcommand{\macro}[1]{#1}                    

\newcommand{\dcc}{LIGO-P1600279-v7}
\newcommand{\aligo}[1][]{aLIGO#1\xspace}
\newcommand{\avirgo}{Advanced Virgo\gdef\avirgo{Adv.~Virgo\xspace}\xspace}
\newcommand{\bh}[1][]{black hole#1 (BH#1)\renewcommand{\bh}[1][]{BH##1\xspace}\xspace}
\newcommand{\bbh}[1][]{binary black hole#1 (BBH#1)\renewcommand{\bbh}[1][]{BBH##1\xspace}\xspace}
\newcommand{\gw}[1][]{gravitational wave#1 (GW#1)\renewcommand{\gw}[1][]{GW##1\xspace}\xspace}
\newcommand{\nr}{numerical relativity (NR)\gdef\nr{NR\xspace}\xspace}
\newcommand{\peaklum}{peak luminosity\xspace}
\newcommand{\peaklums}{peak luminosities\xspace}
\newcommand{\event}{GW150914\xspace}

\newcommand{\xmas}{GW151226\xspace}
\newcommand{\gr}{General Relativity (GR)\gdef\gr{GR\xspace}\xspace}

\newcommand{\smbh}[1][]{super-massive black hole#1 (SMBH#1)\renewcommand{\smbh}[1][]{SMBH##1\xspace}\xspace}
\newcommand{\pade}{Pad{\'e}\xspace}
\newcommand{\ansaetze}{ans{\"a}tze\xspace}
\newcommand{\emri}[1][]{extreme-mass-ratio inspiral#1 (EMRI#1)\renewcommand{\emri}[1][]{EMRI##1\xspace}\xspace}
\newcommand{\emr}{extreme-mass-ratio\xspace}

\newcommand{\emrl}{extreme-mass-ratio limit\xspace}
\newcommand{\wordforhighq}{large\xspace}
\newcommand{\lmr}{\wordforhighq-mass-ratio\xspace}

\newcommand{\lmrs}{\wordforhighq mass ratios\xspace}

\newcommand{\eqmeqS}{equal-mass-equal-spin\xspace}

\newcommand{\BAM}{\texttt{BAM}\xspace}
\newcommand{\SpeC}{Spectral Einstein Code~(SpEC)\gdef\SpeC{SpEC\xspace}\xspace}
\newcommand{\LAZEV}{\texttt{LAZEV}\xspace}
\newcommand{\MAYA}{\texttt{MAYA}\xspace}
\newcommand{\rmse}{root-mean-square error~(RMSE)\gdef\rmse{RMSE\xspace}\xspace}
\newcommand{\PN}{Post-Newtonian~(PN)\xspace\gdef\PN{PN\xspace}}
\newcommand{\ffi}{fixed-frequency-integration~(FFI)\gdef\ffi{FFI\xspace}\xspace}
\newcommand{\RWZ}{Regge-Wheeler-Zerilli~(RWZ)\gdef\RWZ{RWZ\xspace}\xspace}
\newcommand{\imr}{inspiral-merger-ringdown\xspace}
\newcommand{\eob}{effective-one-body (EOB)\gdef\eob{EOB\xspace}\xspace}

\newcommand{\const}{\mathrm{const.}}
\newcommand{\Msun}{M_\odot}
\newcommand{\Lsun}{L_{\odot}}
\newcommand{\chieff}{\chi_{\mathrm{eff}}}
\newcommand{\Seff}{\widehat{S}}
\newcommand{\myS}{\Seff}
\newcommand{\chidiff}{\Delta\chi}
\newcommand{\Lpeak}{L_{\mathrm{peak}}}
\newcommand{\Lo}{L_0}
\newcommand{\Lscaled}{\Lpeak^\prime}
\newcommand{\oneDeta}{\left(\eta,\myS=0\right)}
\newcommand{\oneDS}{\left(\eta=0.25,\myS\right)}
\newcommand{\twoDparams}{\left(\eta,\myS\right)}
\newcommand{\threeDparams}{\left(\eta,\myS,\chidiff\right)}
\newcommand{\Lofeta}{\Lscaled\oneDeta}
\newcommand{\LofS}{\Lscaled\oneDS}
\newcommand{\LtwoD}{\Lscaled\twoDparams}
\newcommand{\LthreeD}{\Lscaled\threeDparams}
\newcommand{\Ldiff}{\Delta\Lscaled}

\newcommand{\lmax}{\ell_{\max}}

\newcommand{\Mf}{M_{\mathrm{f}}}
\newcommand{\Erad}{E_{\mathrm{rad}}}
\newcommand{\Ndata}{N_{\mathrm{data}}}
\newcommand{\Ncoef}{N_{\mathrm{coeffs}}}

\newcommand{\AICc}{\mathrm{AICc}}
\newcommand{\BIC}{\mathrm{BIC}}
\newcommand{\maxlnL}{\mathcal{L}_{\max}}

\newcommand{\poly}[1]{\mathrm{P}(#1)}
\newcommand{\rational}[2]{\mathrm{R}(#1,#2)}

\newcommand{\numNRcodes}{\macro{four}\xspace}				
\newcommand{\NRcodesInOrderAnd}{\macro{\BAM, SXS, GaTech, and RIT}\xspace}	
\newcommand{\NRcountTotal}{\macro{419}\xspace}				
\newcommand{\NRcountBAM}{\macro{47}\xspace}				
\newcommand{\NRcountSXS}{\macro{160}\xspace}				
\newcommand{\NRcountGaT}{\macro{105}\xspace}				
\newcommand{\NRcountRIT}{\macro{107}\xspace}				
\newcommand{\NRcount}{\macro{378}\xspace}				
\newcommand{\NRcountNS}{\macro{81}\xspace}				
\newcommand{\NRcountEqSqone}{\macro{32}\xspace}				
\newcommand{\NRcountEqSrest}{\macro{50}\xspace}				
\newcommand{\NRcountUneqS}{\macro{215}\xspace}				
\newcommand{\NRcountNZSUM}{\macro{265}\xspace}				
\newcommand{\datacount}{\macro{423}\xspace}				
\newcommand{\datacountNS}{\macro{84}\xspace}				
\newcommand{\datacountEqSqone}{\NRcountEqSqone}				
\newcommand{\datacountEqSrest}{\macro{92}\xspace}			
\newcommand{\datacountNZSUM}{\macro{307}\xspace}			
\newcommand{\EMRIcountqonek}{\macro{31}\xspace}

\newcommand{\EMRIcountqtenkWords}{\macro{seven}\xspace}

\newcommand{\NRcountOutliers}{\macro{41}\xspace}			
\newcommand{\NRcountOutliersEqS}{\macro{22}\xspace}			
\newcommand{\NRcountOutliersTuples}{\macro{17}\xspace}			
\newcommand{\NRmaxq}{\macro{18}\xspace}					
\newcommand{\NRmaxchi}{\macro{0.995}\xspace}				
\newcommand{\NRLovalue}{\macro{0.0164}\xspace}				
\newcommand{\NRLostdevrel}{\macro{0.2\%}\xspace}			

\newcommand{\LPSOLRELATIVE}{\macro{\ensuremath{{\sim}10^{23}}}\xspace}	
\newcommand{\LSUNVALUE}{\macro{\ensuremath{3.8\times10^{33}}}\xspace}	
\newcommand{\GRBLum}{\macro{$(4.7\pm{0.2})\times10^{54}$ erg/s}\xspace}	
\newcommand{\GWVSGRB}{\macro{60--90}\xspace}				
\newcommand{\LMW}{\macro{2\times10^{11}\Lsun}\xspace}			
\newcommand{\LPMWRELATIVE}{\macro{$\gtrsim10^{11}$}\xspace}		

\newcommand{\MTWOSCOMPACTNOCALIB}{\macro{\ensuremath{29_{-4}^{+4}}}} 
\newcommand{\MONESCOMPACTNOCALIB}{\macro{\ensuremath{36_{-4}^{+5}}}} 
\newcommand{\LPEAKCOMPACT}{\macro{\ensuremath{3.6_{-0.4}^{+0.5} \times 10^{56}}}} 
\newcommand{\LPEAKSOLCOMPACT}{\macro{\ensuremath{200_{-20}^{+30}}}} 

\begin{document}

\addtolength{\textheight}{-4\baselineskip}


\title{The most powerful astrophysical events: Gravitational-wave peak\\ luminosity of binary black holes as predicted by numerical relativity}

\newcommand{\cardiff}{School of Physics and Astronomy, Cardiff University, The Parade, Cardiff, CF24 3AA, United Kingdom}
\newcommand{\uib}{Universitat de les Illes Balears, IAC3---IEEC, 07122 Palma de Mallorca, Spain}
\newcommand{\icts}{International Centre for Theoretical Sciences,\\ Tata Institute of Fundamental Research, Bangalore 560012, India\\\vspace{\baselineskip}}
\newcommand{\birm}{School of Physics and Astronomy, University of Birmingham, Edgbaston, Birmingham, B15 2TT, United Kingdom}
\newcommand{\golm}{Albert-Einstein-Institut, Am M{\"u}hlenberg 1, D-14476 Potsdam-Golm, Germany}
\newcommand{\hannover}{Albert-Einstein-Institut, Callinstra{\ss}e 38, 30167 Hannover, Germany}
\newcommand{\glasgow}{School of Physics and Astronomy, University of Glasgow, Glasgow G12 8QQ, United Kingdom}
\newcommand{\jena}{Theoretical Physics Institute, University of Jena, 07743 Jena, Germany}
\newcommand{\parma}{Department of Mathematical, Physical and Computer Sciences,\\ University of Parma, I-43124, Parma, Italy}
\newcommand{\infn}{Istituto Nazionale di Fisica Nucleare, Sezione Milano Bicocca,\\ gruppo collegato di Parma, I-43124 Parma, Italy}
\newcommand{\ihes}{Institut des Hautes Etudes Scientifiques, 91440 Bures-sur-Yvette, France}

\author{David~Keitel}\email{david.keitel@ligo.org}\affiliation{\uib}\affiliation{\glasgow}
\author{Xisco~Jim{\'e}nez~Forteza}\email{francisco.forteza@ligo.org}\affiliation{\uib}
\author{Sascha~Husa}\email{sascha.husa@ligo.org}\affiliation{\uib}
\author{Lionel~London}\email{lionel.london@ligo.org}\affiliation{\cardiff}
\author{Sebastiano~Bernuzzi}\affiliation{\parma}\affiliation{\infn}
\author{Enno~Harms}\affiliation{\jena}
\author{Alessandro~Nagar}\affiliation{\ihes}
\author{Mark~Hannam}\affiliation{\cardiff}
\author{Sebastian~Khan}\affiliation{\cardiff}\affiliation{\hannover}
\author{Michael~P{\"u}rrer}\affiliation{\golm}
\author{Geraint~Pratten}\affiliation{\uib}
\author{Vivek~Chaurasia}\affiliation{\icts}\affiliation{\jena}

\begin{abstract}
	For a brief moment,
	a \bbh merger can be the most powerful astrophysical event in the visible Universe.
	Here we present a model fit for this gravitational-wave \textit{\peaklum}
	of nonprecessing quasicircular \bbh systems
	as a function of the masses and spins of the component black holes,
	based on \nr simulations and the hierarchical fitting approach introduced
        by X.~Jim{\'e}nez Forteza \textit{et al.} [\href{https://doi.org/10.1103/PhysRevD.95.064024}{Phys. Rev. D 95, 064024 (2017)}].
	This fit improves over previous results in accuracy and parameter-space coverage
	and can be used to infer posterior distributions
	for the \peaklum of future astrophysical signals like \event and \xmas.
	The model is calibrated to the \mbox{$\ell\leq6$} modes of
	\NRcount nonprecessing \nr simulations
	up to mass ratios of \NRmaxq
	and dimensionless spin magnitudes up to \NRmaxchi,
	and includes unequal-spin effects.
	We also constrain the fit to perturbative numerical results for \lmrs.
	Studies of key contributions to the uncertainty in \nr \peaklums, such as
	(i) mode selection,
	(ii) finite resolution,
	(iii) finite extraction radius,
	and (iv) different methods for converting \nr waveforms to luminosity,
	allow us to use \nr simulations from \numNRcodes different codes
	as a homogeneous calibration set.
	This study of systematic fits to combined \nr and \lmr data,
	including higher modes,
	also paves the way for improved \imr waveform models.

	\vspace{\baselineskip}
	\centering 
	           published journal version: \\[0.5\baselineskip]
	           Phsyical Review D \textbf{96}, 024006 (2017) \\[0.5\baselineskip]
	           \href{https://doi.org/10.1103/PhysRevD.96.024006}{doi:10.1103/PhysRevD.96.024006}\\
	           \vspace{\baselineskip}
\end{abstract}

\preprint{\dcc}
\date{14 July 2017 (\textit{this version}),
      30 December 2016 (\textit{original version}) \\
..LIGO document number: \dcc}

\maketitle

\section{Introduction}

With Advanced LIGO's~\cite{TheLIGOScientific:2014jea,TheLIGOScientific:2016agk} first detections~\cite{Abbott:2016blz,Abbott:2016nmj,TheLIGOScientific:2016pea},
\bbh coalescences have become objects of observational astronomy.
The peak rate at which \bbh[s] radiate \gw energy
makes them the most luminous known phenomena in the Universe.
The source of the first \gw detection \event has been inferred to be consistent with two \bh[s]
of $\MTWOSCOMPACTNOCALIB\,\Msun$ and $\MONESCOMPACTNOCALIB\,\Msun$
inspiraling, merging and ringing down as described by \gr.
Its emission of \gw energy reached,
for a small fraction of a second,
a peak rate of $\LPEAKCOMPACT{}$~erg/s,
equivalent to $\LPEAKSOLCOMPACT\,\Msun\,c^2/\mathrm{s}$~\cite{Abbott:2016blz,TheLIGOScientific:2016wfe,TheLIGOScientific:2016src}.

Though this \textit{\peaklum}, $\Lpeak$, is not electromagnetic, but gravitational,
we can compare its numerical value to the photon luminosity of other astrophysical sources
to illustrate its scale:
\event at its peak emitted as much power as \LPSOLRELATIVE suns,
\LPMWRELATIVE times more than all stars in the Milky Way,
and still \GWVSGRB times more than
the ultra-luminous gamma-ray burst GRB\,110918A~\cite{Frederiks:2013cga}.\footnote{
Assuming $\Lsun=\LSUNVALUE$erg/s,
$L_{\mathrm{MW}}=\LMW$
and the GRB's estimated peak isotropic equivalent luminosity of \GRBLum~\cite{Frederiks:2013cga}.
}

The \peaklums for LIGO's first BBH events were inferred using a fit~\cite{T1600018} to data from numerical relativity (\nr) simulations,
which we will improve upon in this paper through an enhanced fitting method and a significantly larger calibration data set.
Source parameters of \gw events are determined through Bayesian inference~\cite{Aasi:2013jjl,Veitch:2014wba,TheLIGOScientific:2016wfe,TheLIGOScientific:2016pea,Abbott:2016izl},
comparing LIGO data with waveform models,
which are approximate maps between the masses and spins of the binary components and
the \gw signal. 
As of the \aligo O1 run, the state-of-the-art \nr-calibrated \bbh waveform models were PhenomPv2~\cite{Hannam:2013oca,T1500602,PhenomPv2Paper}
(a precessing model based on the aligned-spin PhenomD~\cite{Husa:2015iqa,Khan:2015jqa})
and SEOBNRv2/3~\cite{Pan:2013rra,Purrer:2014fza,Purrer:2015tud,Abbott:2016izl,Babak:2016tgq}.
A detailed recent study~\cite{Abbott:2016wiq} has found these models to be at least sufficiently accurate
in the parameter region corresponding to the first detection.

\addtolength{\parskip}{0.5\baselineskip}

The primary products of this inference are multidimensional sample chains
that approximate posterior probability density functions (PDFs) for the intrinsic and extrinsic \bbh parameters.
Subsequently, such a sampled PDF can be used to infer other quantities,
typically obtained through fitting formulas calibrated to \nr.
Examples include final-state properties~\cite{Healy:2014yta,Husa:2015iqa,Healy:2016lce,Hofmann:2016yih,Jimenez-Forteza:2016oae}:
the final spin and final mass of the merger remnant,
a single Kerr \bh,
which also yield the total radiated energy.
In fact, full \imr waveform models include such final-state \nr fits to describe the ringdown phase,
but due to practical implementation details and for greater flexibility in using updated fits,
the values reported in Refs.~\cite{Abbott:2016blz,Abbott:2016nmj,TheLIGOScientific:2016pea,Abbott:2016izl}
come from stand-alone fitting formulas evaluated on posterior PDFs.

The same approach is used for inference of the \gw \peaklum.
However, a robust $\Lpeak$ model for general \bbh configurations was not available in the literature prior to O1.
An early phenomenological formula~\cite{Baker:2008mj} is limited to nonspinning \bbh[s],
and thus a new fit~\cite{T1600018} had to be developed.
To accurately capture the luminosities of the \nr calibration set,
it also took into account contributions from subdominant harmonics not included in most current waveform models.

In this paper, we present an improved version of that model fit for the \gw \peaklum
from the merger of more general \bbh systems,
including spins on both binary components.
Still, we concentrate on cases where the spin of each \bh is aligned with the system's total angular momentum,
using the dimensionless components
\mbox{$\chi_i=S_i/m_i^2=\tfrac{\vec{S}_i}{m_i^2}\cdot\tfrac{\vec{L}}{|\vec{L}|}$}
of the spins $\vec{S}_i$ projected onto the orbital angular momentum $\vec{L}$.
We use the hierarchical data-driven approach introduced in Ref.~\cite{Jimenez-Forteza:2016oae}
to develop a three-dimensional ansatz
and fit it to a total of \NRcount simulations from \numNRcodes separate \nr codes,
including more subdominant modes than before,
and to independent numerical results for \lmrs
obtained with the perturbative scheme of Refs.~\cite{Nagar:2006xv,Bernuzzi:2011aj,Harms:2014dqa}.
This addition is essential in producing a well-constrained fit at very unequal masses
where \nr coverage is sparse or nonexistent.

Notably, the \gw \peaklum $\Lpeak$ does not depend on the total mass of a \bbh system:
luminosity generally scales with emitted energy over emission time scale,
\mbox{$L\sim\Erad/\Delta t$}.
But for a \bbh,
both the total radiated energy $\Erad$
and the characteristic merger time scale $\Delta t$
are proportional to the total mass,
so that $\Lpeak$ is independent of it.
Hence, the \gw \peaklums even of \smbh binaries,
observable by eLISA-like missions~\cite{AmaroSeoane:2012km,Seoane:2013qna,Klein:2015hvg}
or by pulsar timing arrays (PTAs,~\cite{Rajagopal:1994zj,Sesana:2008xk,Sesana:2010ac}),
are similar to those of stellar-mass \bbh[s].
The results of this paper will be applicable to such systems as well.

Besides using $\Lpeak$ to compare the energetics of \gw[s] and other astrophysical events,
one can also consider its relevance for the effect of \bbh coalescences on their immediate surroundings.
The influence of \smbh mergers on circumbinary accretion disks
(see Ref.~\cite{Schnittman:2013qxa} and references therein)
is determined mostly by the integrated radiated energy of the late-inspiral and merger phase,
though the authors of Refs.~\cite{Kocsis:2008aa,Li:2012dta} suggested weak prompt electromagnetic counterparts
sensitive to $\Lpeak$ and $L_{\mathrm{GW}}(t)$.
For stellar-mass \bbh[s], any significant interaction with surrounding material or fields is highly speculative --
see e.g. the references in Sec.~4 of Ref.~\cite{Connaughton:2016umz}.
Still it is conceivable that an accurate $\Lpeak$ model could be useful in constraining exotic models.

Turning $\Lpeak$ into an independent \gw observable
through direct signal reconstruction~\cite{Cornish:2014kda,Klimenko:2015ypf,Lynch:2015yin,Becsy:2016ofp,TheLIGOScientific:2016uux}
will require improved detector sensitivity and calibration accuracy,
so that the peak strain can be measured to high absolute accuracy
and that degeneracies between the distance estimate and other parameters are significantly reduced.
Currently, \nr-calibrated fits are the only accurate method to infer \peaklums from \gw observations.

Another motivation for this improved fit is its role as a test case of the general
fitting method from Ref.~\cite{Jimenez-Forteza:2016oae} for quantities that require
accurate treatment of the higher modes from \nr simulations,
and of combining \nr and perturbative \lmr results.
In these aspects, the present study is a preparation for the development of improved \imr waveform models.

In this paper,
we use geometric units with \mbox{$G=1$}, \mbox{$c=1$} and unit total mass $M$,
so that luminosity is a dimensionless quantity,
corresponding to $1M$ of energy radiated per $1M$ of time.
The conversion factor to Watt is $c^5/G$,
and another factor $10^7$ for the usual astronomical unit of erg/s.
%
We will first review the catalog of \nr simulations of \bbh mergers and perturbative \lmr data
used to construct our model in \autoref{sec:data}.
We discuss the challenges in combining \nr data from different simulation codes,
the steps taken to process the different sets into a single, effectively homogeneous data set,
and how this set is augmented with independent results for \lmr systems.
Given this data set, we discuss the construction and validation of our model fit for \peaklum
in Secs.~\ref{sec:fit} and~\ref{sec:results}.
We also compare our new fit with the previous result of Ref.~\cite{T1600018},
calibrated to a smaller \nr data set,
that was used during O1~\cite{Abbott:2016blz,Abbott:2016nmj,TheLIGOScientific:2016pea,TheLIGOScientific:2016wfe,Abbott:2016izl};
and to another independent, recently published fit~\cite{Healy:2016lce}.
The \hyperref[sec:appendix-nr]{Appendix} includes more details on investigations of the \nr data.

\addtolength{\parskip}{-0.5\baselineskip}
\clearpage

\section{Input data}
\label{sec:data}

\subsection{Numerical relativity data sets}
\label{sec:data-NR}

\begin{figure}[t]
 \includegraphics[width=\columnwidth]{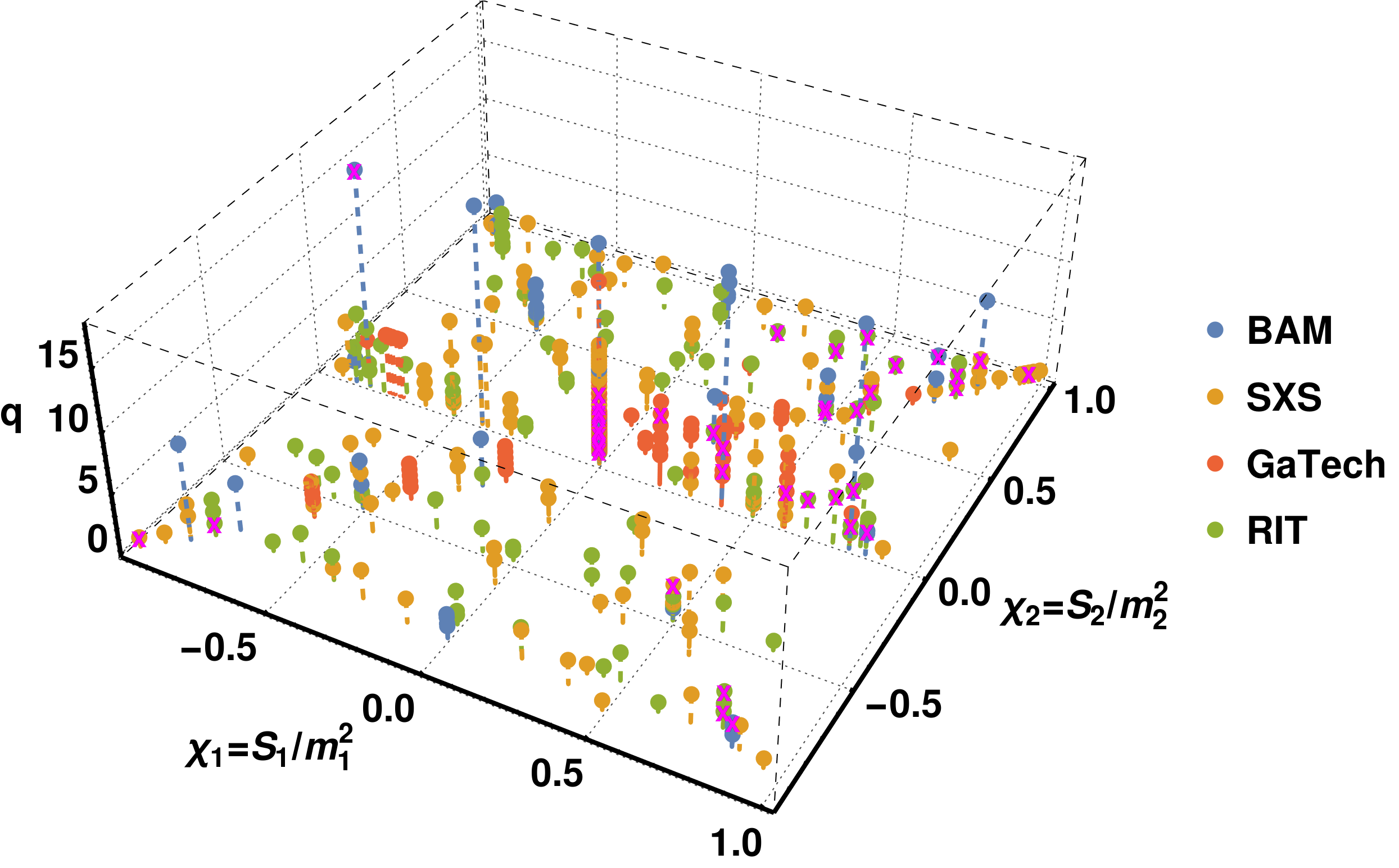}
 \caption{
  \label{fig:NR_eta_chi1_chi2}
  Parameter-space coverage of the combined \nr data set from \NRcodesInOrderAnd,
  shown against the individual \bh spins and the mass ratio \mbox{$q=m_1/m_2$} of the system.
  Simulations not used in the fit (see Table~\ref{tbl:outliers}) are marked with magenta crosses.
 }
\end{figure}

\begin{figure}[t]
 \includegraphics[width=\columnwidth]{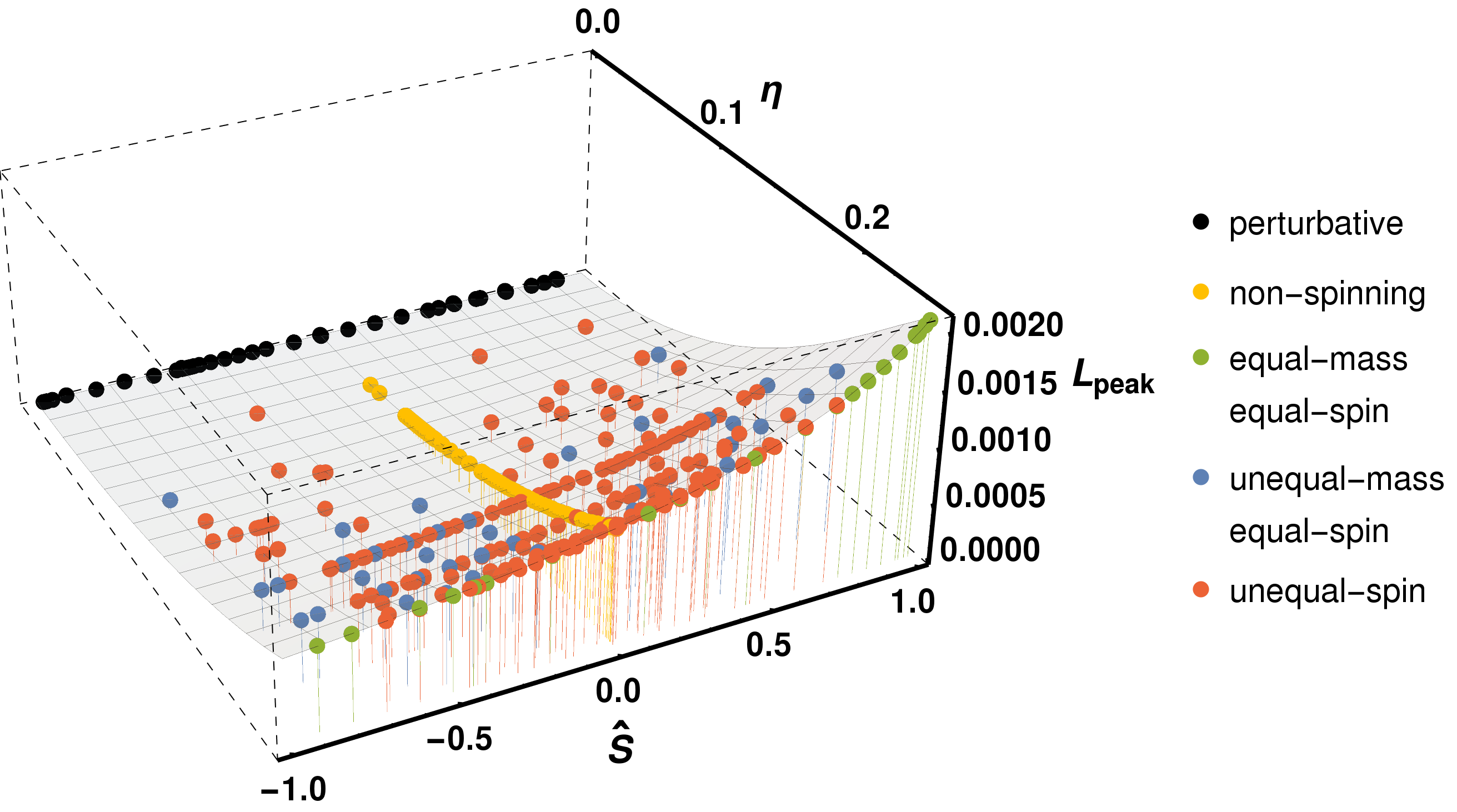}
 \includegraphics[width=\columnwidth]{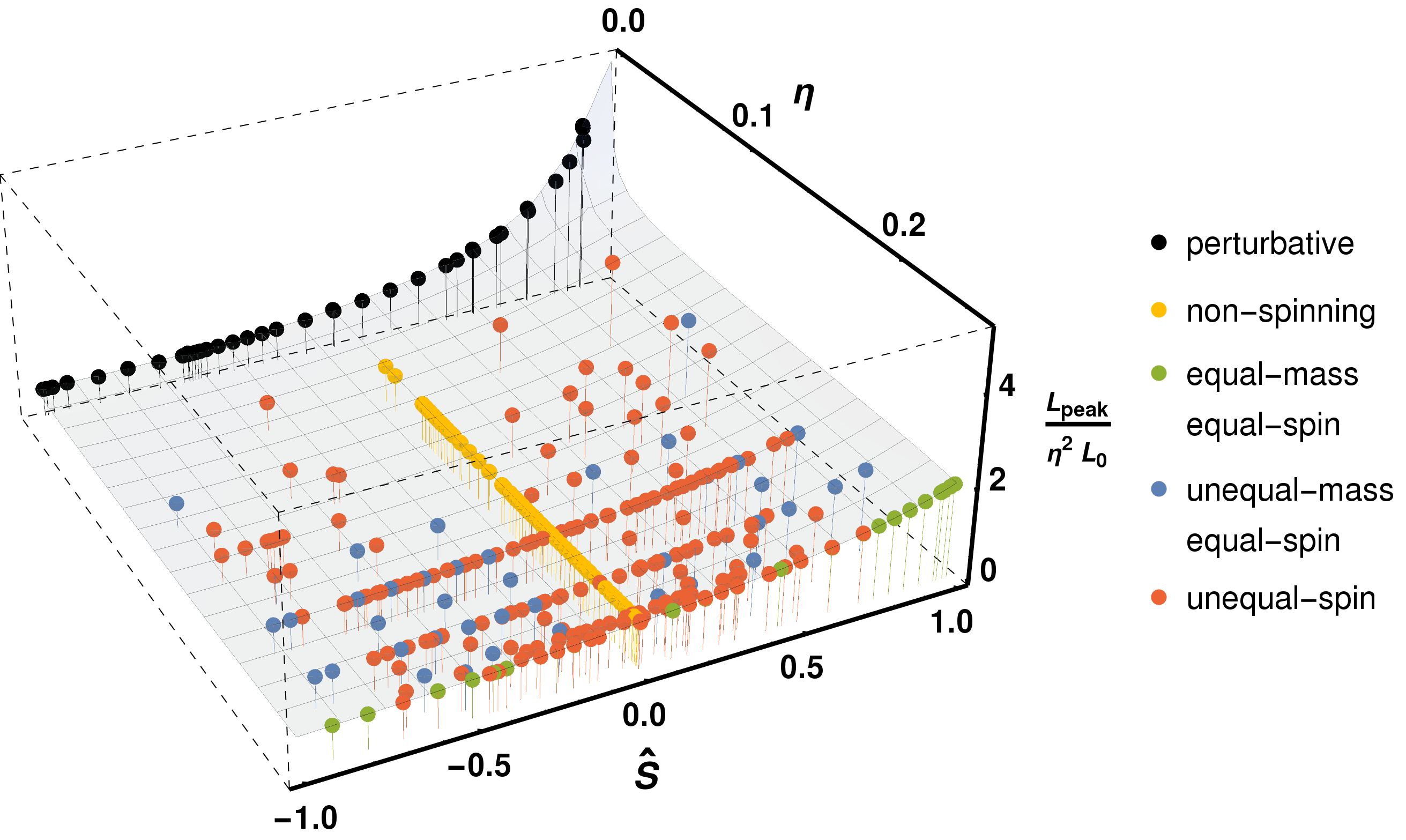}
 \caption{
  \label{fig:NR_eta_S_Lpeak}
  Combined data set over the two-dimensional space spanned by symmetric mass ratio $\eta$
  and effective spin $\myS$, defined in \autoref{eq:Seff}.
  Top panel: \peaklum $\Lpeak$.
  Lower panel: rescaled as \mbox{$\Lpeak/\eta^2 \Lo$}.
  Subsets used in the various steps of \autoref{sec:fit}
  are highlighted by colors.
  The shaded surface is added here to guide the eye,
  but is in fact the 2D projection of the new fit developed in this paper.
 }
\end{figure}

We begin by considering \NRcountTotal nonprecessing \nr simulations from \numNRcodes sources,
with their coverage of the three-dimensional \bbh parameter space
illustrated in Figs.~\ref{fig:NR_eta_chi1_chi2} and~\ref{fig:NR_eta_S_Lpeak}:
\begin{enumerate}[(i)]
 \item \NRcountBAM simulations performed by the authors
                   with the \BAM code~\cite{Bruegmann:2006at,Husa:2007hp},
                   including those first used in Refs.~\cite{Husa:2015iqa,Jimenez-Forteza:2016oae}.
 \item \NRcountSXS simulations from the public SXS catalog~\cite{Mroue:2013xna,SXScatalog}
                   performed with the \SpeC~\cite{SpEC}.
 \item \NRcountGaT simulations from the public GaTech catalog~\cite{Jani:2016wkt,GaTechcatalog},
                   performed with the \MAYA code~\cite{Herrmann:2006ks,Vaishnav:2007nm,Healy:2009zm,Pekowsky:2013ska}.
 \item \NRcountRIT simulations~\cite{Healy:2014yta,Healy:2016lce,Healy:2017psd,RITcatalog}
                   with the \LAZEV code~\cite{Zlochower:2005bj},
                   labeled ``RIT'' in the following.
\end{enumerate}
\BAM, \MAYA and \LAZEV are finite difference codes
to solve the Baumgarte-Shapiro-Shibata-Nakamura formulation of the \gr initial value problem~\cite{Baumgarte:1998te}
with a singularity-avoiding slicing condition following the moving puncture approach~\cite{Campanelli:2005dd,Baker:2005vv},
whereas the simulations of the SXS Collaboration have been performed with the pseudospectral \SpeC code~\cite{SpEC}
which employs the generalized harmonic gauge (GHG) combined with black-hole excision.

We use mass and spin parameters of the component \bh[s] after equilibration and the initial burst of ``junk`` radiation.
To compute the luminosity for \BAM, SXS and GaTech simulations,
we begin with the Weyl scalar $\psi_4$ decomposed into its spin-two spherical harmonic multipoles,
\begin{equation}
 \label{eq:psi4}
 \psi_{\ell m}(t) \; = \; \frac{1}{r} \, \int_{\Omega} \, _{-2}\overline{Y}_{\ell m}(\theta,\phi)\,\psi_4(t,\theta,\phi) \, \mathrm{d}\Omega \;.
\end{equation}
For SXS $\psi_4$ data, we have applied corrections for center-of-mass drifts~\cite{Boyle:2013nka,Boyle:2014ioa,Boyle:2015nqa},
which remove some unphysical oscillations in higher modes.
From these spherical harmonic multipoles,
we calculate the \gw strain-rate multipoles $\dot{h}_{\ell m}(t)$
via the \ffi method described in Ref.~\cite{Reisswig:2010di}. 
We then compute the \peaklum according to
\begin{align}
 \label{eq:Lpeaksum}
 \Lpeak = \mathop{\mathrm{max}}_{t} \; \lim_{r\rightarrow \infty} \frac{r^2}{16\,\pi} \sum_{\ell=2}^{\lmax}\sum_{m=-\ell}^{+\ell} \left|\,\dot{h}_{\ell m}(t)\,\right|^2 \,,
\end{align}
truncating the sum over $\ell$ at \mbox{$\lmax=6$}.
For RIT simulations, we use directly the \peaklum values as given in Ref.~\cite{Healy:2016lce},
which again include all modes up to \mbox{$\lmax=6$}.

\nr simulation results always have finite accuracy,
and the post-processing from the initial $\psi_4(t,\theta,\phi)$ to the final product $\Lpeak$
can lead to additional sources of inaccuracies that could substantially affect
the accuracy of individual data points and of any \nr-calibrated fits.
As we aim to fit small subdominant effects, such as those of unequal spins,
possible error sources must be carefully analyzed.
Thus, we have considered the impact of the following effects on $\Lpeak$,
with details on each aspect given in the \hyperref[sec:appendix-nr]{Appendix}:

\begin{enumerate}[(i)]
 \item Conversion between strain $h(t)$, $\psi_{4}(t)$ and luminosity $L(t)$:
       The \ffi approach is known to be accurate at the $1\%$ level or better.~\cite{Reisswig:2010di}
       We have also tested its validity by verifying that it agrees with a newly developed alternative technique
       to compute $h(t)$ and $L(t)$ from the time-domain $\psi_{4}(t)$.
 \item Extrapolation effects:
       Gauge invariability of the waveforms is only well defined for an observer at null infinity.
       We have extrapolated $\Lpeak$,
       as computed from waveforms available at a range of finite extraction radii,
       to infinity and estimated the extrapolation uncertainty.
 \item Finite resolution:
       Convergence tests are only available for a small subset of \nr configurations,
       with estimates indicating that errors due to finite grid resolution
       are a nondominant but non-negligible contribution to the total error budget.
 \item Peak accuracy and discreteness:
       The peak in luminosity can be quite steep,
       but we have verified that with a sampling of 0.1M in time
       the difference between two points next to the peak is only on the order of 0.01--0.2\%.
 \item Mode selection:
       While the \mbox{$\ell=\vert m\vert = 2$} mode has the largest peak amplitude for all mass ratios,
       the importance of other spherical harmonics monotonically increases toward the \emrl.
       We can thus bound higher-mode contributions by comparison with the \lmr results,
       where neglecting modes with \mbox{$\ell>8$} incurs an error below 1\% even for mass ratios of $10^5$
       (see below in Sec.~\ref{sec:data-extreme}).
       For the available \nr waveforms,
       we determine that it is generally sufficient to consider modes up to \mbox{$\lmax=6$},
       with contributions of the \mbox{$\ell=7,8$} modes rising to only 1\% for mass ratio 18 nonspinning cases.
\end{enumerate}

We remove \NRcountOutliers cases from the initial catalog for reasons as discussed in the \hyperref[sec:appendix-nr]{Appendix},
e.g. because they are inconsistent with equivalent or nearby configurations from the same or other codes.
Thus, we perform our fit with a final set of \NRcount \nr results.

\vspace{\baselineskip}

\subsection{Perturbative \lmr data}
\label{sec:data-extreme}

As the computational cost of \nr simulations increases rapidly for unequal mass ratios,
no data for \bbh systems with mass ratios \mbox{$q>\NRmaxq$} are currently available from any of the simulation codes discussed above.
( \mbox{$q=m_1/m_2$} with the convention \mbox{$m_1>m_2$}.)
However, constraining fits to some known behavior in or close to the \emrl
is essential in ensuring a sane extrapolation towards that limit,
and also to reduce uncertainty in the intermediate-mass-ratio region where there is some \nr coverage,
but it is still very sparse.

For the final-state quantities studied in Ref.~\cite{Jimenez-Forteza:2016oae},
we used analytical expressions from Ref.~\cite{Bardeen:1972fi}
for the limiting case of a test particle orbiting a Kerr black hole.
For the peak luminosity, it is known~\cite{Teukolsky:1974yv,Fujita:2014eta}
that the leading-order term as \mbox{$\eta\rightarrow0$}
must be \mbox{$\Lpeak \propto \eta^2$},
with the symmetric mass ratio \mbox{$\eta = (m_1 m_2)/(m_1 + m_2)^2  = q/(1+q)^2$}.
However, no fully analytical results for the spin dependence in the \emrl exist.
Instead, here we constrain our fit by numerical results for finite, but very \wordforhighq mass ratios.

The simulation method for \bbh mergers in the test-mass (large-mass-ratio) limit
developed in Refs.~\cite{Nagar:2006xv,Bernuzzi:2010ty,Harms:2014dqa} combines a semianalytical description of the dynamics
with a time-domain numerical approach for computing the multipolar waveform based on \bh perturbation theory.
The small \bh['s] dynamics are prescribed using the \eob test-mass dynamics,
i.e. the conservative (geodesic) motion augmented with a linear-in-$\eta$ radiation reaction expression~\cite{Nagar:2006xv,Damour:2007xr}. 
The latter is built from the factorized and resummed circularized waveform
introduced in Ref.~\cite{Damour:2008gu} (and Ref.~\cite{Pan:2010hz} for spin) 
and uses \PN information up to 5.5PN
(see also Refs.~\cite{Fujita:2012cm,Nagar:2016ayt} for extensions up to 22PN).
Waveforms are calculated by solving either the \RWZ 1+1 equations (nonspinning case)
or the Teukolsky 2+1 equation (spinning case). 
Those equations are solved in the time-domain using hyperboloidal coordinates 
to extract the radiation unambiguously at scri (null infinity)~\cite{Bernuzzi:2010xj,Bernuzzi:2011aj,Harms:2014dqa}.

The method has been extensively applied for developing and assessing the
quality of \eob waveforms~\cite{Bernuzzi:2011aj,Harms:2014dqa},
informing the \eob model in the test-mass limit~\cite{Damour:2012ky},
quasinormal mode excitation~\cite{Bernuzzi:2010ty,Harms:2014dqa},
and computing gravitational recoils~\cite{Nagar:2013sga,Nagar:2014kha}.

The \lmr waveforms employed here were produced 
in Refs.~\cite{Bernuzzi:2011aj} and~\cite{Harms:2014dqa} (RWZ and Teukolsky data, respectively).
These waveforms are approximate because $O(\eta)$ effects
are not taken into account in the conservative dynamics.
However, the consistency of the method was  proven in the nonspinning case
by showing that, for \mbox{$\eta\to0$}, the analytical mechanical flux assumed for the small \bh['s] motion
agrees with the numerical \gw fluxes to a few percent up to the last stable orbits~\cite{Bernuzzi:2011aj}.
The same check has been performed for the spinning case
where, instead, significant deviations were found where the spin of the central \bh is large ($\chi_1\gtrsim0.7$)
and aligned with the orbital angular momentum~\cite{Harms:2014dqa}. 
The discrepancy originates from poor performance of the straight, 5PN-accurate, 
\eob-resummed analytical multipolar waveform,
from which the radiation reaction force is built, for large spins~\cite{Pan:2010hz}.
An iterative method to produce consistent $O(\eta)$ spinning waveforms has been proposed~\cite{Harms:2014dqa};
and two such waveforms at \mbox{$\chi_1=\pm0.9$} are available for consistency tests.
The method proposed in Ref.~\cite{Harms:2014dqa} is very expensive,
since several iterations are needed to find good consistency between the radiation reaction
that is used to drive the dynamics and the waveform.
The authors of Ref.~\cite{Nagar:2016ayt} proposed an additional factorization
and resummation of the residual waveform amplitudes of Ref.~\cite{Pan:2010hz}
that delivers a more accurate analytical waveform amplitude up to the last stable orbit (or even the light-ring)
when the \bh dimensionless spin tends to 1.
The additional resummation discussed in Ref.~\cite{Nagar:2016ayt} (or minimal variations of it),
once incorporated in the radiation reaction,
is expected to strongly improve the self-consistency of the Teukolsky waveforms computed as in Ref.~\cite{Harms:2014dqa}
without need for the iteration procedure.

We use only the Teukolsky results at \mbox{$q=10^3$} (\EMRIcountqonek data points)
and the \RWZ results at \mbox{$q=10^4$} and \mbox{$q=10^5$} (\EMRIcountqtenkWords data points each),
as the \RWZ at \mbox{$q=10^3$} are expected to be less accurate,
and indeed their luminosities deviate at negative $\chi_1$.
In \autoref{fig:NR_eta_S_Lpeak} we compare the qualitative behavior of \peaklums from \nr and perturbative data,
and the spin dependence is analyzed in \autoref{sec:fit-emri}.

\begin{figure}[t]
 \includegraphics[width=\columnwidth]{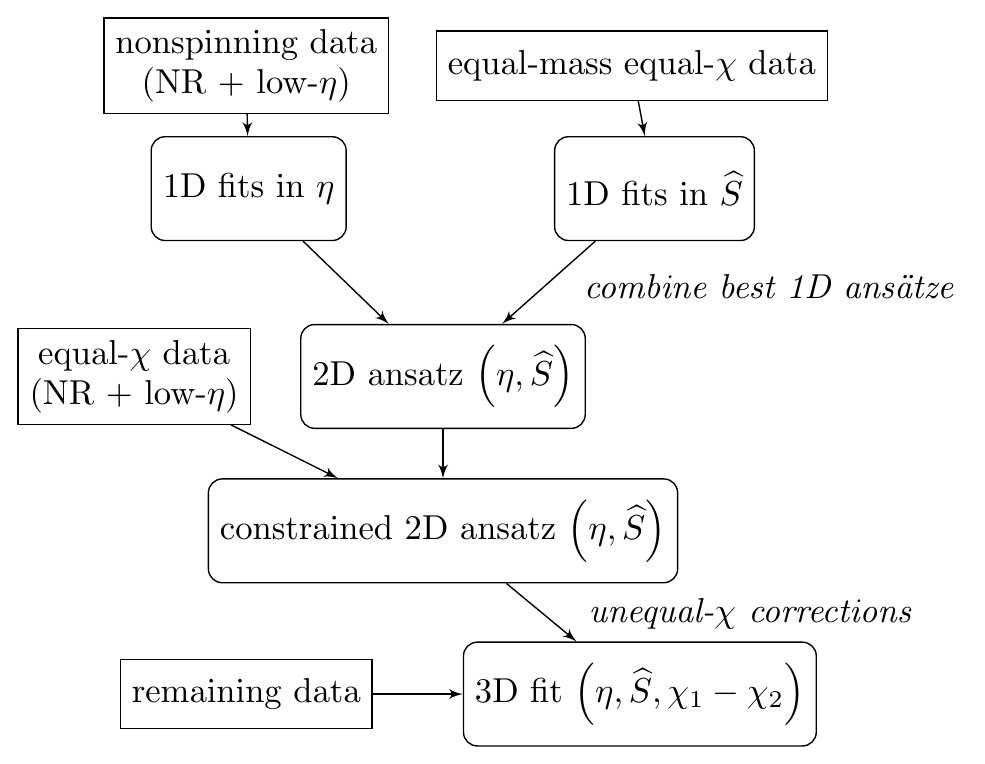}
 \caption{
  \label{fig:flowchart}
  Flow chart of the hierarchical step-by-step construction leading to the final 3D ansatz,
  as introduced in Ref.~\cite{Jimenez-Forteza:2016oae},
  but adjusted for the different handling of \lmr (low-$\eta$) information.
 }
\end{figure}

\vspace{2\baselineskip}

\section{Constructing the phenomenological fit}
\label{sec:fit}

\vspace{\baselineskip}

We apply the hierarchical modeling scheme for the three-dimensional nonprecessing \bbh parameter space
that was introduced in Ref.~\cite{Jimenez-Forteza:2016oae} and is summarized in~\autoref{fig:flowchart}.
The general idea is to construct a fit ansatz that matches the structure actually seen in the data set,
and to model effects in order of their importance:
first fit well-constrained subspaces as functions of the dominant parameters,
and then add subdominant effects only to the degree that they are supported by the data.

The parameter-space dimensionality is the same for \peaklum as for final spin or radiated energy:
just like the final (dimensionless) spin,
the \peaklum is manifestly independent of the total mass of the system,
while for radiated energy the total mass is still only an overall scale factor.
Hence, for nonprecessing quasicircular \bbh[s],
this leaves a three-dimensional parameter space:
the mass ratio and two spin parameters $\chi_1$ and $\chi_2$.

As expected, and obvious visually in \autoref{fig:NR_eta_S_Lpeak},
one principal direction of curvature in the data set is given
by the mass ratio, equivalently expressed as $q$ or the symmetric $\eta$.
We can then perform a three-dimensional (3D) hierarchical fit by changing the spin parametrization
from the two component spins \mbox{$\chi_i=S_i/m_i^2$}
to a dominant symmetric component
\begin{equation}
 \label{eq:Seff}
 \Seff   = (m_1^2 \, \chi_1 + m_2^2 \, \chi_2)/(m_1^2 + m_2^2)
\end{equation}
and a subdominant antisymmetric component \mbox{$\chidiff = \chi_1 - \chi_2$}.
This is the same choice of effective spin parameter as in Refs.~\cite{Husa:2015iqa,Jimenez-Forteza:2016oae}.
Tests in Appendix C of Ref.~\cite{Jimenez-Forteza:2016oae}
show that the fitting method is robust under changing to different parametrizations
like the $\chieff$ parameter previously used in Ref.~\cite{T1600018}.

We perform our fits not on the \peaklum $\Lpeak$ itself,
but on the rescaled quantity \mbox{$\Lscaled=\Lpeak/\left(\eta^2 \Lo\right)$}.
This removes the expected dominant $\eta^2$ dependence
(known analytically for \lmrs~\cite{Teukolsky:1974yv,Fujita:2014eta}
and found as the dominant term in previous fits~\cite{Baker:2008mj,T1600018})
to make small subdominant corrections easier to identify.
Here we have also scaled out the equal-mass, zero-spin value \mbox{$\Lo = \Lpeak\left(\eta=0.25,\,\chi_1=\chi_2=0\right)\,/\,0.25^2$}
to get typical values of order unity.
We use $\Lo\approx\NRLovalue$, the average of the SXS, GaTech and RIT results at this configuration.
(These three simulations agree within \NRLostdevrel.)

All fits are performed with Mathematica's \texttt{NonlinearModelFit} function.
To avoid overfitting, our model selection is guided by
the Akaike and Bayesian information criteria (AICc and BIC,~\cite{Akaike:1974,Schwarz:1978}),
which not only help to choose between fits based on the overall goodness of fit,
as measured e.g. by the \rmse,
but also penalize excessively high numbers of free coefficients.
The AICc is defined as
\begin{equation}
 \label{eq:AICc}
 \AICc = -2 \ln \maxlnL + 2\Ncoef + \frac{2\Ncoef(\Ncoef+1)}{\Ndata-\Ncoef-1} \,,
\end{equation}
with the maximum log-likelihood $\maxlnL$ (assumed Gaussian).
This is the standard AIC proposed by Akaike~\cite{Akaike:1974} plus a correction for low $\Ndata$.
Schwarz' alternative criterion~\cite{Schwarz:1978}
\begin{equation}
 \label{eq:BIC}
 \BIC = -2 \ln \maxlnL + \Ncoef \ln(\Ndata) \,,
\end{equation}
despite its name, generally cannot be understood directly as a Bayesian evidence.
For specific advantages and disadvantages of these two criteria,
their mathematical and philosophical basics and other alternatives
see e.g. Ref.~\cite{Liddle:2007fy} and references therein.
For both, the model with the \emph{lowest} value is preferred.
The BIC tends to impose a slightly stronger penalty on extra parameters than the AICc,
and we use it as a default ranking of fits,
though in practice we do not find disagreements between the two criteria.

\vspace{\baselineskip}

\subsection{One-dimensional nonspinning fit}
\label{sec:fit-1d-eta}

First, we analyze \datacountNS nonspinning cases,
including \NRcountNS \nr simulations as well as the nonspinning \lmr data points.
As in Ref.~\cite{Jimenez-Forteza:2016oae}, we consider several ansatz choices for the one-dimensional function $\Lscaled(\eta)$:
polynomials up to seventh order, denoted as $\poly{m}$,
as well as rational functions, denoted as \mbox{$\rational{m}{k}$} for polynomial orders $m$ and $k$ in the numerator and denominator, respectively.
We construct the latter as \pade approximants from an initial polynomial fit
to simplify the handling of initial values in the fitting algorithm.

\begin{figure}[t]
 \includegraphics[width=\columnwidth]{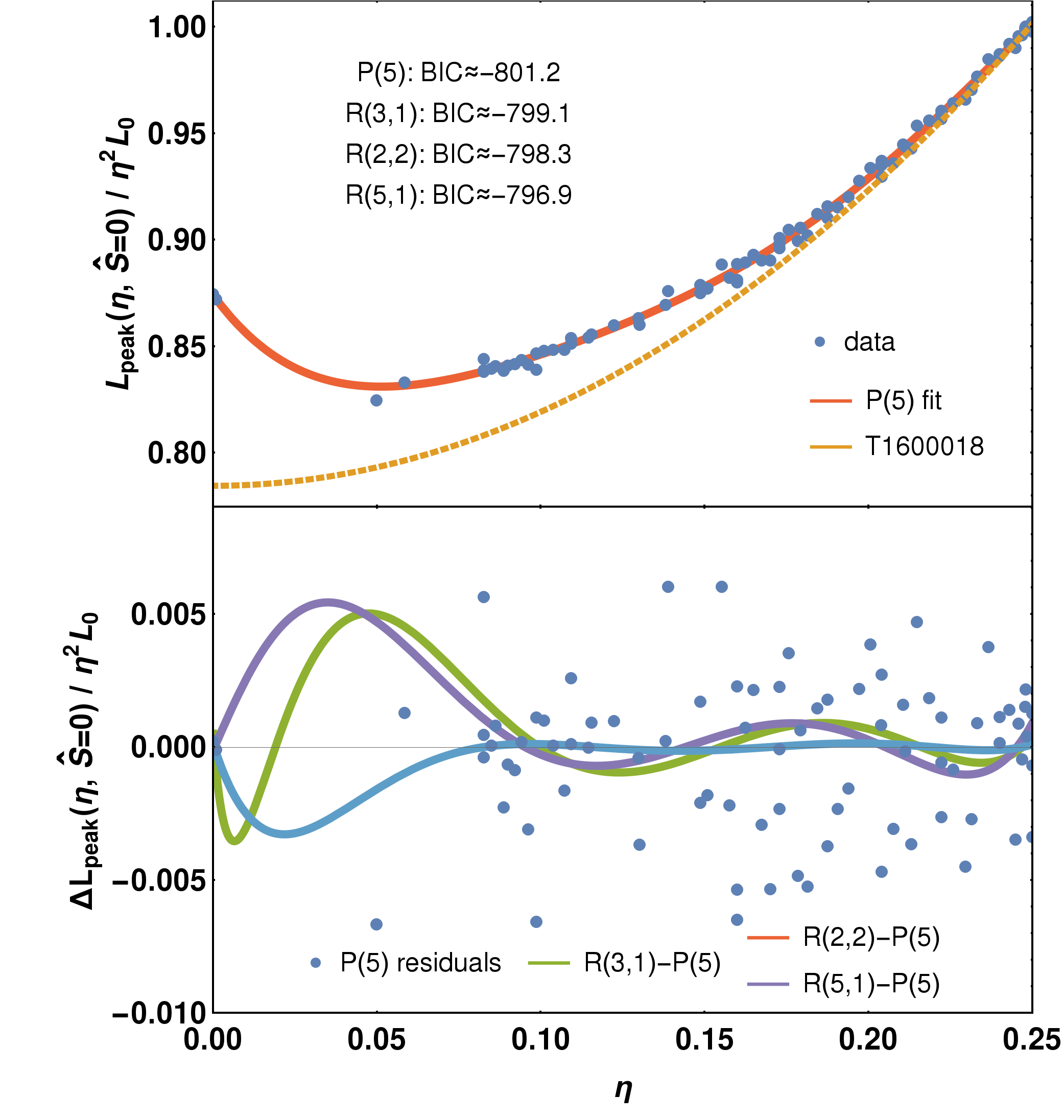}
 \caption{
  \label{fig:etafit}
  One-dimensional fits of the rescaled nonspinning \peaklum \mbox{$\Lofeta$}.
  Top panel: The preferred fifth-order polynomial, see \autoref{eq:etaansatz},
  and comparison with the previous fit from Ref.~\cite{T1600018}.
  Lower panel: Residuals of this fit (points)
  and differences from the three next-highest-ranking fits in terms of BIC (lines).
 }
 \vspace{2\baselineskip}
 \begin{tabular}{lrrr}\hline\hline
  &$ \text{Estimate} $&$ \text{Standard error} $&$ \text{Relative error [$\%$]} $\\\hline$
 a_0 $&$      0.8742            $&$   0.0010            $&$  0.1 $\\$
 a_1 $&$     -2.11\hphantom{34} $&$   0.28\hphantom{34} $&$ 13.3 $\\$
 a_2 $&$    35.2\hphantom{234}  $&$   7.0\hphantom{234} $&$ 19.9 $\\$
 a_3 $&$  -245\hphantom{.1234}  $&$  64\hphantom{.1234} $&$ 26.0 $\\$
 a_4 $&$   877\hphantom{.1234}  $&$ 248\hphantom{.1234} $&$ 28.3 $\\$
 a_5 $&$ -1173\hphantom{.1234}  $&$ 354\hphantom{.1234} $&$ 30.2 $\\
\hline\hline\end{tabular}
 \vspace{\baselineskip}
 \captionof{table}{
  \label{tbl:eta_fit_coeffs}
  Fit coefficients for the one-dimensional non-spinning \mbox{$\Lofeta$} fit
  over \datacountNS data points,
  along with their uncertainties (standard errors)
  and relative errors (std.err./estimate).
 }
\end{figure}

With the dominant $\eta^2$ dependence already scaled out,
fitting the higher-order corrections allows us to achieve subpercent accuracy,
though the additional fit coefficients are not very tightly constrained.
The top-ranked fit both by BIC and AICc (with marginally significant differences)
is a fifth-order polynomial
\begin{equation}
 \label{eq:etaansatz}
 \Lofeta =
 a_5 \eta ^5+a_4 \eta ^4+a_3 \eta ^3+a_2 \eta ^2+a_1 \eta +a_0
\end{equation}
with its fit coefficients and their uncertainties given in Table~\ref{tbl:eta_fit_coeffs}.

Figure~\ref{fig:etafit} shows this fit, its residuals
and comparisons with both the previous fit from Ref.~\cite{T1600018} (``T1600018'')
and the next-highest-ranked alternatives.
These next-best alternatives are all rational functions,
with the next-simpler polynomial P(4) disfavored by 7 in BIC and 20\% in \rmse
and the next-higher-order P(6) marginally disfavored by 4 in BIC with almost identical \rmse.
We find a clear upwards correction over the T1600018 result at low $\eta$,
and differences between high-ranking fits that are much smaller than this correction or the typical residuals.
In the data-less region between the lowest-$\eta$ \nr case (\mbox{$q=\NRmaxq$}) and the perturbative results,
differences between the highest-ranking fits are larger,
but still at most at the same level as the typical fit residuals at higher $\eta$,
corresponding to relative errors below 0.6\%.
As another comparison, refitting the simple \mbox{$\Lpeak(\eta)=a_2\eta^2+a_4\eta^4$} ansatz that we used in Ref.~\cite{T1600018}
(which in $\Lscaled$ corresponds to just \mbox{$\const+\eta^2$})
is disfavored by over 280 in BIC over this data set,
and has a four times higher \rmse.

All highly ranked fits agree that the \nr data
cannot be connected to the \lmr regime with a simple monotonic function.
This behavior might seem surprising, but can be explained by studying the individual modes:
the observed behavior of the total \peaklum results from competing trends
of modes that either fall or rise towards \mbox{$\eta\rightarrow0$}.
(See Appendix~\ref{sec:appendix-nr-modesel} and \autoref{fig:lLumdependence} for details,
and Refs.~\cite{Berti:2007fi,Baker:2008mj,Kelly:2011bp,Pekowsky:2012sr,Bustillo:2015qty} for previous studies of higher-mode amplitudes.)
Also we recall that the full \mbox{$\Lpeak\oneDeta$} is of course monotonic
after the dominant $\eta^2$ term has been factored back in.

\vspace{\baselineskip}

\subsection{One-dimensional equal-mass-equal-spin fit}
\label{sec:fit-1d-spin}

\begin{figure}[t]
 \includegraphics[width=\columnwidth]{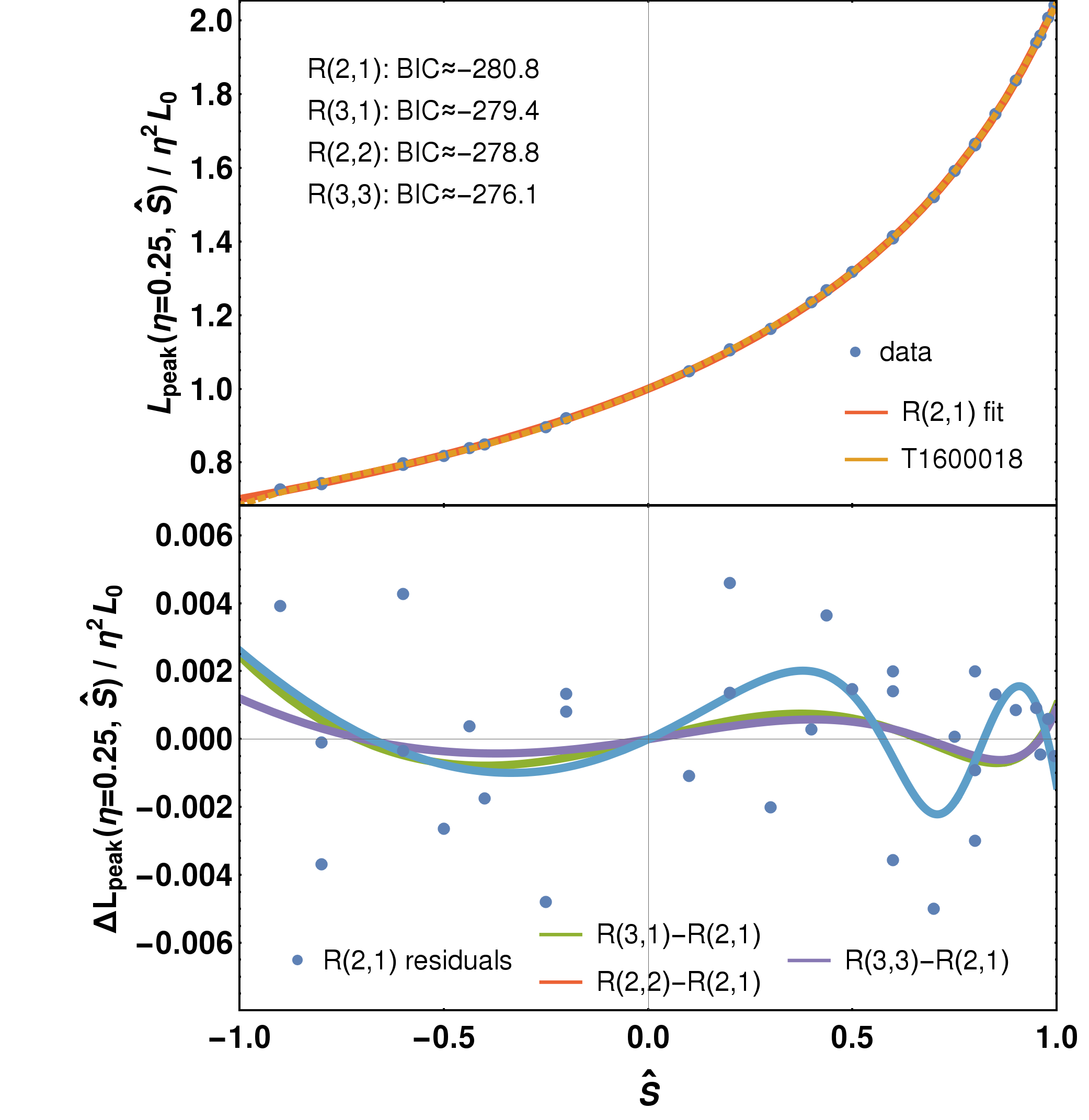}
 \caption{
  \label{fig:spinfit}
  One-dimensional fits of the rescaled \eqmeqS ($\chi_1=\chi_2$) \peaklum \mbox{$\LofS$}. \\
  Top panel: Best fit in terms of BIC,
  a rational function R(2,1), see \autoref{eq:Sansatz},
  and the almost indistinguishable P(5) from Ref.~\cite{T1600018}. \\
  Lower panel: Residuals of this fit (points)
  and differences from three next-best-ranked fits by BIC (lines).
 }
 \vspace{2\baselineskip}
 \begin{tabular}{lrrr}\hline\hline
  &$ \text{Estimate} $&$ \text{Standard error} $&$ \text{Relative error [$\%$]} $\\\hline$
 b_1 $&$  0.9800            $&$ 0.0023            $&$  0.2 $\\$
 b_2 $&$ -0.178\hphantom{4} $&$ 0.028\hphantom{4} $&$ 15.5 $\\$
 b_4 $&$  1.786\hphantom{4} $&$ 0.014\hphantom{4} $&$  0.6 $\\
\hline\hline\end{tabular}
 \vspace{\baselineskip}
 \captionof{table}{
  \label{tbl:S_fit_coeffs}
  Fit coefficients for the one-dimensional \eqmeqS \mbox{$\LofS$} fit
  over \datacountEqSqone data points.
 }
\end{figure}

Next, we consider \NRcountEqSqone equal-mass and equal-spin \nr simulations,
i.e. configurations with \mbox{$\eta=0.25$} and \mbox{$\chi_1=\chi_2\neq0$},
fitting the one-dimensional function $\LofS$.
We use a similar set of polynomial and rational \ansaetze,
with the intercept fixed by requiring consistency
with the $\eta$ fit in the nonspinning case,
\mbox{$\Lscaled\left(\eta=0.25,\,\myS=0\right)$}.
The curvature of this spin dependence at equal masses is relatively mild
and can be best fit by a three-coefficient rational function ansatz
\begin{equation}
 \label{eq:Sansatz}
 \LofS =
 \frac{0.107 b_2 \widehat{S}^2+0.465 b_1 \widehat{S}}{1-0.328 b_4 \widehat{S}}+1.00095 \,,
\end{equation}
with the numerical prefactors being due to constructing the ansatz as a \pade approximant
to simplify the handling of initial values in the fitting code.
This fit is marginally top-ranked by both AICc and BIC;
it is shown in \autoref{fig:spinfit}
and the coefficients $b_i$ are given in Table~\ref{tbl:S_fit_coeffs}.
Low-order rational functions are clearly preferred over polynomials,
with the P(5) we used in Ref.~\cite{T1600018} disfavored by +14 in BIC
and having 12\% higher \rmse,
and the simple R(2,1) ansatz is fully sufficient to describe the data
to similar subpercent accuracy as the nonspinning set.
Adding another term in either the numerator or denominator is possible,
but does not improve the statistics;
while adding too many terms tends to induce unconstrained coefficients
or singularities within the fitting region.

\vspace{\baselineskip}

\subsection{Spin dependence at \lmrs}
\label{sec:fit-emri}

\begin{figure}[t]
 \includegraphics[width=\columnwidth]{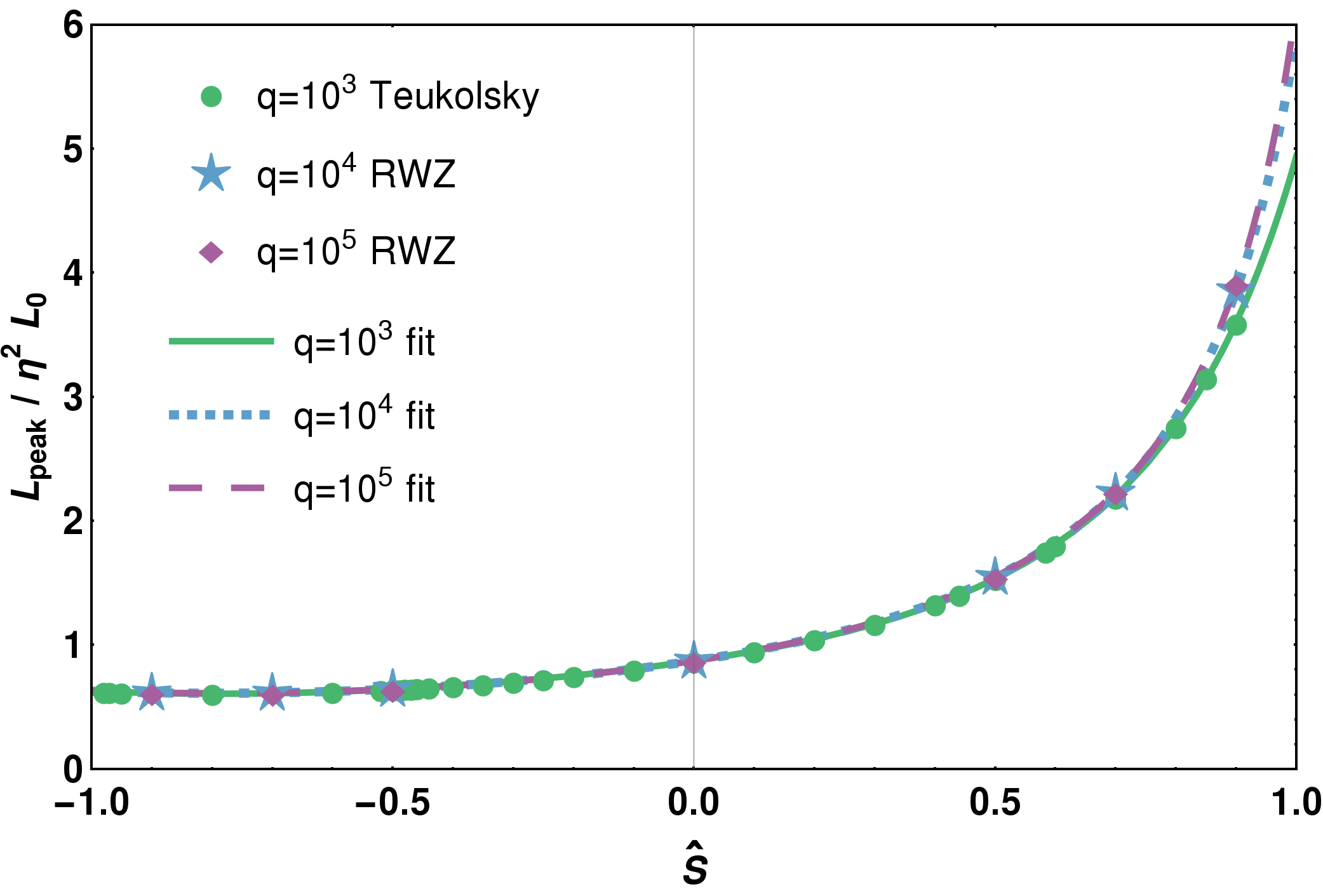}
 \caption{
  \label{fig:emridata}
  Numerical data from two perturbative codes
  (circles, stars and diamonds for mass ratios \mbox{$q=\{10^3,10^4,10^5\}$}),
  together with separate fits of the form \mbox{$\const + \rational{4}{1}$}
  at each $q$.
 }
\end{figure}

Analyzing the perturbative data at \lmrs,
we verify that the mass-ratio dependence is completely dominated by the leading-order $\eta^2$ scaling in this regime,
with the rescaled $\Lscaled$ equal to within 0.2\% for the three nonspinning data points at mass ratios
\mbox{$q\in\left\{10^3,10^4,10^5\right\}$}.

We treat the single-spin perturbative data
as equivalent to results with \mbox{$\chi_1=\chi_2$} which\footnote{
Here and in the following,
we always use ``equal-spin'' to refer to equal dimensionless spin components \mbox{$\chi_1=\chi_2$}.}
is easily accurate enough as
\mbox{$\myS(q=10^3,\,\chi_1=1.0,\,\chi_2=1.0)-\myS(10^3,1.0,0.0)\approx10^{-6}$}
and the spin-difference terms (to be fitted below in \autoref{sec:fit-3d})
are expected to be suppressed at least by $\eta^2$.

We find that the spin dependence at each of these mass-ratio steps is much steeper than for equal masses,
requiring a higher-order spin ansatz.
A spin term at least as complex as R(4,1) is clearly preferred over any lower-order alternatives,
with the \mbox{$q=10^3$} data yielding
\mbox{$\Delta\BIC\approx44$} and a factor $>2$ in \rmse in favor of R(4,1) against the R(2,1) found at equal masses,
and similar preference even for the two highest mass ratios for which we have only \EMRIcountqtenkWords data points available each.

Since this analysis of the \lmr data alone is only used to guide the ansatz choice in the next section,
but not used directly as a constraint, we do not list the detailed results of these fits here.
Instead, the final 3D fit using \nr and perturbative data will be compared with the high-$q$ data in \autoref{sec:results-extreme}.

\begin{figure*}[t]
 \resizebox{\textwidth}{!}{
  \includegraphics[height=5\baselineskip]{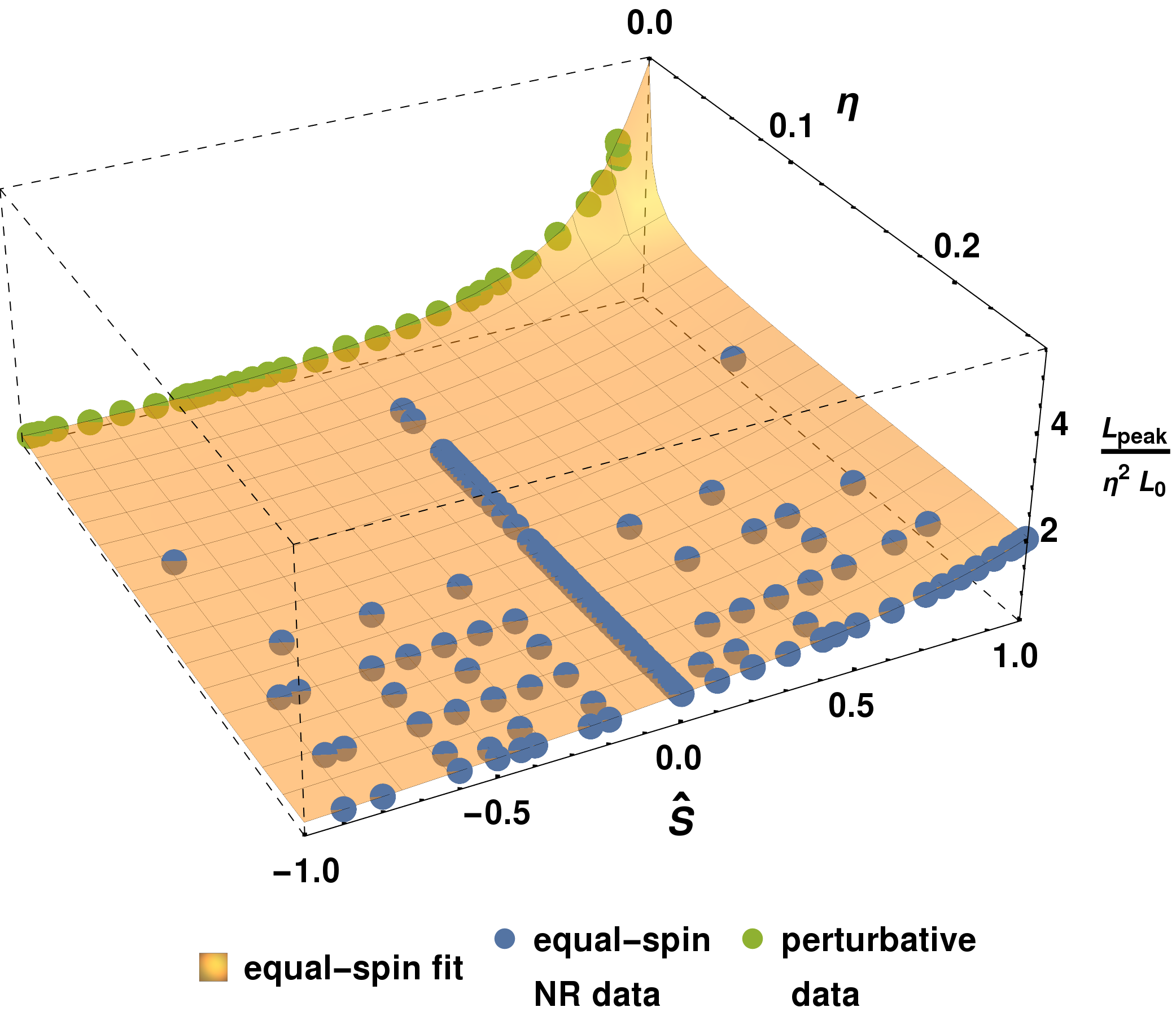}
  \includegraphics[height=5\baselineskip]{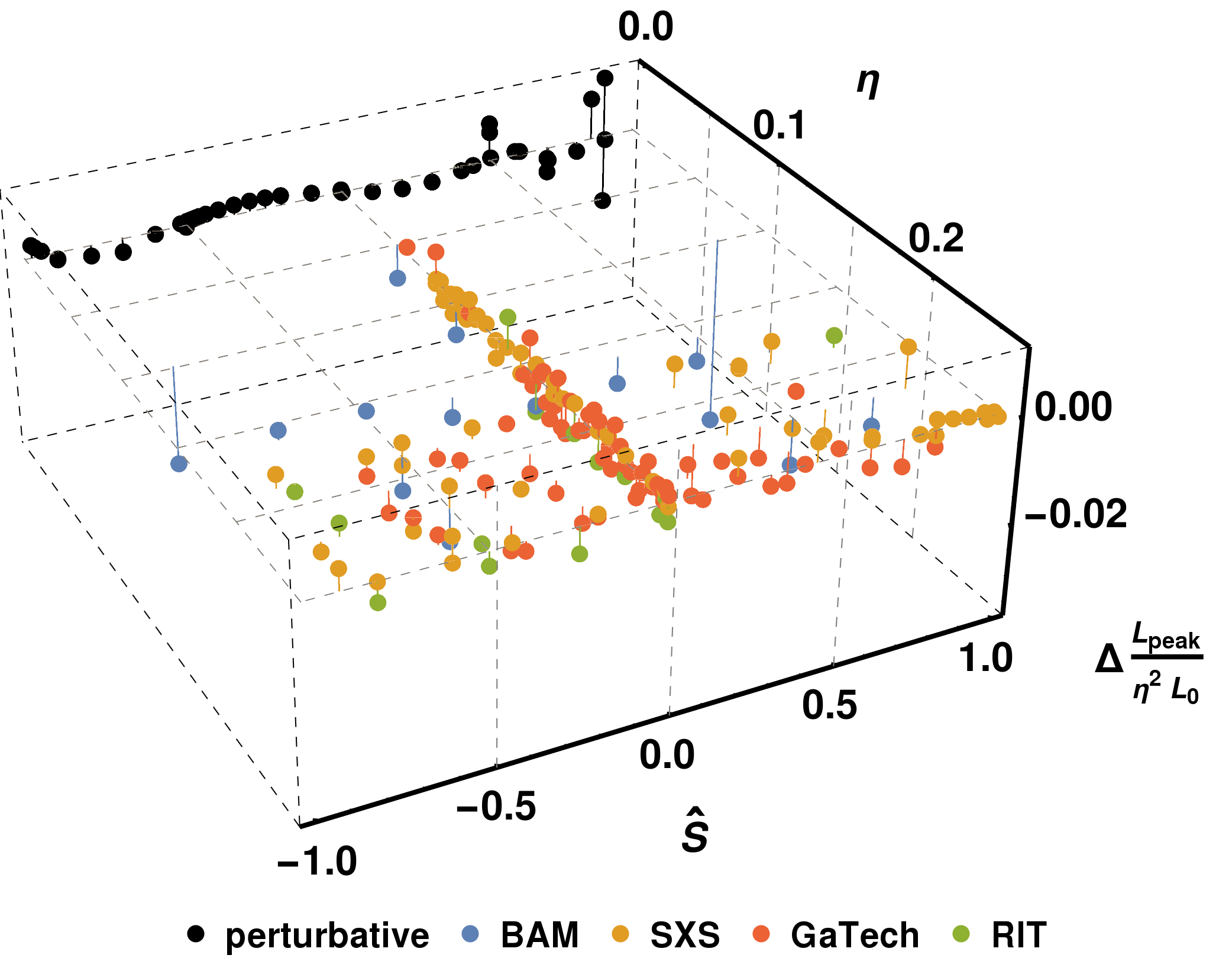}
 }
 \caption{
  \label{fig:fit2d}
  Results of the two-dimensional equal-spin $\LtwoD$ fit.
  First panel:
  Comparison of the smooth fit surface with the equal-spin \nr data and perturbative results.
  Second panel:
  Residuals over the parameter space, color-coded by data provenance.
 }
\end{figure*}

\vspace{\baselineskip}

\subsection{Two-dimensional fits}
\label{sec:fit-2d}

In proceeding with the hierarchical modeling approach,
we can now make a two-dimensional (2D) equal-spin ansatz informed and constrained by the previous one-dimensional (1D) steps
and the \lmr information.
In Ref.~\cite{Jimenez-Forteza:2016oae}, we constructed 2D final-state \ansaetze by 
first simply adding the two one-dimensional fits
and then generalizing each spin coefficient by a polynomial in $\eta$.
This time, we find that we need to introduce additional $\eta$-dependent higher-order terms in $\myS$,
as the curvature of $\Lscaled$ along the spin dimension increases
from equal masses towards the largest mass ratios.

We thus consider a 2D ansatz of the general form
\begin{equation}
 \label{eq:2dansatz}
 \LtwoD = \Lofeta \, + \, \rational{m}{k} \twoDparams
\end{equation}
with the $\eta$ fit from \autoref{eq:etaansatz}
and the rational function $\rational{m}{k}$ in $\myS$ inheriting the coefficients $b_i$ from Table~\ref{tbl:S_fit_coeffs}
and filled up with \mbox{$b_i=1.0$} for orders not present in \mbox{$\LofS$} from \autoref{eq:Sansatz}.
We then introduce the required freedom to change the curvature along the $\eta$ dimension through the substitution
\begin{equation}
 \label{eq:2D_substitution}
 b_i \rightarrow b_i \sum_{j=0}^{j=J} f_{ij} \, \eta^j \,,
\end{equation}
with a maximum expansion order $J$.

On the other hand, the number of free coefficients is reduced again by consistency constraints with the 1D fits:
\begin{subequations}
 \label{eq:2dconstraints}
 \begin{align}
  f_{i2} &=& 16&-16 f_{i0}-4 f_{i1} \quad \text{for } b_i \text{ from } \eta=0.25 \text{ fit}\,, \\
  f_{i2} &=&   &-16 f_{i0}-4 f_{i1} \quad \text{for other } b_i \,.
 \end{align}
\end{subequations}
In practice, we use R(4,2) to match the \mbox{$q=10^3$} result,
thus introducing one extra power of $\myS$ in both the numerator and denominator
compared to \mbox{$\LofS$} in \autoref{eq:Sansatz}.

With \datacountEqSrest equal-spin data points not yet used in the two one-dimensional subspace fits
(including \NRcountEqSrest \nr simulations
and the single-spin \lmr results,
which as discussed above can be considered as effectively equal-spin),
we can easily expand the polynomials in $\eta$ from \autoref{eq:2D_substitution} to order \mbox{$J=2$},
\mbox{$b_i \rightarrow b_i \, \left( f_{i0} + f_{i1}\eta + f_{i2}\eta^2 \right)$},
and still obtain a well-constrained fit.
The only further constraint is that we set the remaining highest-order coefficient in the denominator, $f_{71}$,
to zero to avoid a singularity within the physical $\twoDparams$ region,
leaving \macro{11} free coefficients.

The resulting fit and its residuals over the equal-spin data set are plotted in \autoref{fig:fit2d}.
We again find subpercent relative errors over most of the calibration set,
with an \rmse of $\approx\macro{0.0057}$
and only \macro{two} cases over 1\% relative error (both \mbox{$q=8$} from \BAM).
There is no apparent curvature or oscillatory feature except for the \lmr region,
where the $\Lscaled$ quantity plotted in \autoref{fig:fit2d} overemphasizes any remaining features
and the corresponding relative errors are below \macro{0.5\%}.
This accuracy is similar to that of the \lmr-only fits in \autoref{sec:fit-emri},
thus proving that the combined two-dimensional fit successfully captures both
the shallow spin slope at similar masses
and the steep slope in the perturbative regime.
As discussed in Appendix~\ref{sec:appendix-nr-outliers},
several outliers have been removed before the fit;
the 2D fit still matches all equal-spin outliers to below \macro{4\%} relative error.

As this equal-spin part of the full $\LthreeD$ will be refitted,
together with unequal-spin corrections,
in the next and final step of the hierarchical procedure,
we do not tabulate its best-fit coefficients at this point.

\vspace{\baselineskip}

\subsection{Unequal-spin contributions and 3D fit}
\label{sec:fit-3d}

\begin{figure*}[thbp]
 \includegraphics[width=\columnwidth]{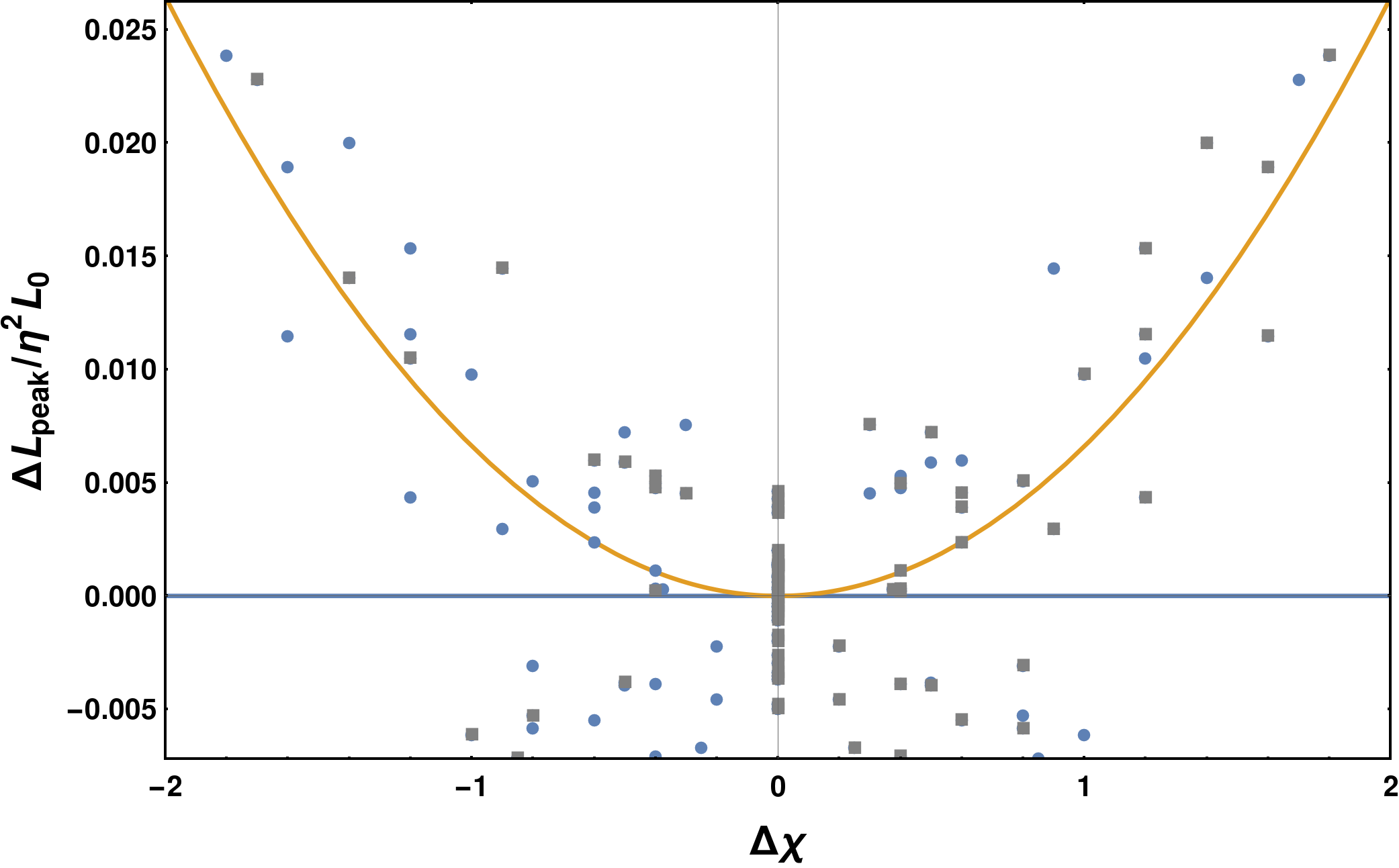} 
 \includegraphics[width=\columnwidth]{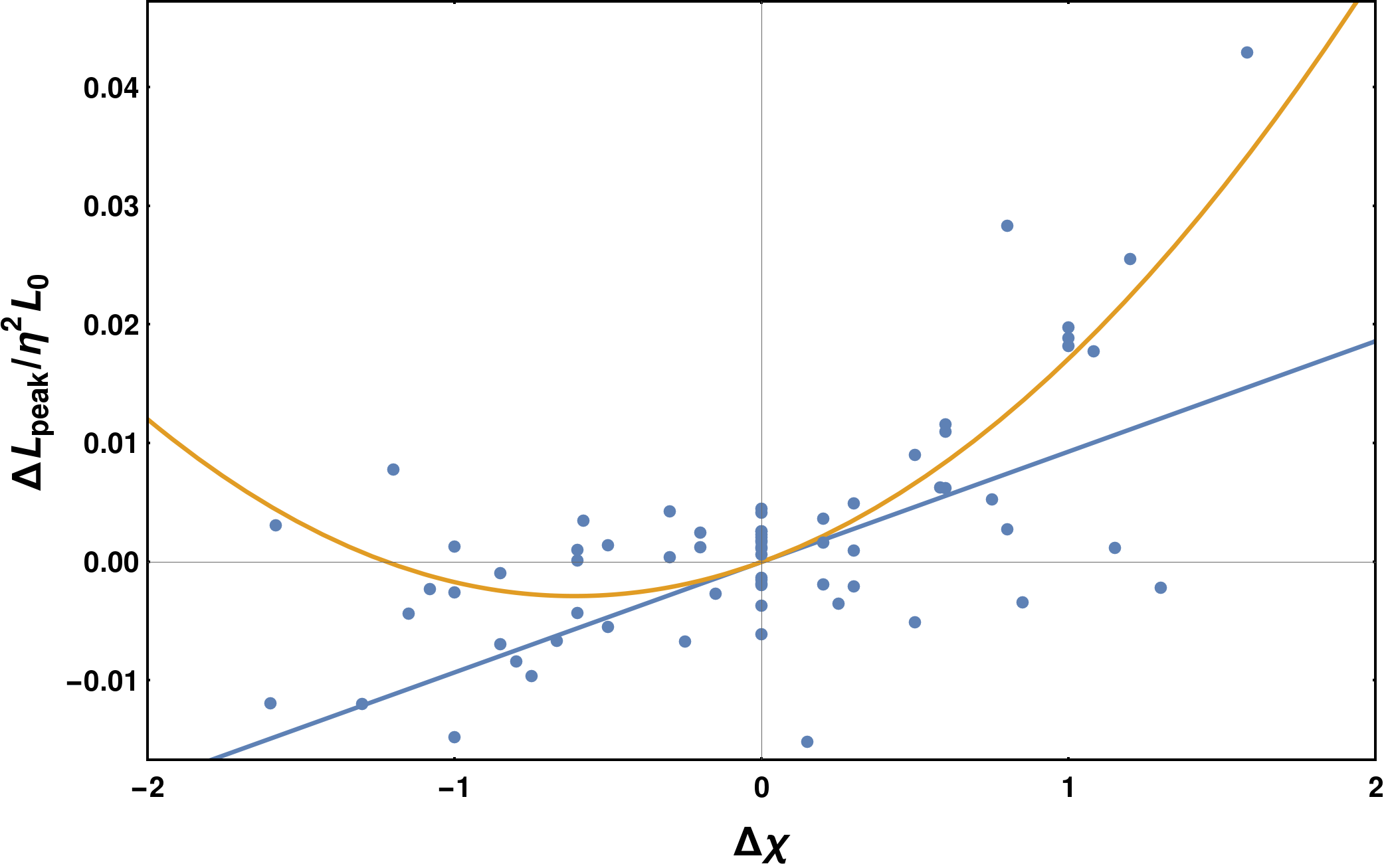}
 \caption{
  \label{fig:spin_diff_per_q}
  Examples of spin-difference behavior at fixed mass ratios,
  for scaled \nr data $\Lscaled$ after subtraction of the 2D $\twoDparams$ fit.
  First panel: \mbox{$q=1$} with the linear term vanishing due to symmetry and mainly quadratic dependence;
  points in gray are mirror duplicates exploiting the \mbox{$\chi_1\leftrightarrow\chi_2$} symmetry at equal masses.
  There is residual scatter in the \mbox{$\left|\chidiff\right| \lesssim 1$} range
  not captured by the quadratic fit,
  which is however not much larger than the scatter in equal-spin residuals,
  and hence probably related to the general uncertainties in \nr data quality for $\Lpeak$.
  Second panel: \mbox{$q=3$} where the linear term dominates and the apparent quadratic dependence likely is noise dominated.
 }
\end{figure*}

Simply extending the 2D fit to the full 3D parameter space either
by evaluating fit errors of the equal-spin-only calibrated fit over the whole data set,
or by refitting the 2D ansatz from \autoref{eq:2dansatz},
more than doubles the \rmse and induces oscillations at high $\left|\myS\right|$.
But even for such a naive approach, relative errors are still limited to below \macro{10\%},
so that the effects of unequal spins ($\chi_1\neq\chi_2$)
can evidently be treated as subdominant corrections.
We follow here the same approach as in Ref.~\cite{Jimenez-Forteza:2016oae} to model spin-difference effects,
constructing a 3D ansatz as
\begin{equation}
 \label{eq:final_ansatz}
 \LthreeD = \LtwoD + \Ldiff(\eta,\myS,\chidiff) \,.
\end{equation}
We choose the correction terms $\Ldiff$ with guidance from
(i)  \PN analytical results and
(ii) an analysis of the residuals of unequal-spin \nr simulations under the 2D equal-spin fit.

Though \PN cannot be expected to be quantitatively accurate for the late-inspiral and merger stages of \bbh coalescence --
where the \peaklum emanates from --
it can still give some intuition on the qualitative shape of spin and spin-difference effects.
The \PN spin-orbit flux terms as given in
Eq.~(3.13) of Ref.~\cite{Bohe:2013cla} and Eq.~(4.9) of Ref.~\cite{Marsat:2013caa}
include linear terms in $\chidiff$
with an $\eta$-dependent prefactor that can be expressed as
$\poly{\eta}\sqrt{1 - 4 \eta}$
with a polynomial $\poly{\eta}$.
The next-to-leading-order contributions would be quadratic in $\chidiff$
and a mixed term proportional to $\myS\chidiff$.

At equal masses (\mbox{$\eta=0.25$}) \bbh configurations are symmetric under relabeling of the component \bh[s],
so that terms linear in $\chidiff$ must vanish;
this is ensured by the $\sqrt{1 - 4 \eta}$ factor,
which we therefore expect both in the linear and the mixture term,
but not in the quadratic term.
Hence, we make the general spin-difference ansatz
\begin{equation}
 \label{eq:chidiff-ansatz}
  \Ldiff\,\threeDparams =   A_1(\eta) \, \chidiff
                          + A_2(\eta) \, \chidiff^2
                          + A_3(\eta) \, \myS \chidiff
\end{equation}
with a simple polynomial for $A_2(\eta)$ and $A_1(\eta)$, $A_3(\eta)$ both being a polynomial multiplied by the symmetry factor.

\begin{figure*}[thbp]
 \includegraphics[width=0.9\columnwidth]{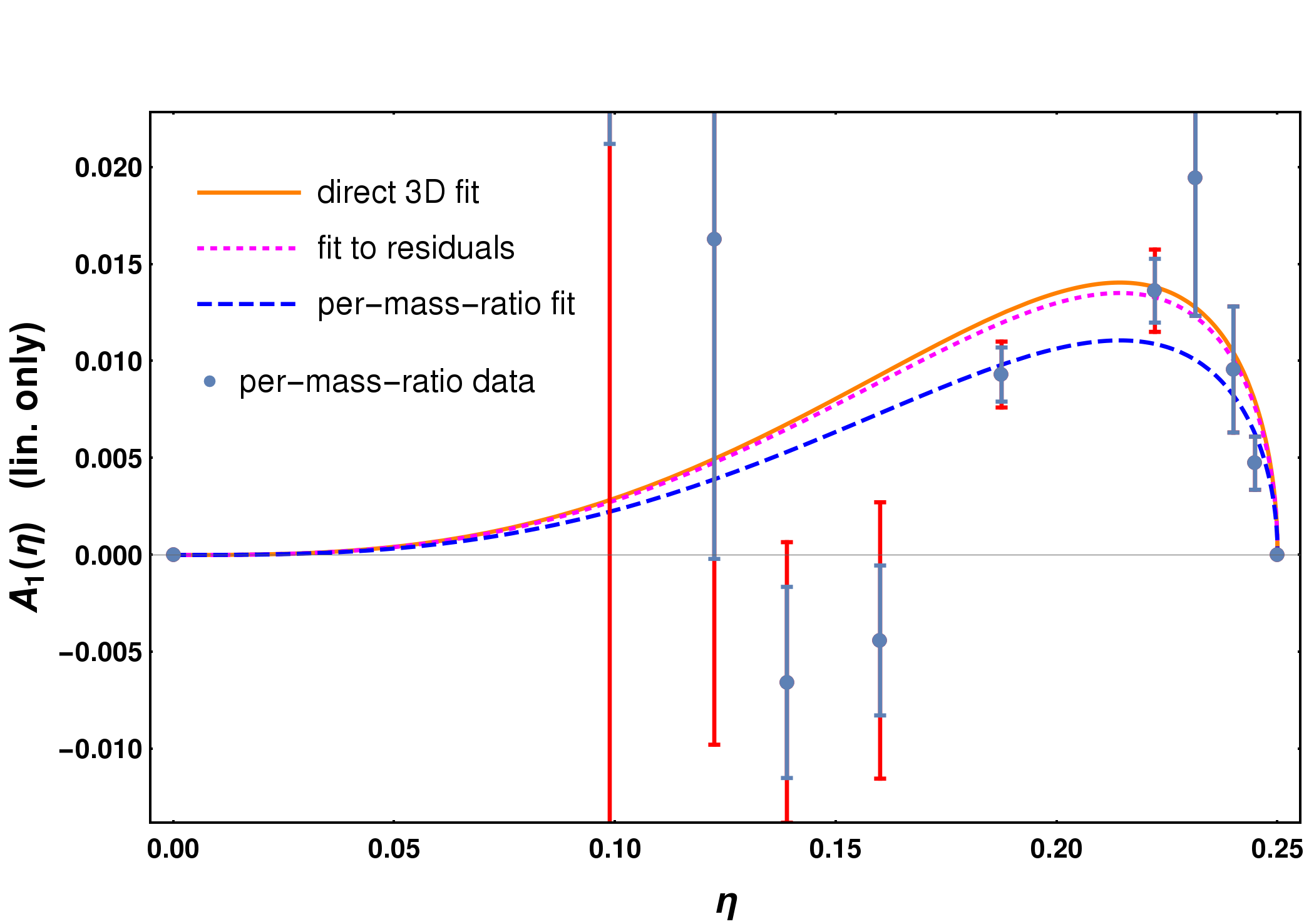} \hspace{1cm}
 \includegraphics[width=0.9\columnwidth]{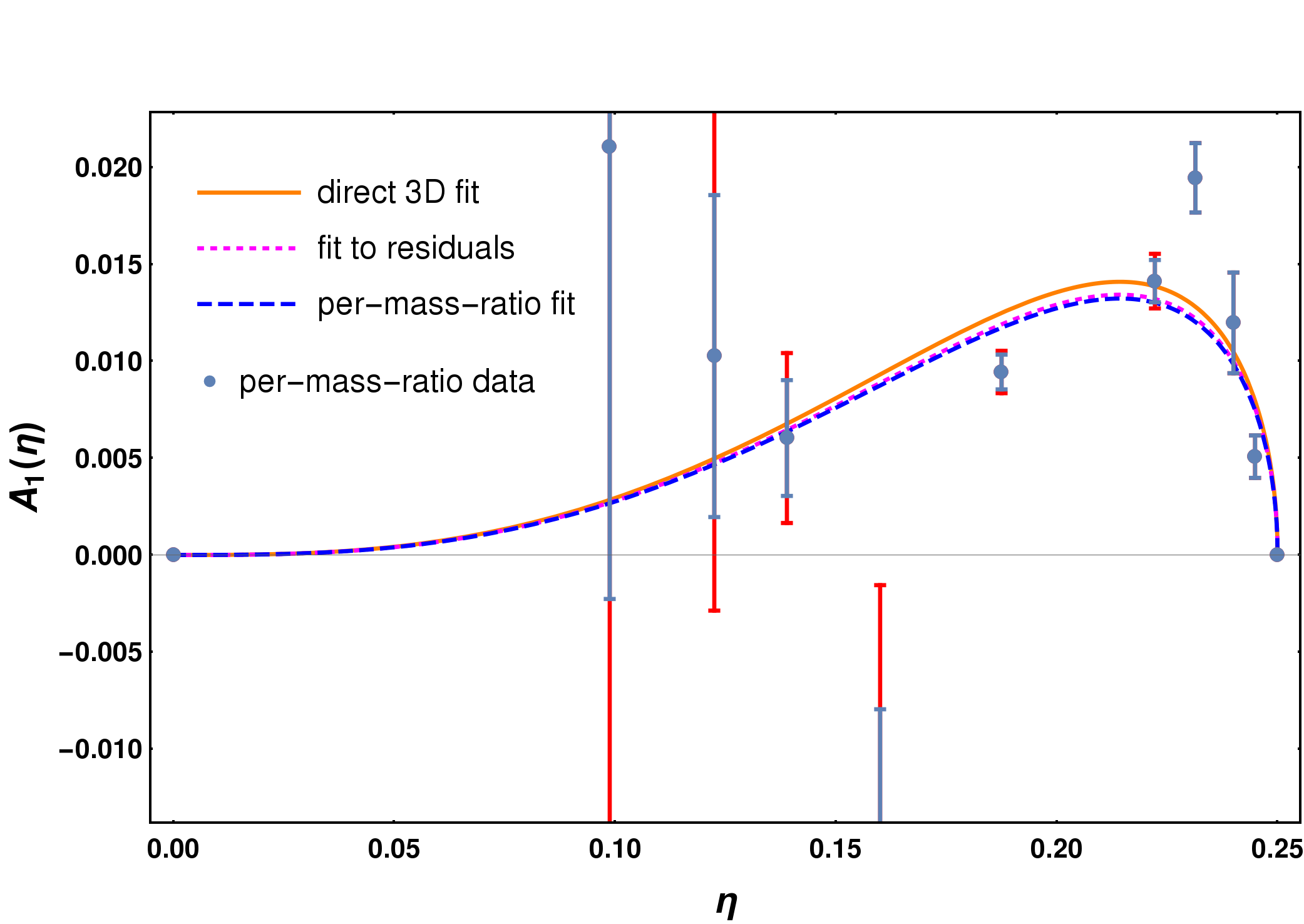} \\[-\baselineskip]
 \includegraphics[width=0.9\columnwidth]{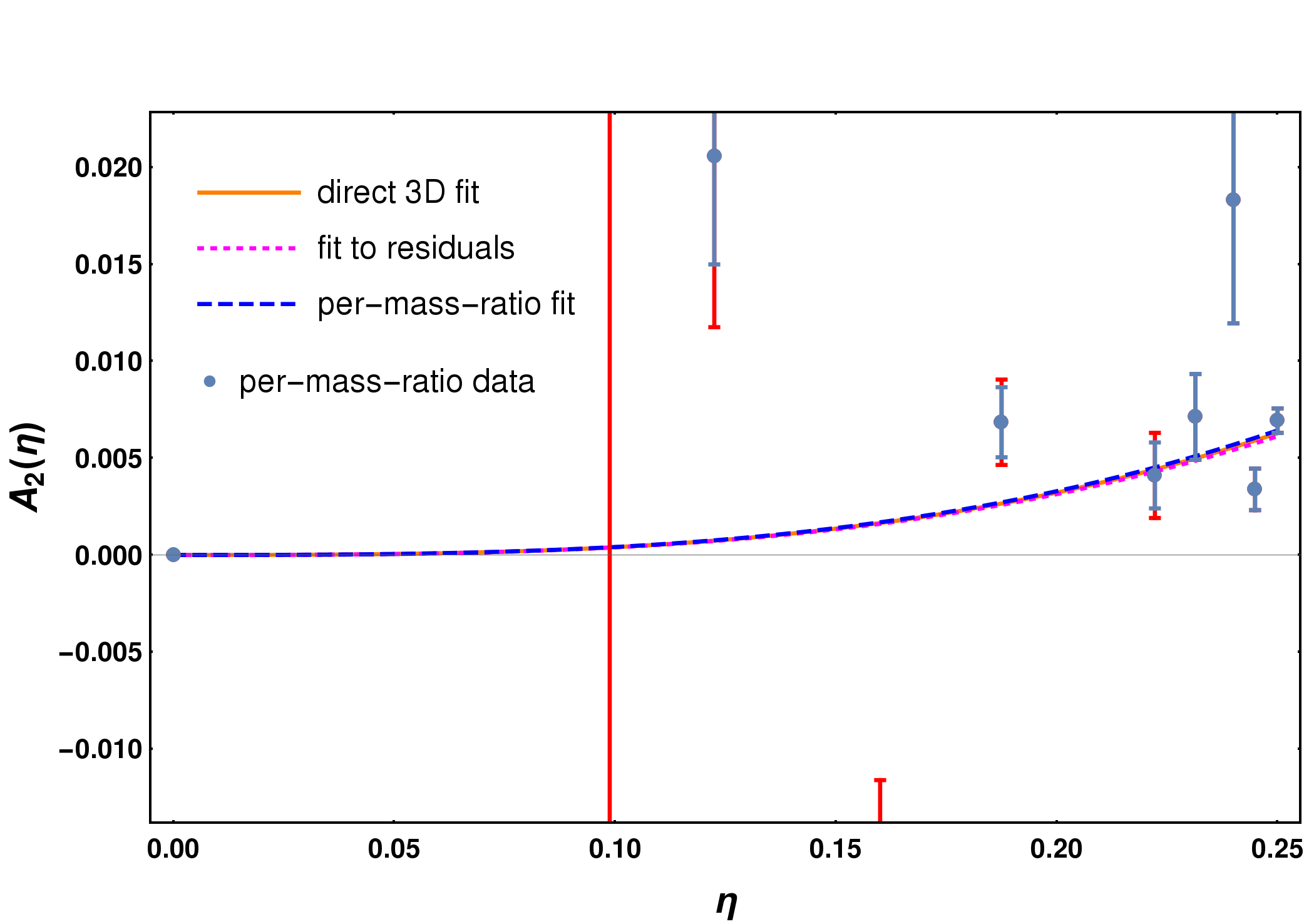} \hspace{1cm}
 \includegraphics[width=0.9\columnwidth]{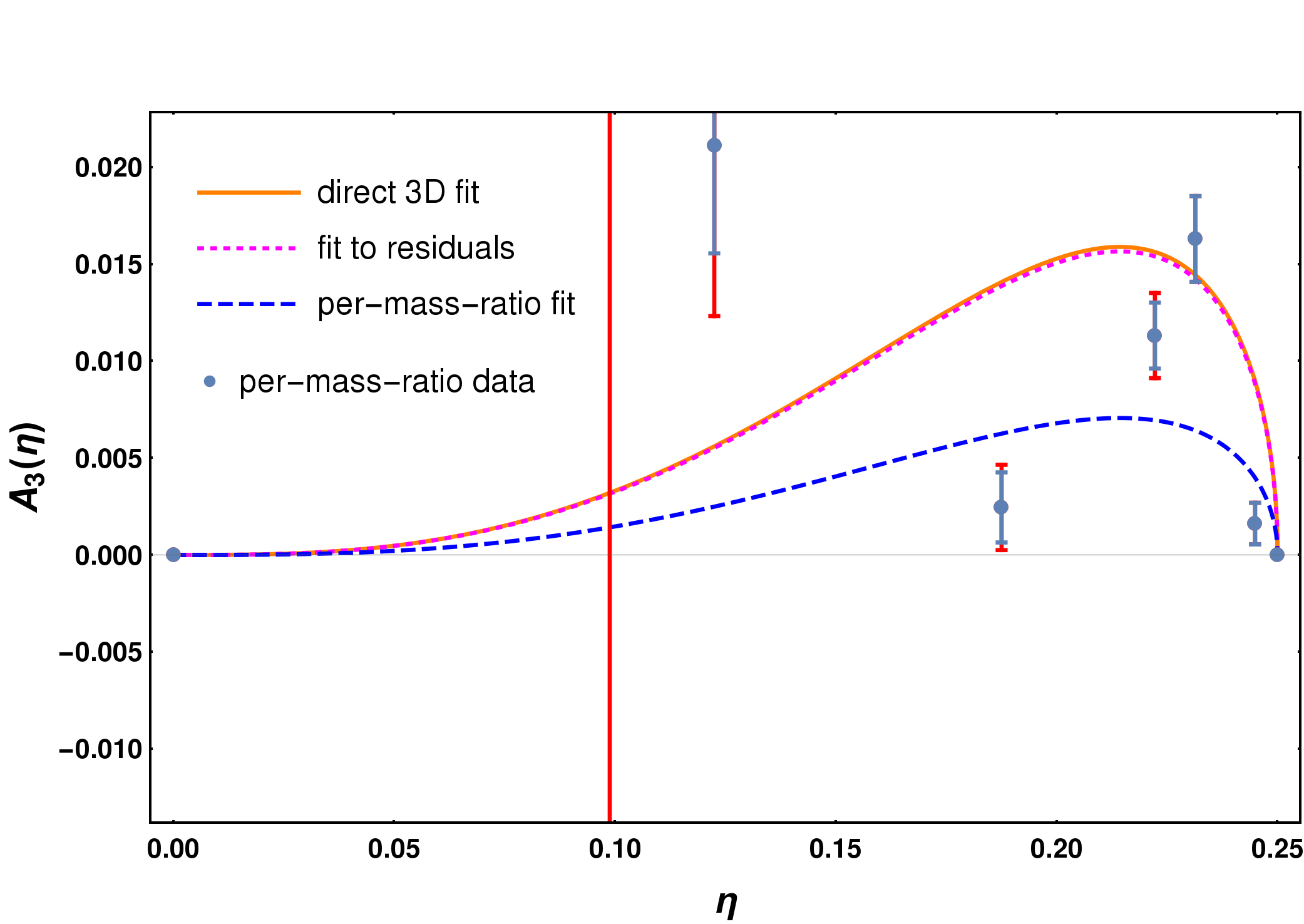}
 \caption{
  \label{fig:spin_diff_fits}
  Spin-difference behavior of the scaled \nr luminosities $\Lscaled$ after subtraction of the 2D $\twoDparams$ fit over mass ratio $\eta$,
  showing the results of fits as in \autoref{fig:spin_diff_per_q} at $\eta$ steps corresponding to \mbox{$q=\{1,\,1.33,\,1.5,\,1.75,\,2,\,3,\,4,\,5,\,6,\,8\}$}
  and three estimates for the ansatz functions $A_i(\eta)$ from \autoref{eq:spindiff_terms}:
  (i) unequal-spin part of the final 3D fit from \autoref{eq:3dansatz} (``direct 3D fit''),
  (ii) fit of the unequal-spin terms from \autoref{eq:spindiff_terms} (``fit to residuals'')
  to the residuals of the 2D fit from \autoref{eq:2dansatz} over all mass ratios,
  and (iii) fits of \autoref{eq:spindiff_terms} to the per-mass-ratio results.
  Top-left panel: Linear term $A_1$ only.
  The remaining panels are for the combined linear+quadratic+mixture fit, in clockwise order:
  linear term $A_1$,
  quadratic term $A_2$ and
  mixture term $A_3$.
  $A_1(\eta)$ from the combined ansatz is very similar to the linear-only fit,
  demonstrating its robustness.
  Error bars for the per-mass-ratio points include components from
  the fit uncertainty at that ratio (blue)
  and the average data weight of the contributing \nr cases (red).
  At the lowest $\eta$, some points lie outside the plot range,
  but are so uncertain that they do not contribute significantly to the total fit.
  The direct-3D and residuals-only results are consistent,
  while the per-mass-ratio analysis only matches them qualitatively,
  which is however sufficient since it was only used to investigate the possible shapes of $A_i(\eta)$.
 }
\end{figure*}

To check that these up to three terms accurately describe our available set of \NRcountUneqS unequal-spin \nr cases,
and to get a handle on the functions $A_i(\eta)$,
we visually inspect the data in steps of fixed mass ratio with sufficient numbers of data points.
Examples for \mbox{$q=1$} and \mbox{$q=3$} are shown in \autoref{fig:spin_diff_per_q}.
The unequal-spin data set appears more noisy for luminosity
than for the final-state quantities studied in Ref.~\cite{Jimenez-Forteza:2016oae},
yet can still be analyzed following the same procedure.
For each mass ratio step,
\mbox{$q=\{1,\,1.33,\,1.5,\,1.75,\,2,\,3,\,4,\,5,\,6,\,8\}$},
we compute the residuals under the nonspinning fit from \autoref{eq:2dansatz},
then perform four fits in $\chidiff$:
linear, linear+quadratic, linear+mixed, or the sum of all three terms.
Fits of the collected coefficients, as functions of $\eta$,
give estimates of the functions $A_i(\eta)$,
as displayed with the ``per-mass-ratio data'' points and ``per-mass-ratio fit'' lines in \autoref{fig:spin_diff_fits}.

The scatter of fit coefficients at individual mass-ratio steps is again larger than that
found for final spin and radiated energy in Ref.~\cite{Jimenez-Forteza:2016oae},
but this procedure still yields sufficient evidence for the existence and shape of a linear spin-difference term
and some preference for including both second-order terms,
though the data is too noisy to constrain their $\eta$-dependent shape very well.
For example, there is an apparent sign switch in the linear term at mass ratio \mbox{$q=4$} (\mbox{$\eta=0.16$}),
which is most likely due to a combination of
the 2D fit being relatively weakly constrained in this region
and non-negligible errors in some of the unequal-spin data points,
which however cannot easily be discarded as outliers.

The overall fits in $\eta$ are reasonably robust against such problems,
and in the next step we will use not this step-by-step analysis,
but a more robust fit of the full 3D ansatz to the full data set,
to judge the overall significance of spin-difference terms.
A full model selection of $A_i(\eta)$ is clearly not feasible at this point
without a more detailed understanding of the point-by-point data quality.
Hence, we make very simple choices for the $A_i(\eta)$ with just one power of $\eta$ each:
\begin{subequations}
 \label{eq:spindiff_terms}
 \begin{align}
  A_1(\eta)&=d_{10} \sqrt{1-4 \eta } \, \eta ^3 \\
A_2(\eta)&=d_{20} \eta ^3 \\
A_3(\eta)&=d_{30} \sqrt{1-4 \eta } \, \eta ^3 \,,
 \end{align}
\end{subequations}
and investigate how much improvement this can yield over the 2D fit.

We now use the full data set except for the 1D subspaces
(\datacountNZSUM data points, including \NRcountNZSUM \nr simulations)
to fit the full 3D ansatz from \autoref{eq:final_ansatz},
with the equal-spin and spin-difference contributions from Eqs.~\ref{eq:2dansatz} and~\ref{eq:chidiff-ansatz}+\ref{eq:spindiff_terms}, respectively.
The sets of coefficients $a_{i}$, $b_{i}$ and $f_{i2}$ are already fixed from the 1D fits and consistency constraints
(see Tables~\ref{tbl:eta_fit_coeffs}, \ref{tbl:S_fit_coeffs} and \autoref{eq:2dconstraints}),
leaving between \macro{11} and \macro{14} free coefficients in this final 3D stage.
When including all three spin-difference terms, the full ansatz
(with the constraints from \autoref{eq:2dconstraints} for the $f_{i2}$ still to be applied)
is:
\begin{widetext}
\begin{align}
 \label{eq:3dansatz}
 &\LthreeD =
    a_5 \eta ^5+a_4 \eta ^4+a_3 \eta ^3+a_2 \eta ^2+a_1 \eta +a_0  \\
& + \frac{ 0.107 b_2 \widehat{S}^2 \left(f_{22} \eta ^2+f_{21} \eta +f_{20}\right)
          +0.465 b_1 \widehat{S} \left(f_{12} \eta ^2+f_{11} \eta +f_{10}\right)
          + \widehat{S}^4 \left(f_{42} \eta ^2+f_{41} \eta +f_{40}\right)
          + \widehat{S}^3 \left(f_{32} \eta ^2+f_{31} \eta +f_{30}\right)}
         {-0.328 b_4 \widehat{S} \left(f_{62} \eta ^2+f_{61} \eta +f_{60}\right)
          + \widehat{S}^2 \left(f_{72} \eta ^2 +f_{70}\right)+1.0
         }\nonumber \\
& + d_{20} \, \eta ^3 \left(\chi_1-\chi_2\right)^2
  + d_{10} \sqrt{1-4 \eta } \, \eta ^3 \left(\chi_1-\chi_2\right)
  + d_{30} \sqrt{1-4 \eta } \, \eta ^3 \, \widehat{S} \left(\chi_1-\chi_2\right)
 \nonumber \,.
\end{align}
\end{widetext}

We consider residuals and information criteria,
summarized in Table~\ref{tbl:stats_summary},
to check which spin-difference terms are actually supported by the data.
These rankings depend on the specific choice of $A_i(\eta)$,
but with the current parameter-space coverage and understanding of \nr data quality,
the main goal is to find general evidence for spin-difference effects
and a general idea of their shape,
not to exactly characterize them.
With the choices made in \autoref{eq:spindiff_terms},
we find a \macro{14}-coefficient fit with linear+quadratic+mixture corrections
that has well-constrained coefficients (see Table~\ref{tbl:final_fit_coeffs}),
is evidently preferred in terms of AICc and BIC,
and reduces overall residuals by about \macro{20\%} in \rmse.
Different choices for the powers of $\eta$ in \autoref{eq:spindiff_terms}
yield compatible results,
while polynomials in $\eta$ with several free coefficients
tend to produce underconstrained fits.

\begin{table}[h!]
 \begin{tabular}{lrrcrr}\hline\hline
  &$ N_{\text{data}} $&$ N_{\text{coeff}} $&$ \text{RMSE} $&$ \text{AICc} $&$ \text{BIC} $\\\hline$
 \text{1D $\eta $}                       $&$  84 $&$  6 $&$ 2.81\times 10^{-3} $&$  -817.1 $&$  -801.2 $\\$
 \text{1D }\hat{S}                       $&$  32 $&$  3 $&$ 2.42\times 10^{-3} $&$  -285.8 $&$  -280.8 $\\$
 \left.\text{2D (}\chi _1=\chi _2\right) $&$  92 $&$ 11 $&$ 5.65\times 10^{-3} $&$  -751.7 $&$  -724.8 $\\$
 \text{2D all}                           $&$ 307 $&$ 11 $&$ 1.67\times 10^{-2} $&$ -1914.2 $&$ -1870.4 $\\$
 \text{3D lin}                           $&$ 307 $&$ 12 $&$ 1.51\times 10^{-2} $&$ -2008.0 $&$ -1960.6 $\\$
 \text{3D lin+quad}                      $&$ 307 $&$ 13 $&$ 1.39\times 10^{-2} $&$ -2134.2 $&$ -2083.3 $\\$
 \text{3D lin+mix}                       $&$ 307 $&$ 13 $&$ 1.41\times 10^{-2} $&$ -2082.6 $&$ -2031.7 $\\$
 \text{3D lin+quad+mix}                  $&$ 307 $&$ 14 $&$ 1.36\times 10^{-2} $&$ -2157.8 $&$ -2103.3 $\\
\hline\hline\end{tabular}
 \vspace{\baselineskip}
 \caption{
  \label{tbl:stats_summary}
  Summary statistics for the various steps of the hierarchical fit.
  Note that it is not meaningful to compare AICc and BIC between data sets of different sizes.
  There is preference for the 3D fit including all three linear+mixture+quadratic terms,
  although many different choices of the $A_i(\eta)$ ansatz functions yield similar results
  with just $\pm$ a few percent in \rmse and $\pm$ a few in AICc/BIC,
  so that the shape of these terms is not yet strongly constrained.
  \vspace{-\baselineskip}
 }
\end{table}

\begin{table}[thpb]
 \begin{tabular}{lrrr}\hline\hline
  &$ \text{Estimate} $&$ \text{Standard error} $&$ \text{Relative error [$\%$]} $\\\hline$
 d_{10} $&$    3.79\hphantom{3} $&$  0.28\hphantom{3} $&$  7.5 $\\$
 d_{20} $&$    0.402            $&$  0.044            $&$ 10.9 $\\$
 d_{30} $&$    4.27\hphantom{3} $&$  0.84\hphantom{3} $&$ 19.7 $\\$
 f_{10} $&$    1.628            $&$  0.012            $&$  0.7 $\\$
 f_{11} $&$   -3.63\hphantom{3} $&$  0.23\hphantom{3} $&$  6.3 $\\$
 f_{20} $&$   31.7\hphantom{23} $&$  1.3\hphantom{23} $&$  4.2 $\\$
 f_{21} $&$ -274\hphantom{.123} $&$ 29\hphantom{.123} $&$ 10.4 $\\$
 f_{30} $&$ -0.235              $&$  0.011            $&$  4.7 $\\$
 f_{31} $&$ 6.96\hphantom{3}    $&$  0.44\hphantom{3} $&$  6.3 $\\$
 f_{40} $&$ 0.211               $&$  0.022            $&$ 10.6 $\\$
 f_{41} $&$ 1.53\hphantom{3}    $&$  0.45\hphantom{3} $&$ 29.6 $\\$
 f_{60} $&$ 3.090               $&$  0.044            $&$  1.4 $\\$
 f_{61} $&$ -16.7\hphantom{23}  $&$  1.7\hphantom{23} $&$ 10.0 $\\$
 f_{70} $&$ 0.836               $&$  0.023            $&$  2.8 $\\
\hline\hline\end{tabular}
 \vspace{\baselineskip}
 \caption{
  \label{tbl:final_fit_coeffs}
  Fit coefficients for the final 3D fit stage, cf. \autoref{eq:3dansatz}.
  \vspace{-\baselineskip}
 }
\end{table}

\vspace{\baselineskip}

\section{Fit assessment}
\label{sec:results}

In this section, we assess in some detail the properties and statistical quality of
the new three-dimensional \peaklum fit,
with the actual nonrescaled luminosity (in geometric units of \mbox{$G=c=M=1$})
obtained as \mbox{$\eta^2 \, \Lo \, \LthreeD$}.

We compare with our previous fit~\cite{T1600018} used for LIGO parameter estimation during
O1~\cite{Abbott:2016blz,Abbott:2016nmj,TheLIGOScientific:2016pea,TheLIGOScientific:2016wfe,Abbott:2016izl},
which used a much smaller calibration set of 89 \BAM and SXS simulations,
only modes up to \mbox{$\lmax=4$}
and no \emr constraints;
and with the recent Healy\&Lousto fit~\cite{Healy:2016lce} based on \NRcountRIT RIT simulations,
using modes up to \mbox{$\lmax=6$}.
We attempt to present a fair comparison by analyzing \nr and perturbative \lmr results separately,
and also consider the improvement from refitting the unmodified \ansaetze of Refs.~\cite{T1600018,Healy:2016lce}
to the present \nr data set.

\subsection{Residuals and information criteria}
\label{sec:results-resid}

\addtolength{\parskip}{0.5\baselineskip}

\begin{figure}[b]
 \includegraphics[width=\columnwidth]{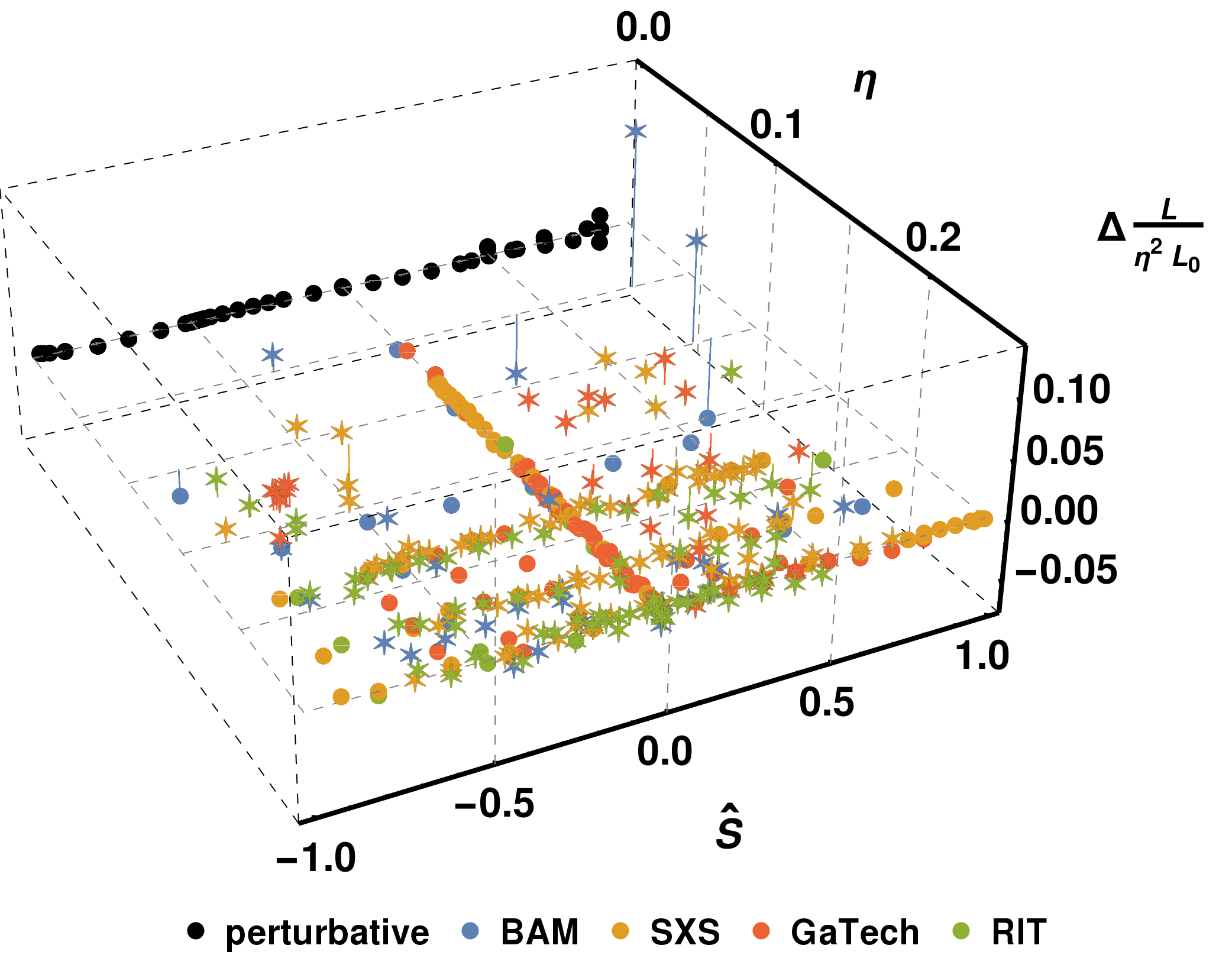}
 \caption{
  \label{fig:3d_residuals_paramspace}
  Residuals of the final 3D \peaklum fit,
  for \nr and perturbative \lmr data,
  projected to the 2D parameter space of $\eta$ and $\myS$.
  The data sets are distinguished by colors,
  and unequal-spin points are highlighted with stars.
  }
\end{figure}

In \autoref{fig:3d_residuals_paramspace} we show the distribution of residuals for the 3D fit in $\Lscaled$
projected to the $\twoDparams$ parameter space,
so that it can be compared to the 2D results in \autoref{fig:fit2d}.
The strongest visible outliers in this scaling are at low $\eta$ and correspond to mild actual deviations;
of at most a \macro{7\%} relative error at \mbox{$q=18$},
with \macro{417} of the \datacount data points below \macro{3\%} relative error.

\begin{figure}[t!]
 \vspace{0.5\baselineskip}
 \includegraphics[width=\columnwidth]{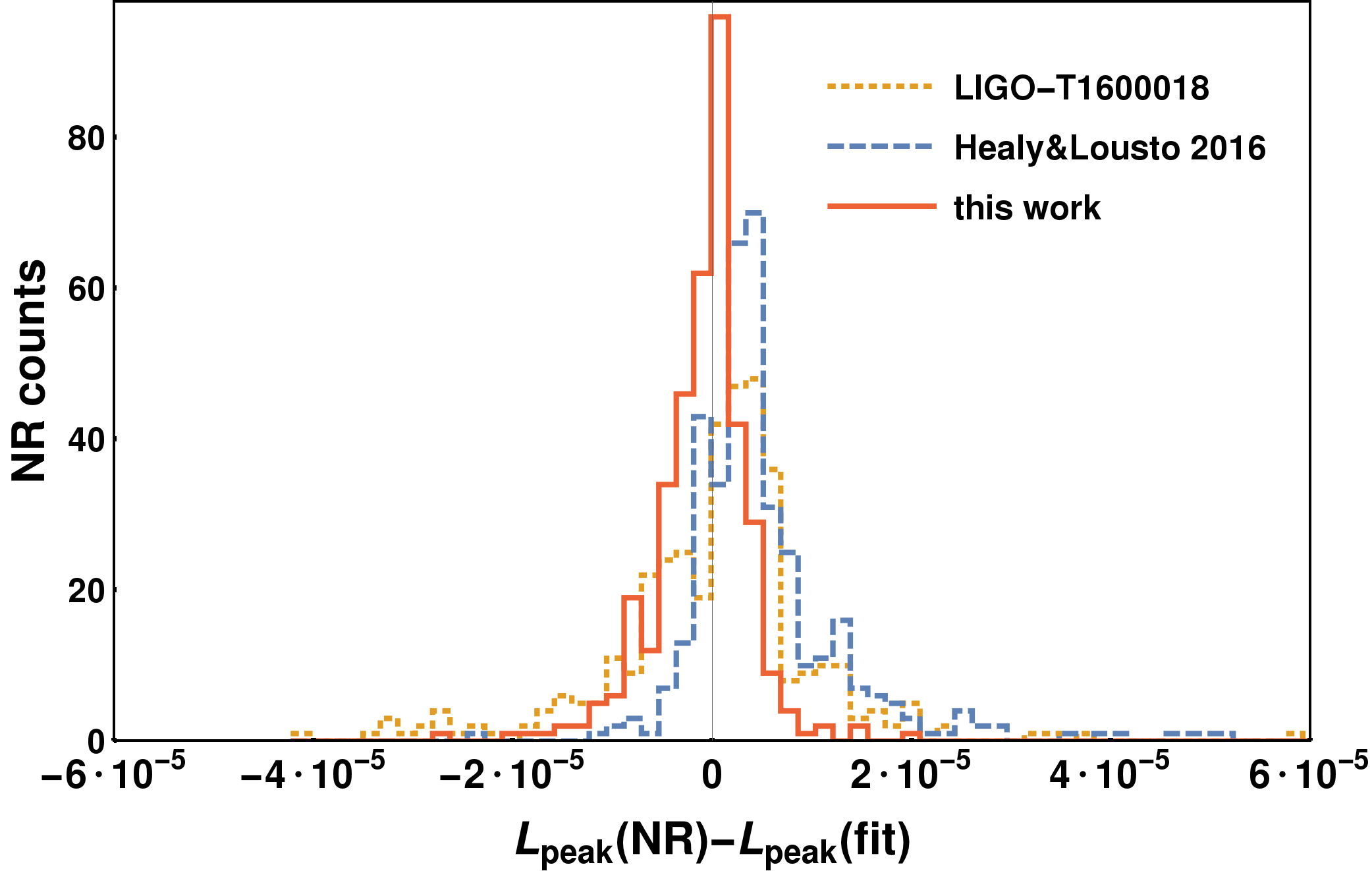}
 \captionof{figure}{
  \label{fig:3d_residuals_hist}
  Fit residuals of the final 3D \peaklum fit compared with the previous fits of
  LIGO-T160018~\cite{T1600018}
  and Healy\&Lousto2016~\cite{Healy:2016lce},
  evaluated over the set of \NRcount NR simulations shown in \autoref{fig:NR_eta_chi1_chi2}.
  \macro{Six} outliers for Healy\&Lousto with \mbox{$|\text{NR}-\text{fit}|>0.00006$} are outside of the plot range.
 }
 \vspace{2\baselineskip}
 \begin{tabular}{lcrccc}\hline\hline
  &$ N_{\text{coef}} $&$ \text{mean} $&$ \text{stdev} $&$ \text{AICc} $&$ \text{BIC} $\\\hline$
 \text{T1600018}     $&$ 11 $&$  3.0\times 10^{-7} $&$ 1.0\times 10^{-5} $&$ -7732.1 $&$ -7685.6 $\\$
 \text{ (refit)}     $&$ 11 $&$ -1.8\times 10^{-6} $&$ 4.0\times 10^{-5} $&$ -6706.0 $&$ -6659.5 $\\$
 \text{HL2016}       $&$ 19 $&$  6.9\times 10^{-6} $&$ 1.5\times 10^{-5} $&$ -7225.5 $&$ -7148.9 $\\$
 \text{ (refit)}     $&$ 19 $&$ -4.9\times 10^{-7} $&$ 1.0\times 10^{-5} $&$ -7708.3 $&$ -7631.7 $\\$
 \text{this work}    $&$ 23 $&$ -9.8\times 10^{-7} $&$ 4.8\times 10^{-6} $&$ -8298.1 $&$ -8206.7 $\\$
 \text{ (refit)}     $&$ 23 $&$ -5.5\times 10^{-7} $&$ 4.8\times 10^{-6} $&$ -8323.6 $&$ -8232.3 $\\
\hline\hline\end{tabular}
 \vspace{\baselineskip}
 \captionof{table}{
  \label{tbl:3D_residuals}
  Summary statistics for the final 3D \peaklum fit
  compared with previous fits with the previous fits of
  LIGO-T160018~\cite{T1600018}
  and Healy\&Lousto2016~\cite{Healy:2016lce},
  evaluated over the \NRcount \nr simulations shown in \autoref{fig:NR_eta_chi1_chi2}.
  The new fit has a total of \macro{23} free coefficients,
  corresponding to tables~\ref{tbl:eta_fit_coeffs}, \ref{tbl:S_fit_coeffs} and \ref{tbl:final_fit_coeffs}.
  We also show results for re-fitting the three \ansaetze to the full \nr + \lmr data set,
  again evaluating the statistics over \nr only.
 }
\end{figure}

For a comparison with the two previous fits,
we first concentrate on the \NRcount \nr simulations only
and revisit \lmrs in \autoref{sec:results-extreme}.
In \autoref{fig:3d_residuals_hist} we show histograms of the residuals in $\Lpeak$
for the three fits over this data set,
demonstrating that the new fit achieves a narrower distribution.
As listed in Table~\ref{tbl:3D_residuals},
the standard deviation of residuals is only half of that for our previous fit
and three times lower than for the RIT fit.
With a mean offset by only a ninth of a standard deviation, there is no evidence for bias,
though that notion is notoriously ambiguous for a data set that samples the parameter space nonuniformly.

The same table contains AICc and BIC values evaluated over the same \nr-only data set,
which both find a very significant preference for the new fit.
Note that, being computed over a different data selection and for $\Lpeak$ instead of $\Lscaled$,
these values are not directly comparable with the previous Table~\ref{tbl:stats_summary}.
Since we have removed \NRcountOutliers \nr cases from the full available data set (see Appendix~\ref{sec:appendix-nr-outliers}),
it is advisable to check that the statistical preference still holds when including these in the evaluation set.
Indeed, the reduction in standard deviations of residuals is then less against the T1600018 and RIT fits,
but still roughly \macro{20\%} and \macro{30\%},
and there is still a preference of several hundreds in both information criteria.

We also show results for refitting the T1600018 and RIT \ansaetze
to the present \nr + perturbative data set,
with the statistics then again evaluated over \nr data only.
Our old ansatz with only 11 coefficients is not well suited to matching the \lmr region
and the large unequal-spin population in the \nr data set,
and the refitted version of this 11-coefficient ansatz performs worse than the original.
On the other hand, the RIT ansatz with 19 coefficients
was only weakly constrained in the original version~\cite{Healy:2016lce} fitted to \NRcountRIT simulations,
with large errors on several fit coefficients,
but improves now significantly through the refit.
Yet, it does not achieve the same level of accuracy as the new ansatz and fit developed in this paper.

As a test of robustness,
we also perform a refit of our final hierarchically obtained ansatz directly using the full data set,
instead of using the constraints from the 1D subsets.
This produces somewhat better summary statistics,
but it also allows uncertainties
from less well-controlled unequal-spin data to influence the nonspinning part of the fit.
The more conservative approach is to calibrate the nonspinning part of the fit only to the corresponding data subset,
as done in \autoref{sec:fit-1d-eta}.
Hence we recommend the stepwise fit,
with coefficients as reported in Tables~\ref{tbl:eta_fit_coeffs}, \ref{tbl:S_fit_coeffs} and~\ref{tbl:final_fit_coeffs},
for further applications.

\subsection{Large-mass-ratio and extremal-spin limits}
\label{sec:results-extreme}

\begin{figure}[t]
 \includegraphics[width=\columnwidth]{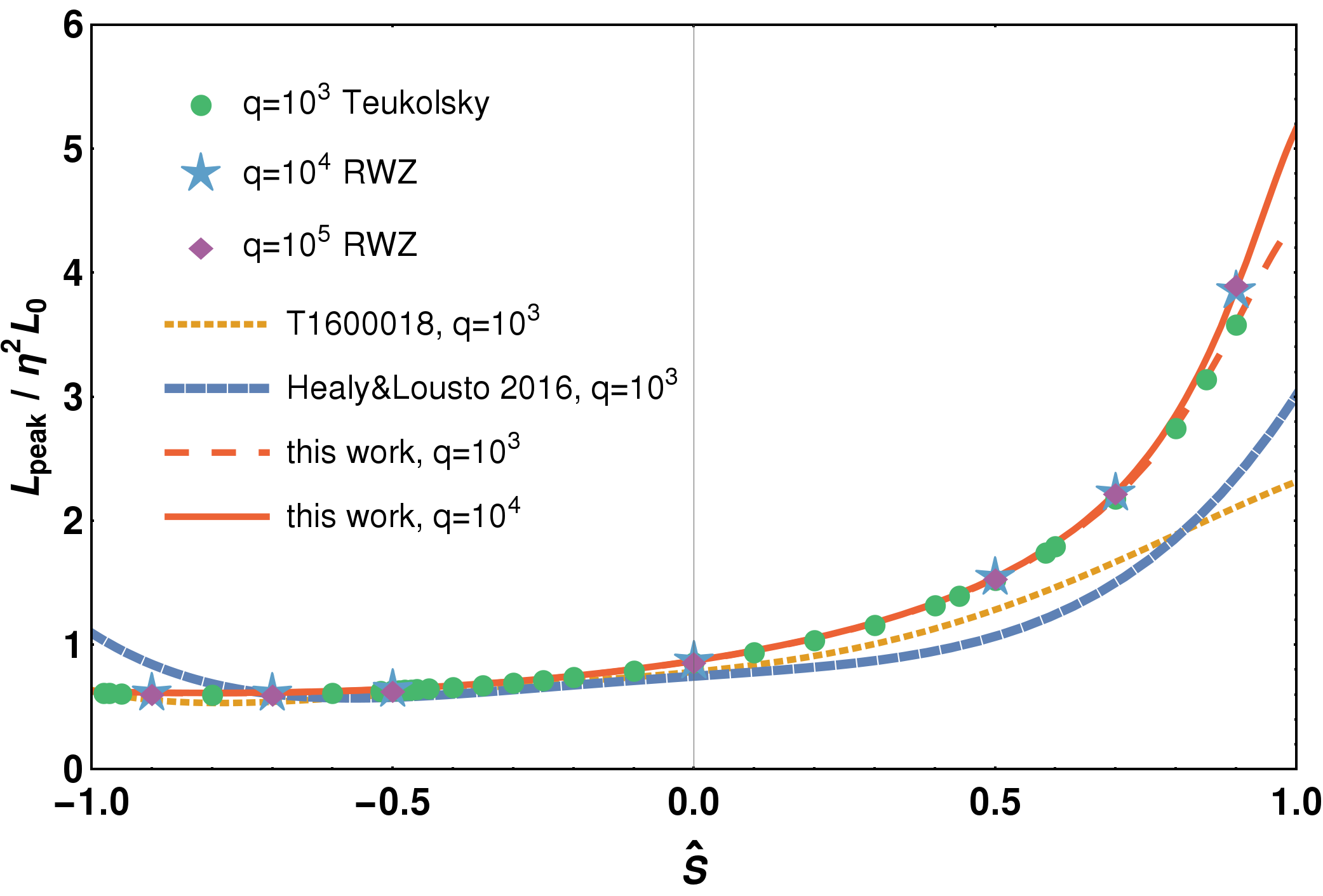}
 \caption{
  \label{fig:emri_fits_comparison}
  Full \nr-calibrated fits from this work and from Refs.~\cite{T1600018,Healy:2016lce}
  evaluated at \lmrs,
  compared with the same perturbative data
  (circles, stars and diamonds for mass ratios \mbox{$q=\{10^3,10^4,10^5\}$})
  as in \autoref{fig:emridata}.
  The T160018 and RIT fits are essentially converged at \mbox{$q=10^3$}
  (e.g. 0.4\% change at \mbox{$S=1.0$} for the RIT fit going to \mbox{$q=10^4$}),
  and the visually identical lines for higher $q$ are not shown;
  our new fit still matches the data at higher $q$.
 }
\end{figure}

In \autoref{fig:emri_fits_comparison}, we compare our full 3D fit with the perturbative \lmr data
and find that it correctly reproduces the behavior it is meant to be constrained to.
The T1600018 fit did not predict the steep rise for positive spins,
and while at negative spins it matches the shape roughly,
it is still off by about 10\% in that region.
The RIT fit disagrees with the perturbative data at high spin magnitudes, either negative or positive,
and does not reproduce the increasing steepness for even higher mass ratios.

\begin{figure}[t]
 \includegraphics[width=\columnwidth]{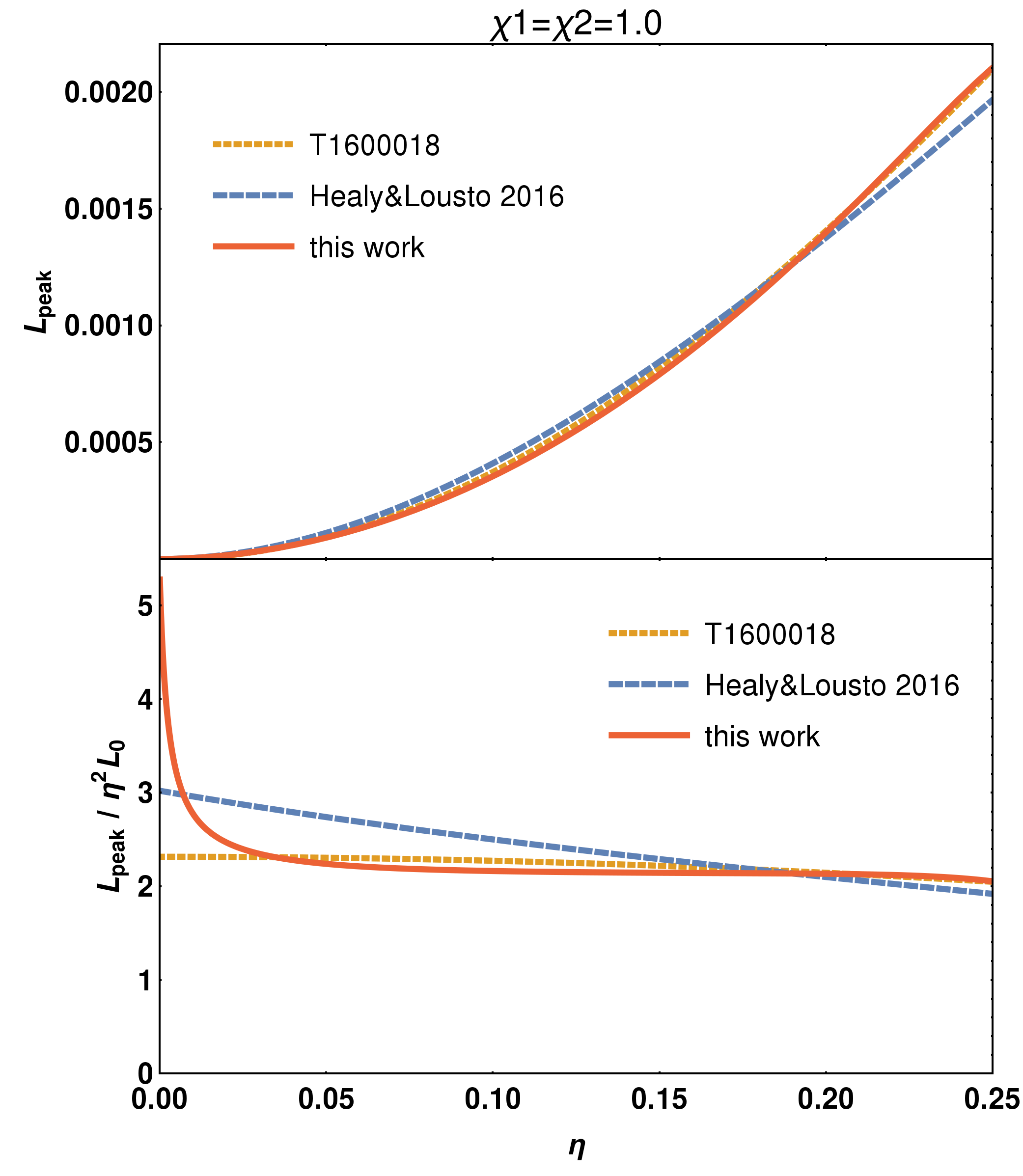}
 \caption{
  \label{fig:extreme_spins_comparison}
  Behavior of the full 3D fit~\eqref{eq:final_ansatz} in the extremal-spin limit,
  \mbox{$\chi_1=\chi_2=\myS=1$},
  where there is no data available.
  Both panels give functions of mass ratio $\eta$,
  and we again compare with the fits from Refs.~\cite{T1600018,Healy:2016lce}.
  Top panel: In terms of physical \peaklum $\Lpeak$.
  Lower panel: In terms of rescaled \mbox{$\Lscaled=\Lpeak/\eta^2\Lo$}.
 }
\end{figure}

The clearest difference between this fit and the previous ones in the \nr-dominated region
is for high aligned spins,
which is shown in  \autoref{fig:extreme_spins_comparison} for the extremal spin limit,
\mbox{$\chi_1=\chi_2=\myS=1$}.
The RIT fit estimates a lower luminosity at equal masses,
but higher values at \mbox{$\eta<0.16$} before approaching the \mbox{$\eta\rightarrow0$} limit rather flatly, as discussed before.
Our older fit and the new one roughly agree at similar masses,
but in the lower panel with the rescaled $\Lscaled$ it is obvious that the previous fit did not anticipate
the steep \mbox{$\eta\rightarrow0$} limit that we are now implementing through fitting the perturbative data.

\vspace{\baselineskip}

\section{Conclusions}
\label{sec:conclusions}

Using the hierarchical analysis approach to the three-dimensional parameter space
of nonprecessing quasicircular binary black hole (BBH) coalescences
introduced in Ref.~\cite{Jimenez-Forteza:2016oae},
we have developed a new model for the peak of the gravitational-wave luminosity of \bbh coalescence events, $\Lpeak$.
This model fit is based on the largest-yet combined set of numerical relativity (NR) results from \numNRcodes independent simulation codes,
as well as on perturbative numerical data for the \lmr regime not currently probed by \nr.

The result that \bbh[s] are, for a brief moment during their merger,
the most powerful astrophysical events
is already clear from dimensional analysis and simplified order-of-magnitude estimates\footnote{
Under some simplifying assumptions,
\mbox{$L_{\mathrm{GW}} \sim \tfrac{c^5}{G}\left(\tfrac{GM}{c^2R}\right)^2\left(\tfrac{v}{c}\right)^6$}
(see Example 3.9 of Ref.~\cite{Creighton:2011zz}),
so that for \event with a ``final`` separation
\mbox{$R \sim R_{\mathrm{S}} = 2GM/c^2$}
and velocity \mbox{$v \sim 0.5c$}~\cite{Abbott:2016blz}
the total mass $M$ scales out and
\mbox{$\Lpeak \sim 4 \times 10^{56}$~erg/s} is reproduced to within a factor of a few,
as it is also with the flux-based argument from Sec. III of Ref.~\cite{TheLIGOScientific:2016wfe}.},
as is the rough scaling of this \peaklum with mass ratio\footnote{
\mbox{$\Lpeak \sim \Lpeak|_{22} \propto \left|\,\dot{h}_{22}(t)\,\right|^2$}, cf.~\autoref{eq:Lpeaksum},
and $\left|\,\dot{h}_{22}(t)\,\right|$ goes to 0 linearly with \mbox{$\eta\rightarrow0$},
so the dominant $\Lpeak$ dependence is $\eta^2$.~\cite{Baker:2008mj}}.
Yet, only detailed \nr-calibrated fits allow for a precise understanding of the parameter-space dependence of $\Lpeak$.
Our new fit significantly reduces the residuals for most available \nr cases
in comparison with a previous version of this fitting procedure~\cite{T1600018} used in LIGO O1 data analysis
and an alternative fit~\cite{Healy:2016lce},
both calibrated to much smaller data sets.

We also characterized the quality of the luminosity data set considering various sources of \nr inaccuracies
and the compatibility between different simulation codes,
finding that the \peaklum's subdominant parameter dependencies are of a similar or even smaller order
than typical discrepancies between simulations.
This limits the level of detail to which we can model spin-difference effects,
though we can still improve over an equal-spin-only fit ($\chi_1=\chi_2$)
and find that the spin-difference dependence qualitatively matches expectations.
These statistical improvements,
wider parameter-space coverage
and systematic understanding of sources of uncertainty
can make the new fit a useful ingredient for future parameter estimation studies of \bbh events.

The final fit ansatz is given in \autoref{eq:3dansatz},
with coefficient estimates listed in Tables~\ref{tbl:eta_fit_coeffs}, \ref{tbl:S_fit_coeffs} and~\ref{tbl:final_fit_coeffs}.
Example implementations of this fit for Mathematica and python are available
as Supplementary Material~\cite{Keitel:2016krm-suppl,*Keitel:2016krm-anc},
along with an ASCII table of the data set.
The python implementation is equivalent to that included in the free software LALInference~\cite{lalsuite} package.

As more \gw detections are made, there will be more opportunities to infer the luminosities of stellar-mass \bbh systems.
In particular, the \aligo observatories in the USA~\cite{TheLIGOScientific:2014jea,TheLIGOScientific:2016agk},
Advanced Virgo~\cite{TheVirgo:2014hva} in Italy,
and forthcoming observatories in Japan~\cite{Somiya:2011np,Aso:2013eba} and India~\cite{Unnikrishnan:2013qwa}
are poised to become pivotal tools for earth-based \gw astronomy,
eventually enabling daily \bbh detections~\cite{Aasi:2013wya} over a wide parameter range.

The accuracy of the fit presented in this paper should be sufficient
for the expected sensitivity at least during the second \aligo observing run
and for ``vanilla'' \bbh events (similar masses, low spins, no strong precession),
with sampling uncertainties in mass ratio and spins still dominating over fit errors.
Still, a continued expansion of the \nr calibration set
and an improved understanding of higher-mode contributions, precession
and the transition from similar-mass to \emr regimes
will be important to improve the understanding of \bbh \peaklums and waveforms.

Meanwhile, this project of fitting \peaklums is an important step in extending
the ``Phenom'' waveform family~\cite{Husa:2015iqa,Khan:2015jqa,Hannam:2013oca,T1500602,PhenomPv2Paper},
as our analysis of higher-mode contributions
and the demonstration of joint calibration to \nr and perturbative \lmr data
can form the basis for improved modeling of full \imr waveforms.

\clearpage

\section*{Acknowledgments}

D.K., X.J. and S.H. were supported by the Spanish Ministry of Economy and Competitiveness grants
FPA2016-76821-P, CSD2009-00064 and FPA2013-41042-P,
the Spanish Agencia Estatal de Investigaci\'on,
European Union FEDER funds,
Vicepresid\`encia i Conselleria d'Innovaci\'o, Recerca i Turisme,
Conselleria d’Educaci\'o i Universitats del Govern de les Illes Balears,
and the Fons Social Europeu;
and D.K. recently also by the EU H2020-MSCA-IF-2015 grant 704094 GRANITE.

L.L. and M.H. were supported by Science and Technology Facilities Council (ST)
grants ST/I001085/1 and ST/H008438/1,
M.H. by European Research Council Consolidator Grant 647839,
S.K. by the STFC,
and M.P. by ST/I001085/1.
V.C. thanks ICTS for support during the S.N. Bhatt Memorial Excellence Fellowship Program 2014.

The authors thankfully acknowledge the computer resources at Advanced Research Computing (ARCCA) at Cardiff,
as part of the European PRACE Research Infrastructure on the clusters Hermit, Curie and SuperMUC,
on the U.K. DiRAC Datacentric cluster and on the BSC MareNostrum computer
under PRACE and RES (Red Espa\~nola de Supercomputaci\'on) grants,
2015133131, AECT-2016-1-0015, AECT-2016-2-0009, AECT-2017-1-0017.

We thank the CBC working group of the LIGO Scientific Collaboration,
and especially Ilya Mandel, Nathan Johnson-McDaniel, Alex Nielsen, Christopher Berry,
Ofek Birnholtz, Aaron Zimmerman and Juan Calderon Bustillo
for discussions of the general fitting method,
the previous results described in Ref.~\cite{T1600018},
as well as the current results and manuscript.
This paper has been assigned document number \mbox{\dcc}.

\addtolength{\parskip}{-0.5\baselineskip}


\vspace{12\baselineskip}

\appendix*

\section{NR data investigations}
\label{sec:appendix-nr}

As a first estimate of the \emph{overall} accuracy of the \peaklum data set,
we study the differences between results from different codes for equal initial parameters.
We then give additional details on the possible error sources listed in \autoref{sec:data-NR}
and on the properties of higher modes,
and discuss the \NRcountOutliers cases not used in the calibration set.

\subsection{Comparison between different codes}
\label{sec:appendix-nr-duplicates}

\begin{figure}[t]
 \includegraphics[width=\columnwidth]{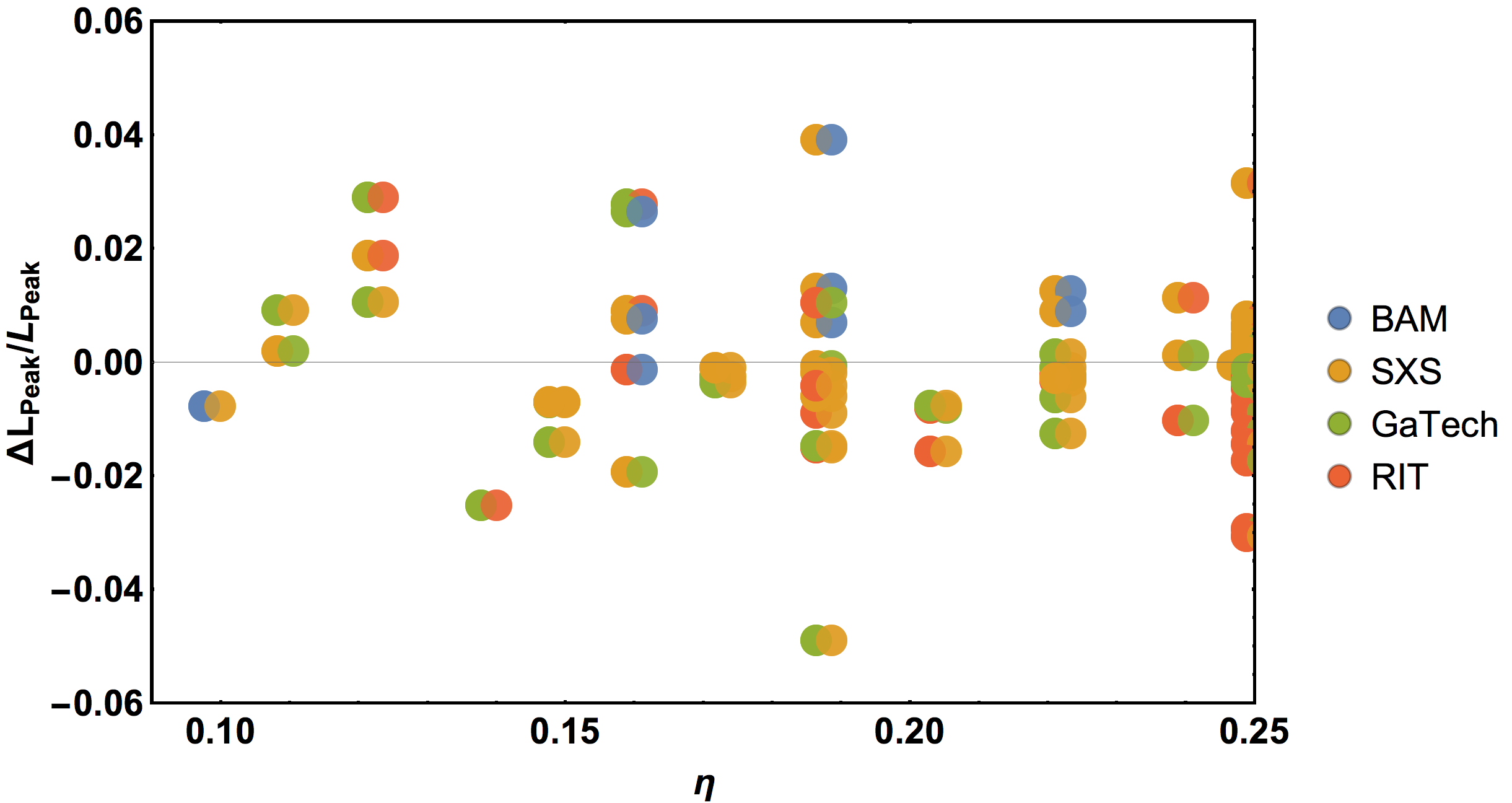}
 \caption{
  \label{fig:duplicates2D}
  Relative differences in the \peaklum for equal-parameter configurations from different \nr codes,
  shown against the symmetric mass ratio $\eta$.
  Pairs of simulations are shown with a small horizontal offset for ease of visual identification.
 }
\end{figure}

To analyze typical deviations between results from different \nr codes,
we identify simulations with initial \bh parameters equal to within numerical accuracy,
with a tolerance criterion
\begin{equation}
\label{eq:NRtolerance}
\left| \lambda_i-\lambda_j \right| \leq \epsilon = 0.0002 \; \text{for} \; \lambda_i=\left\lbrace \eta_i, \, \chi_{1i}, \, \chi_{2i}  \right\rbrace \,.
\end{equation}
This threshold was found in Appendix A of Ref.~\cite{Jimenez-Forteza:2016oae} to be strict enough to reliably identify equivalent initial configurations,
and is also tolerant enough to accommodate the minor relaxation of parameters after the initial ``junk'' radiation
which may be different between codes.
In \autoref{fig:duplicates2D} we show the relative difference in $\Lpeak$ between such matching cases,
including the nonspinning \mbox{$q=4$} case where we have results from all four codes
and a few triple coincidences.
The set of these tuples is too sparse for clear conclusions
on the parameter-space dependence of discrepancies between codes,
though there might be some indication of increasing differences at large positive spins,
which are particularly challenging to simulate
due to increased resolution requirements for capturing the larger metric gradients in the near-horizon zone.
We find many pairs with differences below 1\%,
but also several up to a few \% even at not particularly challenging configurations.

This study gives a useful overall estimate of the possible error magnitude on the \nr data set:
while certainly many simulations are accurate to more than the few-\% level,
in general for any given simulation that does not have a paired case from another code,
or at least nearby neighbors in parameter space,
we cannot confidently assume that the errors will be low.
This affects in particular the unequal-spin cases,
where due to the much larger 3D parameter space very few duplicates exist.
On the other hand, for equal spins --
and particularly for the densely covered nonspinning or equal-mass subsets --
we can use the duplicates analysis to make a very strict selection of calibration points,
allowing the subpercent calibration demonstrated in Secs.~\ref{sec:fit-1d-eta} and~\ref{sec:fit-1d-spin}.
The specific decisions are detailed below in Appendix~\ref{sec:appendix-nr-outliers}.

\begin{figure}[t]
\includegraphics[width=\columnwidth]{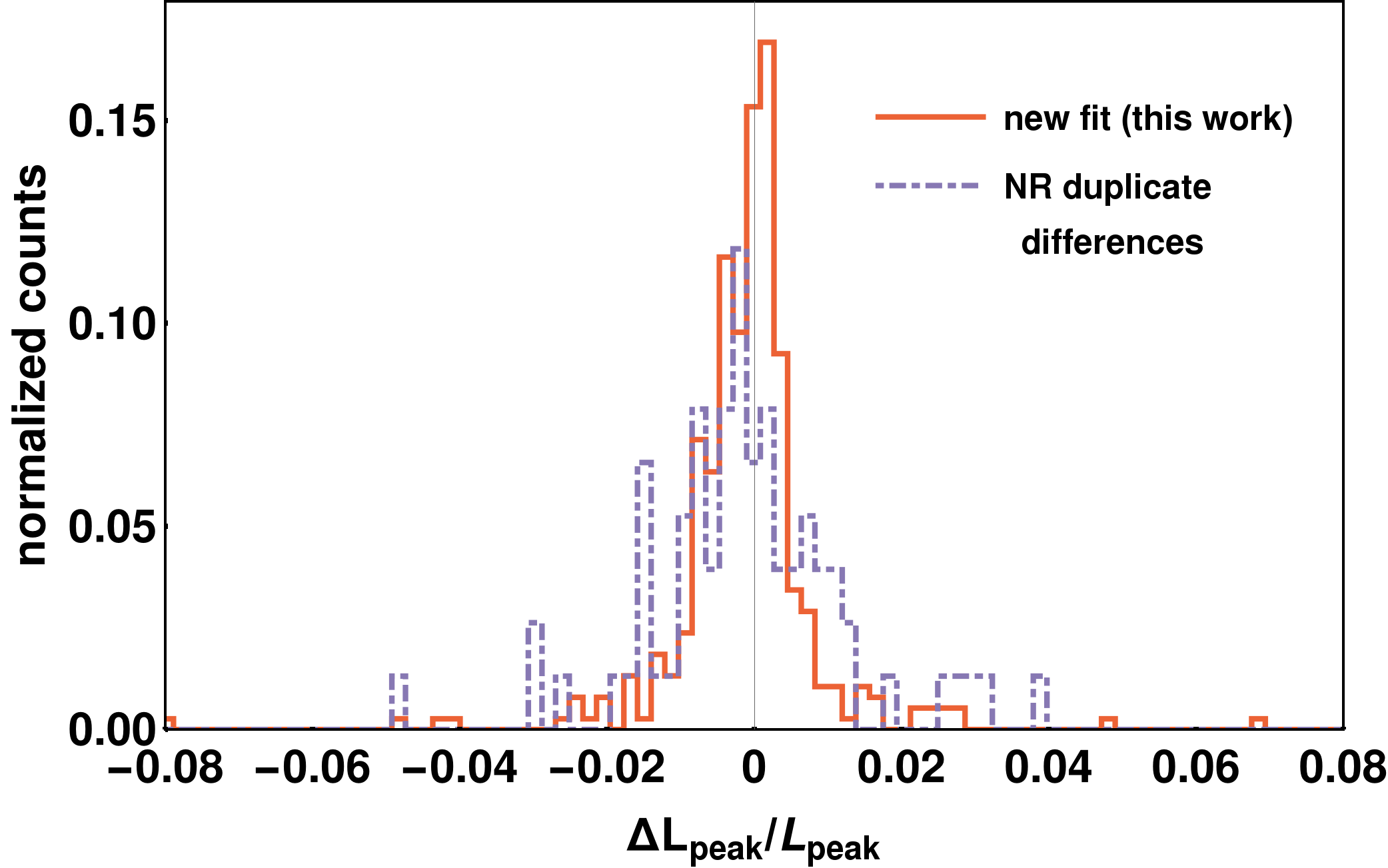}
 \caption{
  \label{fig:duplicates-residuals-histogram}
  Comparison of the distribution of relative fit errors
  (\nr only, same set as in \autoref{fig:3d_residuals_hist})
  and of differences between \nr codes for equivalent parameters.
 }
\end{figure}

As shown in the histograms of \autoref{fig:duplicates-residuals-histogram},
the overall distribution of (relative) differences between equivalent configurations
is of a similar width than that of the fit residuals.
This demonstrates that we are indeed not overfitting the data,
but also that one would need to characterize the accuracy of all \nr cases to a significantly lower level
to extract more information on subdominant effects.

\vspace{\baselineskip}

\subsection{Luminosity computation from \texorpdfstring{$\psi_{4}$}{psi4}}
\label{sec:appendix-nr-psi4tolum}

Equations~\eqref{eq:psi4} and~\eqref{eq:Lpeaksum} describe
the general computation of \peaklums from the Weyl curvature component $\psi_4(t,\vec{r})$.
This conversion is normally performed by either
integrating $\psi_4(t,\vec{r})$ twice in time 
or by first applying a Fourier transform to the data,
both to finally obtain the strain $h(t)$.
However, both strategies for computing $h(t)$
carry the same technical issue:
nonlinear drifts in the final strain
as a consequence of the characteristic low-frequency noise
present when operating on finite segments of data.

\begin{figure}[t]
 \includegraphics[width=\columnwidth]{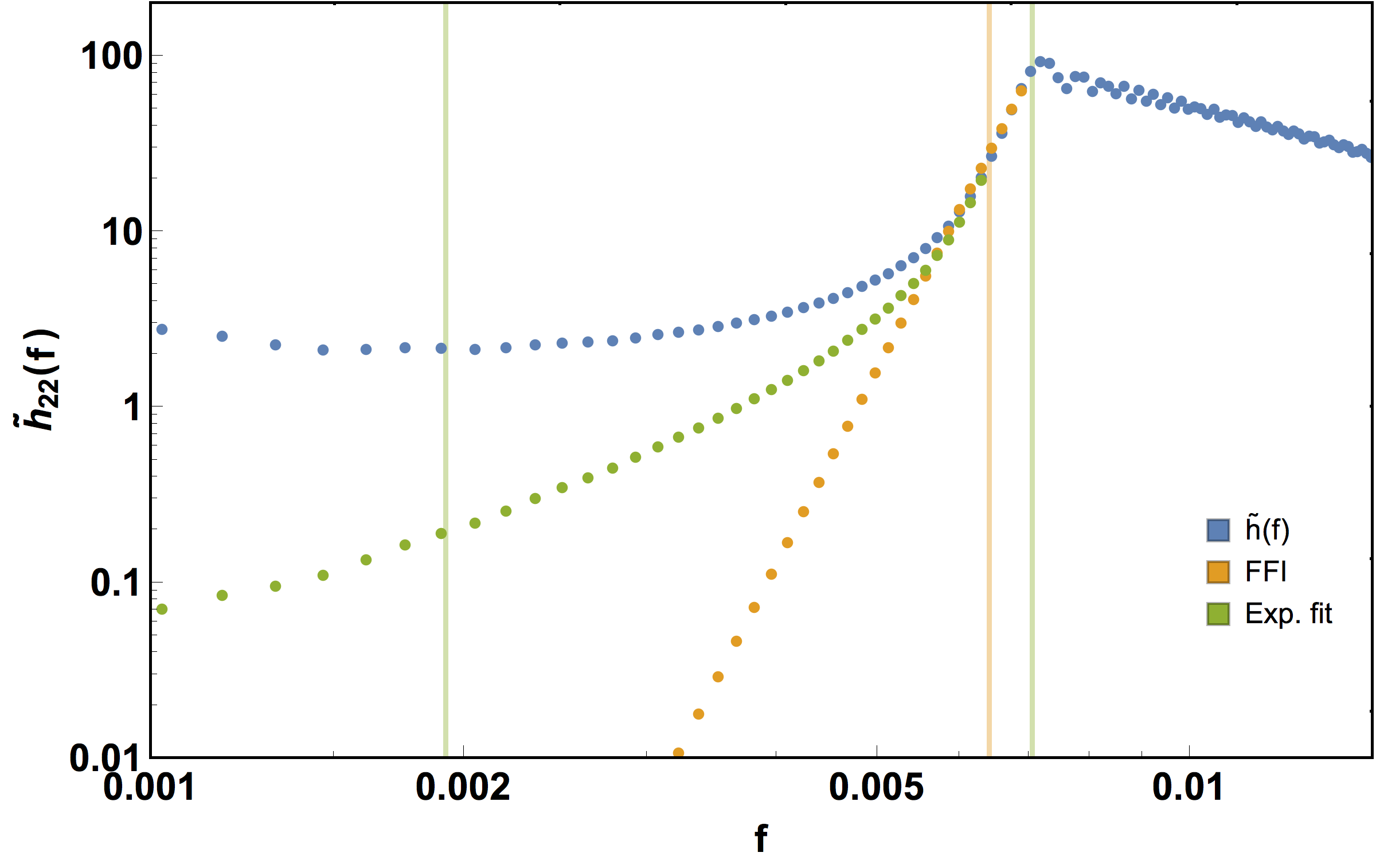}
 \caption{
  \label{fig:ffiflownoise}
  Comparison of the \ffi algorithm and
  the exponential-fit method
  for the low-frequency regime of $\widetilde{h}_{22}(f)$,
  for the example of the \mbox{$q=10$} nonspinning SXS waveform ``SXS:BBH:0185".
  The green vertical lines indicate the local maximum and minimum of $\widetilde{h}_{22}(f)$
  and the yellow line is at the tuned value $f_0$ for the \ffi.
 }
\end{figure}

This problem was already solved in Ref.~\cite{Reisswig:2010di}
by means of the \ffi algorithm, which we briefly describe here.
One takes the Fourier-domain strain as
\vspace{\baselineskip}
\begin{equation}
 \label{eq:FFI}
 \widetilde{h}\,(f)= 
 \begin{cases}
  - \frac{\widetilde{\psi_4}(f)}{f^2},   & \text{if} \quad f\geq f_0 \,, \\\\
  - \frac{\widetilde{\psi_4}(f)}{f_0^2}, & \text{if} \quad f < f_0 \,.
 \end{cases}
\end{equation}
\vspace{0.5\baselineskip}
All physical frequency content must be contained in
$\left[ f_0,f_{\mathrm{QNM}}\right]$
where $f_0$ must be tuned close to the lowest physical frequency for a given mode
and $f_{\mathrm{QNM}}$ is the quasinormal mode frequency of the same mode.
Thus, a proper selection of $f_0$ down-weights contributions from the low-frequency regime,
driving these effects to zero --  see \autoref{fig:ffiflownoise}.

As a consistency check,
we have developed an alternative conversion from $\psi_4(t)$ to $h(t)$
which avoids the step of tuning $f_0$.
In \autoref{fig:ffiflownoise} we show an example of the Fourier transform $\widetilde{h}_{22}(f)$ of the dominant-mode strain.
In general, both local maxima and minima are located in the \mbox{$\left[  f_{\min} \sim 0.5 f_0, f_{\max} \sim 1.2 f_0 \right]$} range.
The plotted low-frequency behavior occurs for any $\widetilde{h}_{22}(f)$
independently of the system's physical parameters,
as a consequence of the finiteness and discreteness of the time-domain waveforms.
Empirically we found that the data in \mbox{$\left[  f_{\min} , f_{\max}  \right]$}
can be well fit with an exponential ansatz,
which is then extended to all data in \mbox{$\left[ 0 , f_{\max}  \right]$}
and combined with the original data above $f_{\max}$:
\begin{equation}
\label{eq:expansatz}
    \widetilde{h}_{\ell m}(f)= 
\begin{cases}
    - \frac{\widetilde{\psi}_{4,\ell m}(f)}{f^2},      & \text{if } f \geq f_{\max}   \,, \\
    a \exp^{-(b -f)} f^c \mathrm{e}^{i \phi_{\ell m}}, & \text{if } f < f_{\max} \,,
\end{cases}
\end{equation}
where $\phi_{\ell m}$ is the original $\psi_4$ phase.
The split in the fit coefficients $a$ (amplitude) and $b$ (peak position) is introduced here
so that good starting values for the fit function can be picked more easily.
With this approach, we smoothly drive the low-frequency noise to zero,
eliminating nonphysical artifacts in the Fourier-domain data.

We find that the difference between \peaklums from the two different algorithms,
when $f_0$ is optimally selected,
is generally negligible,
e.g. it is about $\sim 0.05\%$ in the example of the \mbox{$q=10$} nonspinning SXS waveform,
and no significantly larger discrepancies have been found over the data set.
So this effect is negligible for our analysis
in comparison with other sources of uncertainty.

\vspace{\baselineskip}

\subsection{Extrapolation}
\label{sec:appendix-nr-extrapol}

The \nr waveforms used in this paper are extracted at finite radii,
which implies ambiguities, in particular due to gauge effects.
We therefore extrapolate all waveforms to null infinity,
where unambiguous waveforms can be defined.
This allows us to assemble a consistent set of \peaklum values for different codes,
and to estimate the errors due to finite radius effects.

\begin{figure}[t!]
 \includegraphics[width=\columnwidth]{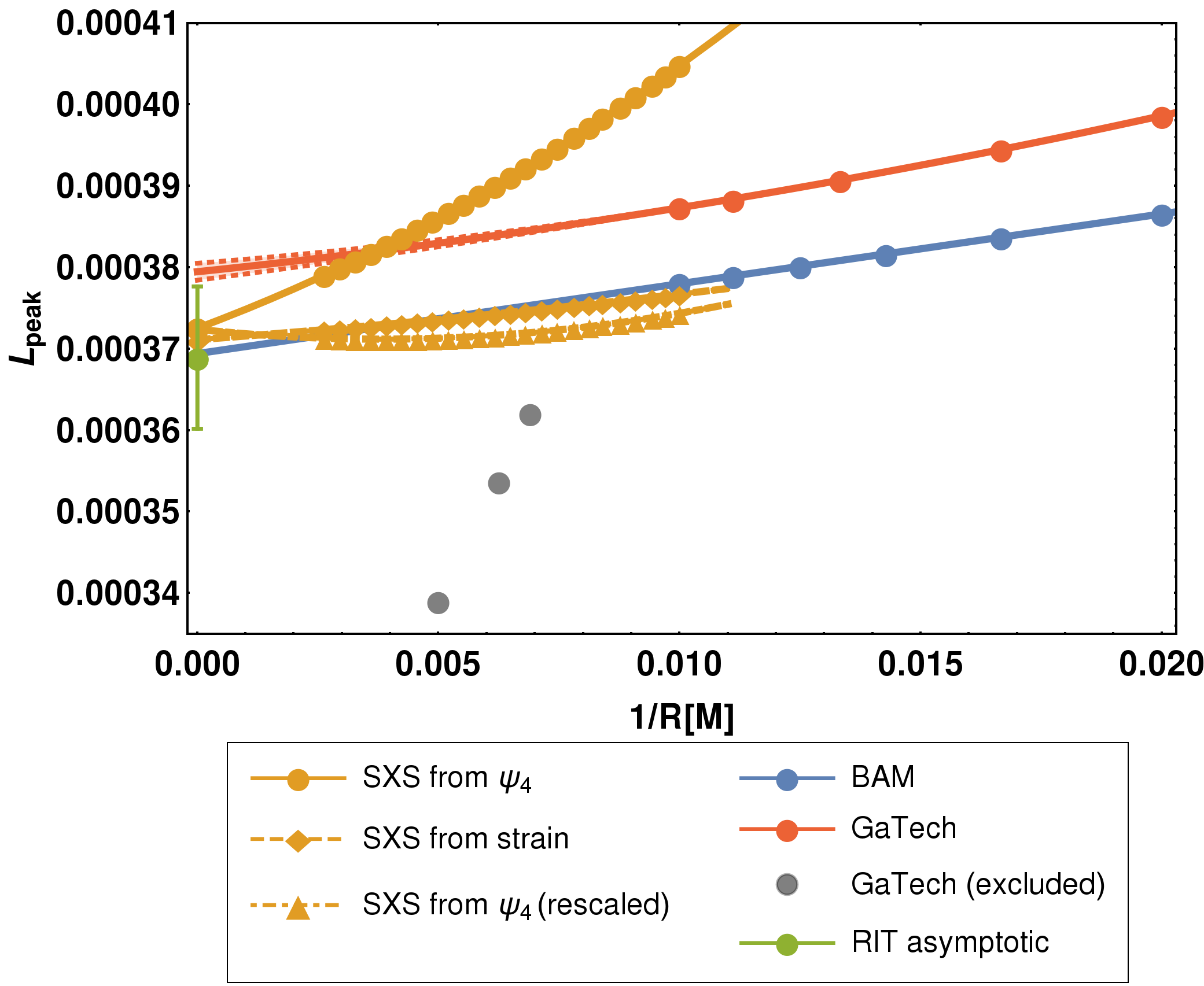}
 \caption{
  \label{fig:extrapolationq4}
  \mbox{$q=4$} nonspinning example of extrapolation from finite radii for \BAM, SXS and GaTech,
  with second-order fits for SXS and GaTech and linear for \BAM;
  as well as the RIT value extrapolated with the method of Ref.~\cite{Nakano:2015pta}
  and its error bar also containing a finite-resolution estimate.
  In this case we find consistent values from \BAM, SXS and RIT,
  with the GaTech case being an outlier.
  The \mbox{$R>100$M} GaTech data would make the trend more inconsistent,
  and are excluded from extrapolation.
  SXS luminosities computed from strain,
  or from $\psi_4$ but with the $\alpha^4$ rescaling discussed in the text,
  show a flatter finite-$R$ behavior more similar to the other codes,
  and extrapolated values consistent with the luminosity from $\psi_4$.
 }
  \vspace{4\baselineskip}
 \includegraphics[width=\columnwidth]{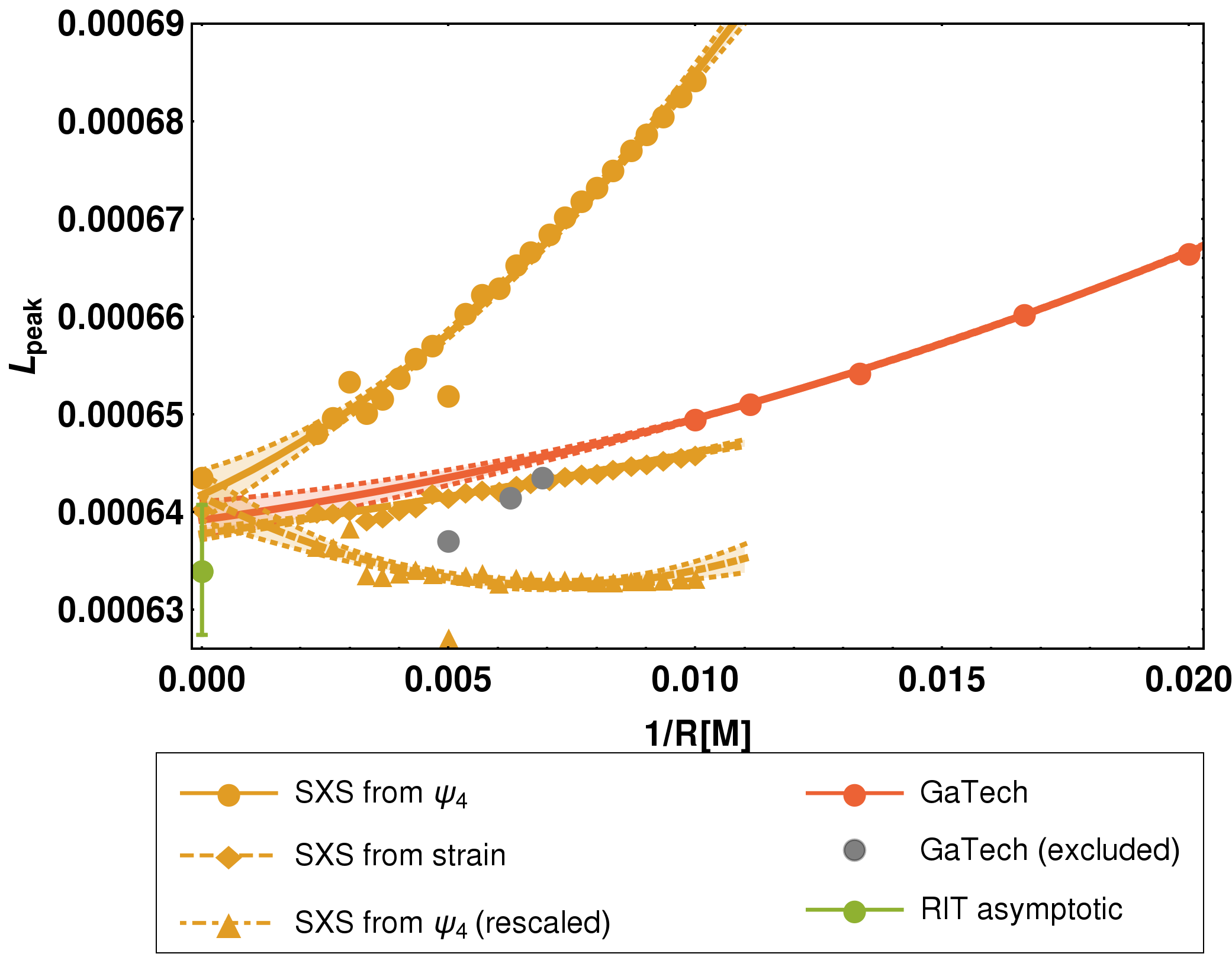}
 \caption{
  \label{fig:extrapolationq2d5}
  \mbox{$q=2.5$} nonspinning example of extrapolation behavior for SXS and GaTech,
  compared with the extrapolated RIT value.
  In this case we find consistent GaTech and RIT values,
  with the asymptotic SXS computed from $\psi_4$ a marginal outlier.
  Extrapolating the finite-radius \peaklums computed from $\psi_4$ with rescaling, or from strain,
  seems to improve consistency in this example, which however needs further study before applying it to the whole data set.
 }
\end{figure}

However, the extraction properties of the codes are not equal,
and thus we have extrapolated the available waveforms following the following prescriptions.
\begin{enumerate}[(i)]
\item \BAM:
      We have calculated $\Lpeak$ at each finite radius
      and then performed a linear-in-$1/R$ extrapolation
      using only the well-resolved extraction radii.
      The maximum used for any case is \mbox{$R \leq 180 M$},
      but for some cases significantly fewer radii can be used for a robust extrapolation,
      depending on simulation grid resolutions.
\item GaTech:
      $\Lpeak$ is again calculated at finite radii
      and then extrapolated with a fit quadratic in $1/R$,
      only using up to \mbox{$R \leq 100 M$}
      because the slope generally changes for higher radii;
      this choice of extrapolation order and radius cut yields
      the most consistent results with other codes in the analysis of equivalent configurations.
\item SXS:
      These waveforms are already provided at second-, third- and fourth-order polynomial extrapolation,
      and we compute $\Lpeak$ from these data products,
      after a correction~\cite{Boyle:2013nka,Boyle:2014ioa,Boyle:2015nqa} for center-of-mass drift,
      using the 2nd order extrapolation as the preferred value following Refs.~\cite{Boyle:2007ft,Bustillo:2015ova}.
      We use waveforms based on the Weyl scalar $\psi_4$,
      but also compare with waveforms based on a computation of the strain.
      The SXS $\psi_4$ data use a definition of null-tetrad which is different from
      their Regge-Wheeler-Zerilli strain data~\cite{Regge:1957td,Zerilli:1970se,Sarbach:2001qq,Rinne:2008vn},
      and from the definition used in other codes.
      For the luminosity this difference corresponds to an overall scaling factor of the lapse function to the fourth power
      as a consequence of the difference between
      Eqs.~(30)--(33) in Ref.~\cite{Bruegmann:2006at}
      and Eqs.~(11)--(12) in Ref.~\cite{Boyle:2007ft}.
      A rough correction for the different tetrad scaling used to compute the Weyl scalar $\psi_4$
      is to multiply it by $\alpha^4$ with \mbox{$\alpha = 1 - 2\Mf/R$}, where $\Mf$ is the final mass
      and $R$ is an approximation to the luminosity distance using the standard relation
      with the
      isotropic radial coordinate for the Schwarzschild spacetime.
      (Compare also with the analysis in Ref.~\cite{Nakano:2015pta}.)
      Comparisons of SXS luminosities computed from $\psi_4$, strain, and heuristically rescaled $\psi_4$
      with data from other codes
      are included in Figs.~\ref{fig:extrapolationq4} and~\ref{fig:extrapolationq2d5}.
\item RIT:
      The luminosity data provided in Ref.~\cite{Healy:2016lce}
      uses the extrapolation method of Ref.~\cite{Nakano:2015pta}.
\end{enumerate}

In \autoref{fig:extrapolationq4} we show the only configuration,
the nonspinning \mbox{$q=4$} case,
for which we have data from all \numNRcodes codes.
This includes \peaklums computed from the finite-radius strain data available as additional data products from SXS
to cross-check the pre-extrapolated value.
We see that extrapolation for \mbox{$R\rightarrow \infty$} reduces discrepancies in $\Lpeak$ between the different codes,
but cannot completely alleviate it in this case.
Another similar example is shown in \autoref{fig:extrapolationq2d5}
for a \mbox{$q=2.5$} nonspinning configuration where we have three simulations from SXS, GaTech and RIT,
with the GaTech and RIT values being more consistent with each other than with SXS in this case.

The uncertainties of extrapolation fits for \BAM, SXS and GaTech
can be estimated by the standard deviation on the intersection parameter
(equivalent to the confidence interval on the extrapolation to \mbox{$1/R=0$}).
For the plotted nonspinning \mbox{$q=4$} case,
these are smaller than the remaining largest difference
between the results from GaTech and other codes,
while for \mbox{$q=2.5$} the uncertainties are almost wide enough to make the results marginally consistent.
For some other cases, these uncertainties can reach up to a few \%,
especially when we want to be conservative and take the maximum of
(i) the statistical uncertainty for the standard extrapolation-order choice
and (ii) the difference between this and the closest alternative order.
In general, such an uncertainty estimate cannot provide information about any systematics
present in the data from different codes,
and indeed for example we find that for \BAM the purely statistical extrapolation uncertainties
are much smaller in some high-$q$ cases than for low-$q$ cases which are generally considered more reliable.

Hence, a study of the extrapolation uncertainties over the whole parameter space
is useful in gaining an understanding of the properties of the different codes,
but cannot directly be used as a measure of total \nr uncertainties.

\vspace{\baselineskip}

\subsection{Finite resolution}
\label{sec:appendix-nr-resolution}

The error contribution from finite numerical resolution can be estimated through convergence tests,
reproducing the same configuration at different resolutions.
This multiplies computational cost
and is hence only practical for a small set of representative simulations.
Comparisons of \nr results at different resolutions have been discussed e.g.
in Refs.~\cite{Husa:2007hp,Husa:2015iqa} for the \BAM code
and in Ref.~\cite{Lovelace:2016uwp} for \event-like SXS and RIT waveforms.
For \peaklums specifically,
multiresolution results are available for some \BAM and RIT simulations.

The error estimates presented for \NRcountRIT RIT simulations in Tables XI--XIII of Ref.~\cite{Healy:2016lce}
combine finite-resolution and finite-radius contributions,
but for four cases at mass ratios \mbox{$q\approx\{1,1.33,2,3\}$} and different spins
we can extract error estimates due to finite resolution only
from Tables XIV--XVII,
by comparing $\Lpeak$ extrapolated to \mbox{$r_{\mathrm{obs}}=\infty$}
for the highest finite resolution with $\Lpeak$ extrapolated to both infinite radius and infinite resolution.
This yields relative error estimates $\Delta\Lpeak/\Lpeak$ of about 0.8--1.6\%.

These estimates fall well within the distributions of
our fit residuals
and of the ''duplicates`` study,
as shown in \autoref{fig:duplicates-residuals-histogram},
and from comparison with the combined RIT error estimates
and with Appendix~\ref{sec:appendix-nr-extrapol} on extrapolation from finite extraction radius
we also see that these two error contributions are typically on a comparable level.

\begin{figure}[thbp]
 \includegraphics[width=\columnwidth]{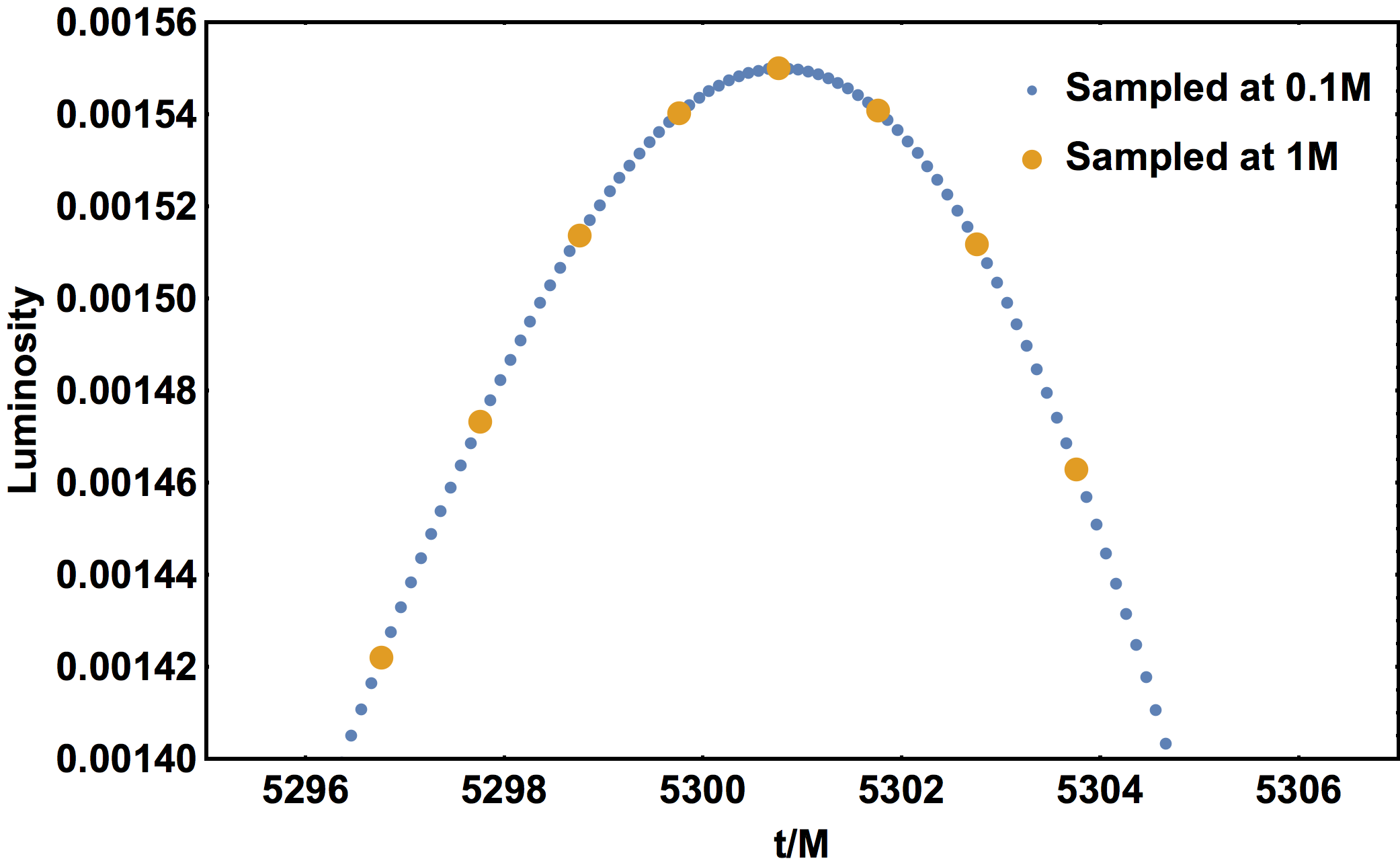}
 \caption{
  \label{fig:discreteness}
  Luminosity computed with different sampling rates in time,
  for the nonspinning \mbox{$q=10$} SXS waveform ``SXS:BBH:0185"
  or ``BBH\_CFMS\_d11d2\_q10\_sA\_0\_0\_0\_sB\_0\_0\_0".
  }
\end{figure}

Convergence testing for BAM runs at the particularly challenging \mbox{$q=18$} mass ratio
has previously been discussed in \cite{Husa:2015iqa},
indicating generally robust behavior.
Estimating the finite-difference error as
the difference between the highest resolution and a Richardson extrapolation
yields $<1\%$ for both $\psi_4$ and $\Lpeak$
in the nonspinning \mbox{$q=18$} case,
and for the \mbox{$\chi_1=0.4$} simulation we find
$\approx1\%$ for $\psi_4$
and $\approx4\%$ for $\Lpeak$.
We already knew that these simulations at high mass ratios
must have wider overall error bars
due to e.g. the higher-mode contributions.

Hence, finite resolution can be conjectured to be a nondominant,
but also non-negligible contribution to the total uncertainty budget,
while a point-by-point evaluation is hindered by the large computational cost.

\vspace{\baselineskip}

\subsection{Peak accuracy}
\label{sec:appendix-nr-peak}

Since we are dealing with discrete numerical data sets,
the peak finding might also be a problem
if the sampling is not fine enough;
particularly for high mass-ratio cases
where the higher modes become more relevant
and it is important to sample each mode accurately
so that the overall peak profile is not washed out.
We have estimated this contribution to \nr uncertainties
by applying two different time samplings to the data:
for the actual $\Lpeak$ values used in this paper, we use \mbox{$\Delta t = 0.1 M$},
while here we compare also with a coarser \mbox{$\Delta t = 1 M$}
to illustrate the possible loss of accuracy.

In \autoref{fig:discreteness} we show,
for an SXS mass-ratio 10 nonspinning case,
that the uncertainty contribution,
measured as the difference between two points bracketing the peak,
would be about 1\% of the total \peaklum with the coarser sampling,
but is only about 0.05\% for the finer sampling that we actually use.
As a worst case, we found 0.2\% for the nonspinning \mbox{$q=18$} \BAM result.

\vspace{\baselineskip}

\subsection{Mode selection}
\label{sec:appendix-nr-modesel}

As introduced in \autoref{eq:Lpeaksum},
we compute \nr \peaklums for \BAM, SXS and GaTech waveforms as sums over all modes up to \mbox{$\lmax = 6$}.
The RIT luminosities from Refs.~\cite{Healy:2014yta,Healy:2016lce} use the same cutoff.
For the perturbative data from Refs.~\cite{Nagar:2006xv,Bernuzzi:2011aj,Harms:2014dqa} at \lmrs,
we use \mbox{$\lmax = 8$}.
These choices are based on studying the individual contribution of each mode to the total luminosity,
finding that \mbox{$\ell>6$} contributions are sufficiently small to be discarded for the \nr data
in comparison with other sources of uncertainty.

\begin{figure}[t!]
 \includegraphics[width=\columnwidth]{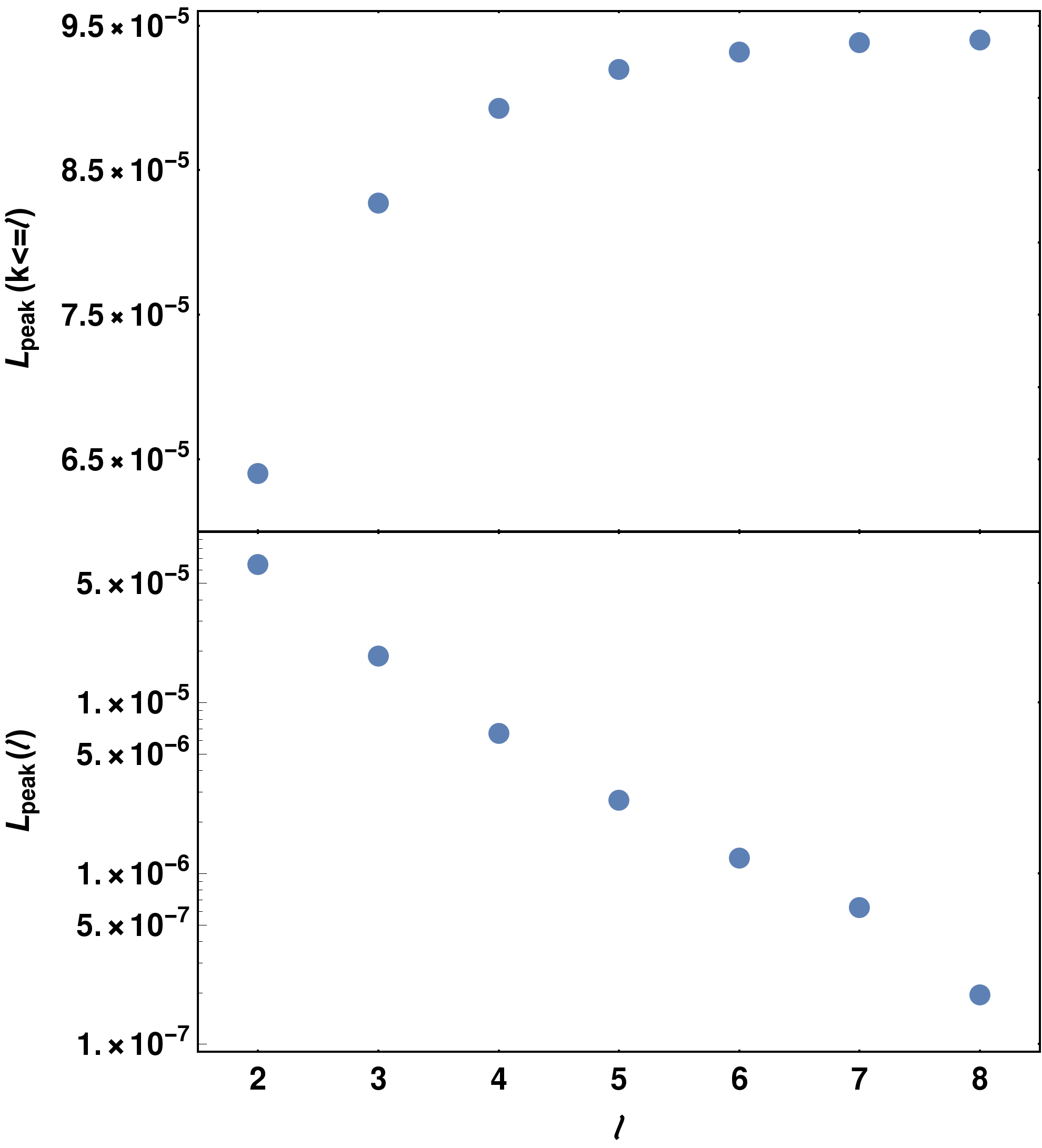}
 \vspace{-2\baselineskip}
 \caption{
  \label{fig:lLumdependence}
  Per-mode contributions to the total \peaklum
  for the same SXS case as in \autoref{fig:discreteness}.
  Top panel: Cumulative sum up to $\ell$.
  Lower panel: Natural logarithm of the luminosity contribution per $\ell$.
  Each point contains all $m$ for the given $\ell$.
  Similar behavior for \lmrs was found in Ref.~\cite{Bernuzzi:2010xj}.
 }
\end{figure}

As an illustrative example, we show in the top panel of \autoref{fig:lLumdependence}
the cumulative \peaklum when adding modes $\ell$ by $\ell$
(including all \mbox{$|m|\leq\ell$} at each step)
for the \mbox{$q=10$} nonspinning SXS waveform,
and the per-$\ell$ contributions in the lower panel.
The falloff of the higher-$\ell$ contributions to the global peak is expected to be exponential,
which is indeed found in this case.

\begin{figure}[t]
 \includegraphics[width=\columnwidth]{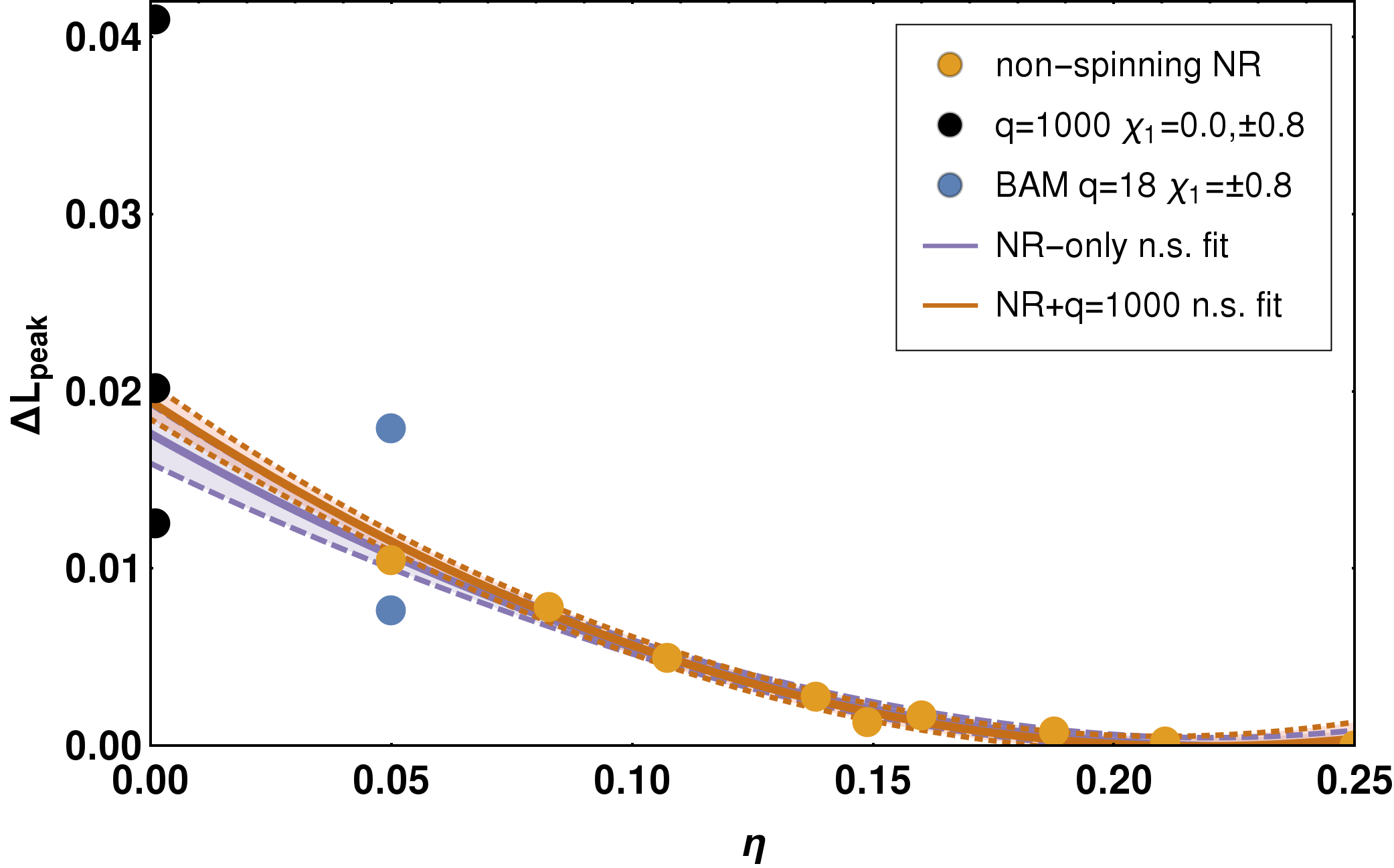}
 \caption{
  \label{fig:peaklosspercent}
  Relative loss in the \peaklum including modes up to \mbox{$\lmax=6$} against \mbox{$\lmax=8$},
  for nonspinning SXS cases up to \mbox{$q=10$}, a nonspinning \BAM case with \mbox{$q=18$}
  and the \mbox{$q=10^3$} Teukolsky result.
  Also shown are \mbox{$q=18$} and \mbox{$q=10^3$} results for \mbox{$\chi_1=+0.8$} (above the nonspinning line)
  and for \mbox{$\chi_1=-0.8$} (below),
  as well as the quadratic nonspinning fit from \autoref{eq:modesloss} to \nr data points only
  and a fit of the same order including the \mbox{$q=10^3$} point,
  with 90\% confidence intervals for both fits.
  }
\vspace{3\baselineskip}
 \includegraphics[width=\columnwidth]{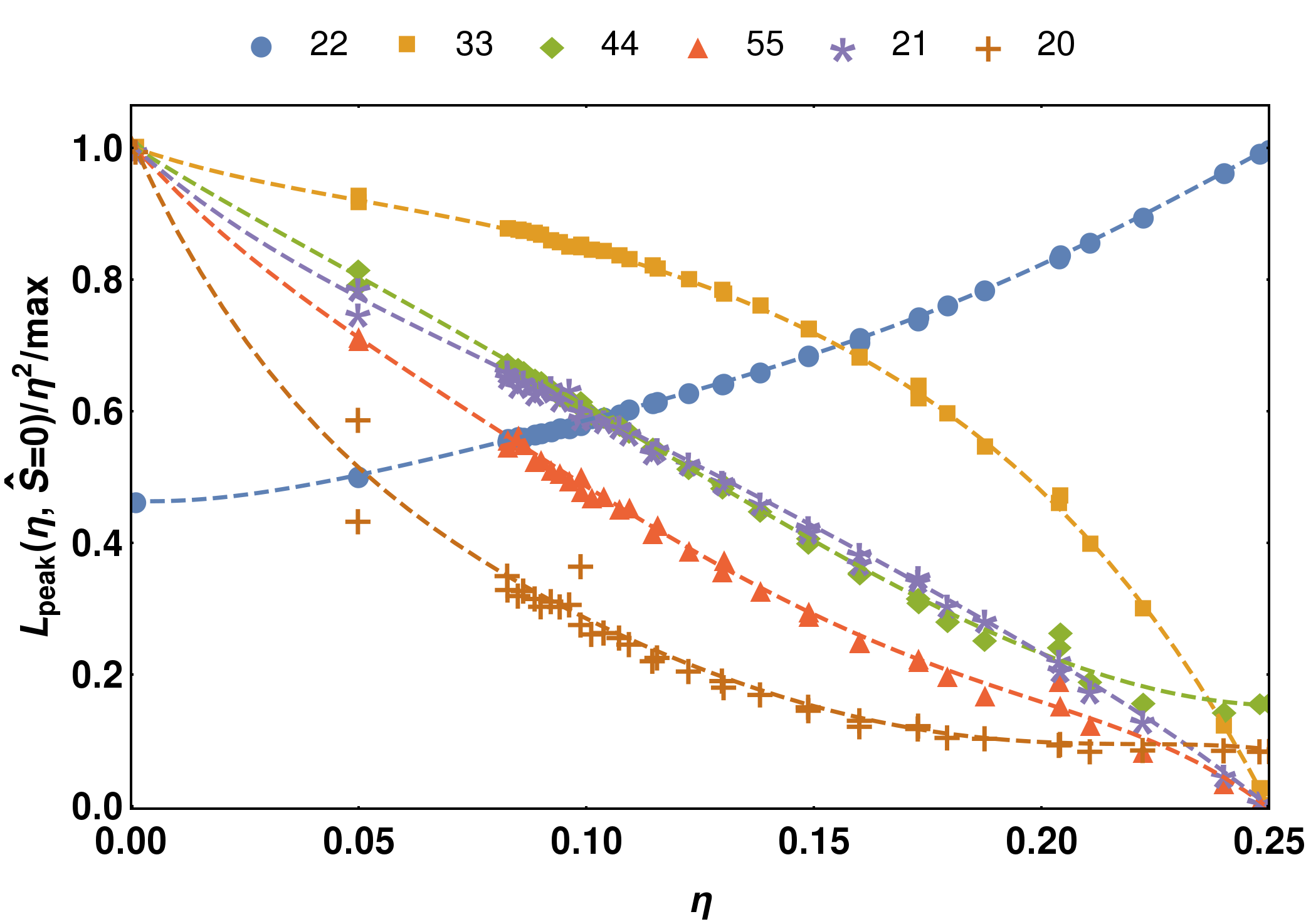}
 \caption{
  \label{fig:NRvsEMRIbymode}
  Comparison of rescaled \peaklums for nonspinning SXS and \BAM and perturbative \lmr data,
  for a small selection of modes.
  The points for each mode have been scaled by the maximum for that mode,
  which is at \mbox{$\eta=0.25$} for the 22 mode
  and at \mbox{$\eta\rightarrow0$} for the other modes.
  The connecting lines are fifth-order polynomial fits,
  which were not statistically optimized and just added to guide the reader's eye.
  This can be compared with the sum over modes in \autoref{fig:etafit}.
  As a guide to the overall strength of the individual modes,
  we list the nonrescaled maxima
  $\max_\eta\Lpeak^{\ell m}(\myS=0)$
  of each of the displayed modes \mbox{$(\ell m)=\{22,33,44,55,21,20\}$}:
  \mbox{$\{1.0\times10^{-3}, 5.9\times10^{-5}, 1.5\times10^{-5}, 5.3\times10^{-6}, 9.8\times10^{-6}, 6.3\times10^{-7}\}$}.
  }
\end{figure}

To quantify and extrapolate the loss generally expected for nonspinning configurations,
we have estimated the relative loss in $\Lpeak$ from not including the \mbox{$\ell = 7,8$} modes
for nonspinning SXS waveforms up to mass ratio \mbox{$q=10$} (maximum loss of 0.6\%)
and the nonspinning \BAM simulation at \mbox{$q=18$} (loss of 1\%),
and fit a quadratic function in $\eta$:
\begin{equation}
\label{eq:modesloss}
 \frac{\Delta \Lpeak}{\Lpeak} = 0.017611 - 0.153760 \eta + 0.334803 \eta^2 \,.
\end{equation}
This result is illustrated in \autoref{fig:peaklosspercent},
together with a marginally consistent fit when including the \mbox{$q=10^3$} Teukolsky result (loss of 2\%).
The \mbox{$\ell > 6$} contributions are smaller for negative spins and larger for positive spins,
as illustrated in the same figure with \mbox{$\chi_1=\pm0.8$} results at \mbox{$q=10^3$} and from \BAM at \mbox{$q=18$}.
The largest loss for any \nr case investigated is $\lesssim2$\% for the \mbox{$q=18$}, \mbox{$\chi_1=+0.8$} \BAM case,
which is a significant contribution to the overall error budget but still on the level of other error sources.
For the perturbative \lmr results,
with a worst-case \mbox{$\ell > 6$} of $\approx5$\%, 
we use \mbox{$\lmax=8$} instead,
so that the loss from \mbox{$\ell > 8$} is limited to $<1$\%.

\begin{table*}[thbp]
  \begin{tabular}{rrrrrrrlc}\hline\hline
  &$ \text{q} $&$ \chi _1 $&$ \chi _2 $&$ L_{\text{peak}} $&$ \text{$\Delta $L}_{\text{peak}} $&$ \text{$\Delta $L}_{\text{peak}}/L_{\text{peak}} $&$ \text{tag} $&$ \text{code} $\\\hline$
 1 $&$ 1.00 $&$ 0.20 $&$ 0.80 $&$ 0.00133540 $&$ -0.00001456 $&$ -0.011 $&$ \text{Q1.00$\_$0.20$\_$0.80} $&$ \text{RIT} $\\$
 2 $&$ 1.00 $&$ 0.25 $&$ 0.25 $&$ 0.00114910 $&$ -0.00001078 $&$ -0.009 $&$ \text{Q1.0000$\_$0.2500$\_$0.2500} $&$ \text{RIT} $\\$
 3 $&$ 1.00 $&$ 0.40 $&$ 0.80 $&$ 0.00143030 $&$ -0.00001617 $&$ -0.011 $&$ \text{Q1.00$\_$0.40$\_$0.80} $&$ \text{RIT} $\\$
 4 $&$ 1.00 $&$ 0.50 $&$ 0.50 $&$ 0.00132610 $&$ -0.00002155 $&$ -0.016 $&$ \text{Q1.0000$\_$0.5000$\_$0.5000} $&$ \text{RIT} $\\$
 5 $&$ 1.00 $&$ 0.80 $&$ 0.80 $&$ 0.00165190 $&$ -0.00005163 $&$ -0.031 $&$ \text{Q1.0000$\_$0.8000$\_$0.8000} $&$ \text{RIT} $\\$
 6 $&$ 1.00 $&$ 0.97 $&$ 0.97 $&$ 0.00185963 $&$ -0.00017055 $&$ -0.092 $&$ \text{d15$\_$q1$\_$sA$\_$0$\_$0$\_$0.97$\_$sB$\_$0$\_$0$\_$0.97$\_$ecc6e-4} $&$ \text{SXS} $\\$
 7 $&$ 1.00 $&$ -0.80 $&$ -0.80 $&$ 0.00075683 $&$ -0.00000522 $&$ -0.007 $&$ \text{d15$\_$q1$\_$sA$\_$0$\_$0$\_$-0.8$\_$sB$\_$0$\_$0$\_$-0.8} $&$ \text{SXS} $\\$
 8 $&$ 1.00 $&$ -0.95 $&$ -0.95 $&$ 0.00071785 $&$ -0.00001083 $&$ -0.015 $&$ \text{d15$\_$q1$\_$sA$\_$0$\_$0$\_$-0.95$\_$sB$\_$0$\_$0$\_$-0.95} $&$ \text{SXS} $\\$
 9 $&$ 1.10 $&$ 0.00 $&$ 0.00 $&$ 0.00102562 $&$ 0.00000646 $&$ 0.006 $&$ \text{D9$\_$q1.1$\_$a0.0$\_$m160} $&$ \text{GaT} $\\$
 10 $&$ 1.33 $&$ 0.50 $&$ 0.50 $&$ 0.00127610 $&$ -0.00001496 $&$ -0.012 $&$ \text{Q0.7500$\_$0.5000$\_$0.5000} $&$ \text{RIT} $\\$
 11 $&$ 1.33 $&$ 0.80 $&$ -0.80 $&$ 0.00113510 $&$ 0.00001336 $&$ 0.012 $&$ \text{Q0.7500$\_$-0.8000$\_$0.8000} $&$ \text{RIT} $\\$
 12 $&$ 1.33 $&$ 0.60 $&$ 0.80 $&$ 0.00144390 $&$ -0.00002266 $&$ -0.016 $&$ \text{Q1.33$\_$0.80$\_$0.60} $&$ \text{RIT} $\\$
 13 $&$ 1.50 $&$ 0.00 $&$ 0.00 $&$ 0.00092086 $&$ -0.00000929 $&$ -0.010 $&$ \text{Q0.6667$\_$0.0000$\_$0.0000} $&$ \text{RIT} $\\$
 14 $&$ 1.67 $&$ 0.00 $&$ 0.00 $&$ 0.00089059 $&$ 0.00001118 $&$ 0.013 $&$ \text{Q0.6000$\_$0.0000$\_$0.0000} $&$ \text{RIT} $\\$
 15 $&$ 2.00 $&$ 0.85 $&$ -0.85 $&$ 0.00104805 $&$ -0.00005372 $&$ -0.051 $&$ \text{q2$\_$-85$\_$85$\_$0.2833$\_$it2$\_$T$\_$96$\_$468} $&$ \text{BAM} $\\$
 16 $&$ 2.00 $&$ 0.60 $&$ 0.60 $&$ 0.00113005 $&$ -0.00001154 $&$ -0.010 $&$ \text{D11$\_$q2.00$\_$a0.60$\_$m200} $&$ \text{GaT} $\\$
 17 $&$ 2.00 $&$ 0.85 $&$ 0.00 $&$ 0.00119969 $&$ -0.00004465 $&$ -0.037 $&$ \text{q2$\_$0$\_$85$\_$0.566667$\_$T$\_$80$\_$360} $&$ \text{BAM} $\\$
 18 $&$ 2.00 $&$ 0.80 $&$ 0.80 $&$ 0.00133220 $&$ -0.00004371 $&$ -0.033 $&$ \text{Q2.00$\_$0.80$\_$0.80} $&$ \text{RIT} $\\$
 19 $&$ 2.00 $&$ 0.60 $&$ 0.50 $&$ 0.00109870 $&$ -0.00002568 $&$ -0.023 $&$ \text{Q0.5000$\_$0.5000$\_$0.6000} $&$ \text{RIT} $\\$
 20 $&$ 2.00 $&$ 0.80 $&$ 0.00 $&$ 0.00115110 $&$ -0.00004828 $&$ -0.042 $&$ \text{Q0.5000$\_$0.0000$\_$0.8000} $&$ \text{RIT} $\\$
 21 $&$ 2.50 $&$ 0.00 $&$ 0.00 $&$ 0.00064369 $&$ 0.00000637 $&$ 0.010 $&$ \text{BBH$\_$CFMS$\_$d16.9$\_$q2.50$\_$sA$\_$0$\_$0$\_$0$\_$sB$\_$0$\_$0$\_$0} $&$ \text{SXS} $\\$
 22 $&$ 3.00 $&$ 0.50 $&$ -0.50 $&$ 0.00067168 $&$ -0.00002270 $&$ -0.034 $&$ \text{q3$\_$-50$\_$50$\_$0.25$\_$T$\_$80$\_$400} $&$ \text{BAM} $\\$
 23 $&$ 3.00 $&$ 0.00 $&$ 0.00 $&$ 0.00051866 $&$ -0.00000761 $&$ -0.015 $&$ \text{D10$\_$q3.00$\_$a0.0$\_$0.0$\_$m240} $&$ \text{GaT} $\\$
 24 $&$ 3.00 $&$ 0.40 $&$ 0.00 $&$ 0.00065030 $&$ -0.00001591 $&$ -0.024 $&$ \text{D10$\_$q3.00$\_$a0.4$\_$0.0$\_$m240} $&$ \text{GaT} $\\$
 25 $&$ 3.00 $&$ 0.50 $&$ 0.80 $&$ 0.00074376 $&$ -0.00001267 $&$ -0.017 $&$ \text{Q0.3333$\_$0.8000$\_$0.5000} $&$ \text{RIT} $\\$
 26 $&$ 3.00 $&$ 0.60 $&$ 0.00 $&$ 0.00074392 $&$ -0.00003003 $&$ -0.040 $&$ \text{D10$\_$q3.00$\_$a0.6$\_$0.0$\_$m240} $&$ \text{GaT} $\\$
 27 $&$ 3.00 $&$ 0.67 $&$ 0.00 $&$ 0.00078909 $&$ -0.00002904 $&$ -0.037 $&$ \text{Q3.00$\_$0.00$\_$0.67} $&$ \text{RIT} $\\$
 28 $&$ 3.00 $&$ 0.80 $&$ -0.80 $&$ 0.00084159 $&$ -0.00002278 $&$ -0.027 $&$ \text{Q3.00$\_$-0.80$\_$0.80} $&$ \text{RIT} $\\$
 29 $&$ 3.00 $&$ 0.85 $&$ 0.85 $&$ 0.00107685 $&$ 0.00003335 $&$ 0.031 $&$ \text{BBH$\_$SKS$\_$d13.9$\_$q3$\_$sA$\_$0$\_$0$\_$0.850$\_$sB$\_$0$\_$0$\_$0.850} $&$ \text{SXS} $\\$
 30 $&$ 4.00 $&$ 0.75 $&$ 0.75 $&$ 0.00069840 $&$ 0.00001188 $&$ 0.017 $&$ \text{q4a075$\_$T$\_$112$\_$448} $&$ \text{BAM} $\\$
 31 $&$ 4.00 $&$ 0.75 $&$ 0.00 $&$ 0.00063280 $&$ -0.00002841 $&$ -0.045 $&$ \text{Q4.00$\_$0.00$\_$0.75} $&$ \text{RIT} $\\$
 32 $&$ 4.00 $&$ 0.00 $&$ 0.00 $&$ 0.00037948 $&$ 0.00000782 $&$ 0.021 $&$ \text{D10$\_$q4.00$\_$a0.0$\_$0.0$\_$m240} $&$ \text{GaT} $\\$
 33 $&$ 4.30 $&$ 0.00 $&$ 0.00 $&$ 0.00034217 $&$ 0.00000421 $&$ 0.012 $&$ \text{D9$\_$q4.3$\_$a0.0$\_$m160} $&$ \text{GaT} $\\$
 34 $&$ 4.50 $&$ 0.00 $&$ 0.00 $&$ 0.00031462 $&$ -0.00000329 $&$ -0.010 $&$ \text{D9$\_$q4.5$\_$a0.0$\_$m160} $&$ \text{GaT} $\\$
 35 $&$ 5.00 $&$ 0.80 $&$ 0.00 $&$ 0.00052483 $&$ -0.00000926 $&$ -0.018 $&$ \text{Q5.00$\_$0.00$\_$0.80} $&$ \text{RIT} $\\$
 36 $&$ 5.00 $&$ 0.00 $&$ 0.00 $&$ 0.00026999 $&$ -0.00000480 $&$ -0.018 $&$ \text{D10$\_$q5.00$\_$a0.0$\_$0.0$\_$m240} $&$ \text{GaT} $\\$
 37 $&$ 5.00 $&$ 0.40 $&$ 0.00 $&$ 0.00034792 $&$ -0.00001784 $&$ -0.051 $&$ \text{D10$\_$q5.00$\_$a0.4$\_$0.0$\_$m240} $&$ \text{GaT} $\\$
 38 $&$ 6.00 $&$ 0.00 $&$ 0.00 $&$ 0.00020707 $&$ -0.00000395 $&$ -0.019 $&$ \text{Q0.1667$\_$0.0000$\_$0.0000} $&$ \text{RIT} $\\$
 39 $&$ 6.00 $&$ 0.00 $&$ 0.00 $&$ 0.00021325 $&$ 0.00000234 $&$ 0.011 $&$ \text{D10$\_$q6.00$\_$a0.00$\_$0.00$\_$m280} $&$ \text{GaT} $\\$
 40 $&$ 6.00 $&$ 0.20 $&$ 0.00 $&$ 0.00023419 $&$ -0.00000829 $&$ -0.035 $&$ \text{D10$\_$q6.00$\_$a0.20$\_$0.00$\_$m280} $&$ \text{GaT} $\\$
 41 $&$ 18.00 $&$ -0.80 $&$ 0.00 $&$ 0.00006179 $&$ 0.00003868 $&$ 0.626 $&$ \text{q18a0aM08c025$\_$96$\_$fine} $&$ \text{BAM} $\\
\hline\hline\end{tabular}
 \vspace{\baselineskip}
 \caption{
  \label{tbl:outliers}
  NR cases from the source catalogs not included in the fit calibration,
  for reasons detailed in the text.
 }
\end{table*}

Another useful investigation is to consider the $\eta$ dependence,
and especially the \mbox{$\eta\rightarrow0$} behavior,
for individual modes.
Fitting $\Lofeta$ in \autoref{sec:fit-1d-eta} we found,
as illustrated in \autoref{fig:etafit},
that the \peaklum of all modes summed up to \mbox{$\lmax=6$},
after scaling out the dominant $\eta^2$ dependence,
is not a monotonic function towards low $\eta$.
The increasing relative amplitudes of higher-order modes at low $\eta$
have been studied with \nr results previously~\cite{Berti:2007fi,Baker:2008mj,Kelly:2011bp,Pekowsky:2012sr,Bustillo:2015qty},
but with our large \peaklum data set we can now investigate the slope more closely.

Repeating the same comparison as in \autoref{fig:etafit}
of rescaled nonspinning \peaklums
between \nr (SXS+\BAM nonspinning) and perturbative \lmr data,
but for individual modes,
we find --
as shown in \autoref{fig:NRvsEMRIbymode} for a subset of modes --
that these are all monotonic as \mbox{$\eta\rightarrow0$};
however, the slopes are very different,
with the dominant 22 mode falling off faster than $\eta^2$
and the subdominant and higher modes falling off much slower,
consistent with the general expectation of stronger contributions at low $\eta$.
This finding of monotonicity in each mode increases our trust in the combination of \nr and perturbative results,
and the nonmonotonicity of the rescaled \peaklums after summing the modes
can thus be explained as a superposition of these counteracting trends in the individual modes.

\vspace{\baselineskip}

\subsection{Cases not used in fit calibration}
\label{sec:appendix-nr-outliers}

Of the full catalog of \NRcountTotal \nr simulations from \numNRcodes codes,
we have only used \NRcount to calibrate our new fit.
Of the \NRcountOutliers removed cases, \NRcountOutliersEqS are non-spinning or equal-spin configurations.
Of these, \NRcountOutliersTuples belong to one of the pairs or groups of equivalent initial parameters
identified in Appendix~\ref{sec:appendix-nr-duplicates},
with differences between the paired results inconsistent at a level
higher than the fit residuals we can otherwise achieve in the corresponding subspace fit;
or they are individual points inconsistent with an otherwise consistent set of direct neighbors.
In these cases we removed from each tuple the case most discrepant with the others and with the global trend.
This includes for example the GaTech \mbox{$q=4$} and SXS \mbox{$q=2.5$} nonspinning cases
shown in the extrapolation comparisons of Figs.~\ref{fig:extrapolationq4} and~\ref{fig:extrapolationq2d5},
or the SXS \mbox{$\left(q=1,\myS=0.97\right)$} point whose luminosity seems inconsistent with other \mbox{$q=1$}, high-spin SXS results.

We emphasize that in the one-dimensional fits for nonspinning and equal-mass-equal-spin \bbh[s]
we calibrate the fits to subpercent accuracies,
so that this is a very strict criterion for removing cases,
which mainly serves to guarantee a very clean calibration of the well-covered subspaces
and dominant effects so that in the later steps we have a better chance of isolating
and extracting subdominant effects from the general, more noisy data set.
In terms of total absolute or relative errors compared with the whole \nr data set,
several of these cases are not overly inaccurate,
and we do not imply that necessarily there are data quality issues with the waveforms from which the luminosities are calculated.

The remaining cases were identified as strong outliers outside of the main distribution in the visual inspection
of the two-dimensional equal-spin fit (\autoref{sec:fit-2d})
and the per-mass-ratio analysis of residuals of unequal-spin cases against the 2D fit (\autoref{sec:fit-3d}).
For these simulations, there are no equivalent or nearby comparison cases,
so it cannot be said with certainty whether they would still be outliers in a more densely covered future data set;
and at the same time a small residual for any given point is no guarantee for its absolute accuracy
when there are no equivalent comparison points.
Hence, we have made much less strict exclusions in the sparsely covered unequal-spin range,
which limits the accuracy to which we can extract the subdominant spin-difference effects
(which are of a similar scale as the remaining scatter in the data set),
but also reduces the risk of overfitting to spurious trends in a more strongly trimmed data set.

\bibliography{../biblio.bib}

\begin{thebibliography}{105}%
\makeatletter
\providecommand \@ifxundefined [1]{%
 \@ifx{#1\undefined}
}%
\providecommand \@ifnum [1]{%
 \ifnum #1\expandafter \@firstoftwo
 \else \expandafter \@secondoftwo
 \fi
}%
\providecommand \@ifx [1]{%
 \ifx #1\expandafter \@firstoftwo
 \else \expandafter \@secondoftwo
 \fi
}%
\providecommand \natexlab [1]{#1}%
\providecommand \enquote  [1]{``#1''}%
\providecommand \bibnamefont  [1]{#1}%
\providecommand \bibfnamefont [1]{#1}%
\providecommand \citenamefont [1]{#1}%
\providecommand \href@noop [0]{\@secondoftwo}%
\providecommand \href [0]{\begingroup \@sanitize@url \@href}%
\providecommand \@href[1]{\@@startlink{#1}\@@href}%
\providecommand \@@href[1]{\endgroup#1\@@endlink}%
\providecommand \@sanitize@url [0]{\catcode `\\12\catcode `\$12\catcode
  `\&12\catcode `\#12\catcode `\^12\catcode `\_12\catcode `\%12\relax}%
\providecommand \@@startlink[1]{}%
\providecommand \@@endlink[0]{}%
\providecommand \url  [0]{\begingroup\@sanitize@url \@url }%
\providecommand \@url [1]{\endgroup\@href {#1}{\urlprefix }}%
\providecommand \urlprefix  [0]{URL }%
\providecommand \Eprint [0]{\href }%
\providecommand \doibase [0]{http://dx.doi.org/}%
\providecommand \selectlanguage [0]{\@gobble}%
\providecommand \bibinfo  [0]{\@secondoftwo}%
\providecommand \bibfield  [0]{\@secondoftwo}%
\providecommand \translation [1]{[#1]}%
\providecommand \BibitemOpen [0]{}%
\providecommand \bibitemStop [0]{}%
\providecommand \bibitemNoStop [0]{.\EOS\space}%
\providecommand \EOS [0]{\spacefactor3000\relax}%
\providecommand \BibitemShut  [1]{\csname bibitem#1\endcsname}%
\let\auto@bib@innerbib\@empty
\bibitem [{\citenamefont {Aasi}\ \emph {et~al.}(2015)\citenamefont {Aasi} \emph
  {et~al.}}]{TheLIGOScientific:2014jea}%
  \BibitemOpen
  \bibfield  {author} {\bibinfo {author} {\bibfnamefont {J.}~\bibnamefont
  {Aasi}} \emph {et~al.} (\bibinfo {collaboration} {LIGO Scientific
  Collaboration}),\ }\href {\doibase 10.1088/0264-9381/32/7/074001} {\bibfield
  {journal} {\bibinfo  {journal} {Class. Quant. Grav.}\ }\textbf {\bibinfo
  {volume} {32}},\ \bibinfo {pages} {074001} (\bibinfo {year} {2015})},\
  \Eprint {http://arxiv.org/abs/1411.4547} {arXiv:1411.4547 [gr-qc]}
  \BibitemShut {NoStop}%
\bibitem [{\citenamefont {Abbott}\ \emph
  {et~al.}(2016{\natexlab{a}})\citenamefont {Abbott} \emph
  {et~al.}}]{TheLIGOScientific:2016agk}%
  \BibitemOpen
  \bibfield  {author} {\bibinfo {author} {\bibfnamefont {B.~P.}\ \bibnamefont
  {Abbott}} \emph {et~al.} (\bibinfo {collaboration} {LIGO Scientific
  Collaboration and Virgo Collaboration}),\ }\href {\doibase
  10.1103/PhysRevLett.116.131103} {\bibfield  {journal} {\bibinfo  {journal}
  {Phys. Rev. Lett.}\ }\textbf {\bibinfo {volume} {116}},\ \bibinfo {pages}
  {131103} (\bibinfo {year} {2016}{\natexlab{a}})},\ \Eprint
  {http://arxiv.org/abs/1602.03838} {arXiv:1602.03838 [gr-qc]} \BibitemShut
  {NoStop}%
\bibitem [{\citenamefont {Abbott}\ \emph
  {et~al.}(2016{\natexlab{b}})\citenamefont {Abbott} \emph
  {et~al.}}]{Abbott:2016blz}%
  \BibitemOpen
  \bibfield  {author} {\bibinfo {author} {\bibfnamefont {B.~P.}\ \bibnamefont
  {Abbott}} \emph {et~al.} (\bibinfo {collaboration} {LIGO Scientific
  Collaboration and Virgo Collaboration}),\ }\href {\doibase
  10.1103/PhysRevLett.116.061102} {\bibfield  {journal} {\bibinfo  {journal}
  {Phys. Rev. Lett.}\ }\textbf {\bibinfo {volume} {116}},\ \bibinfo {pages}
  {061102} (\bibinfo {year} {2016}{\natexlab{b}})},\ \Eprint
  {http://arxiv.org/abs/1602.03837} {arXiv:1602.03837 [gr-qc]} \BibitemShut
  {NoStop}%
\bibitem [{\citenamefont {Abbott}\ \emph
  {et~al.}(2016{\natexlab{c}})\citenamefont {Abbott} \emph
  {et~al.}}]{Abbott:2016nmj}%
  \BibitemOpen
  \bibfield  {author} {\bibinfo {author} {\bibfnamefont {B.~P.}\ \bibnamefont
  {Abbott}} \emph {et~al.} (\bibinfo {collaboration} {LIGO Scientific
  Collaboration and Virgo Collaboration}),\ }\href {\doibase
  10.1103/PhysRevLett.116.241103} {\bibfield  {journal} {\bibinfo  {journal}
  {Phys. Rev. Lett.}\ }\textbf {\bibinfo {volume} {116}},\ \bibinfo {pages}
  {241103} (\bibinfo {year} {2016}{\natexlab{c}})},\ \Eprint
  {http://arxiv.org/abs/1606.04855} {arXiv:1606.04855 [gr-qc]} \BibitemShut
  {NoStop}%
\bibitem [{\citenamefont {Abbott}\ \emph
  {et~al.}(2016{\natexlab{d}})\citenamefont {Abbott} \emph
  {et~al.}}]{TheLIGOScientific:2016pea}%
  \BibitemOpen
  \bibfield  {author} {\bibinfo {author} {\bibfnamefont {B.~P.}\ \bibnamefont
  {Abbott}} \emph {et~al.} (\bibinfo {collaboration} {LIGO Scientific
  Collaboration and Virgo Collaboration}),\ }\href {\doibase
  10.1103/PhysRevX.6.041015} {\bibfield  {journal} {\bibinfo  {journal} {Phys.
  Rev.}\ }\textbf {\bibinfo {volume} {X6}},\ \bibinfo {pages} {041015}
  (\bibinfo {year} {2016}{\natexlab{d}})},\ \Eprint
  {http://arxiv.org/abs/1606.04856} {arXiv:1606.04856 [gr-qc]} \BibitemShut
  {NoStop}%
\bibitem [{\citenamefont {Abbott}\ \emph
  {et~al.}(2016{\natexlab{e}})\citenamefont {Abbott} \emph
  {et~al.}}]{TheLIGOScientific:2016wfe}%
  \BibitemOpen
  \bibfield  {author} {\bibinfo {author} {\bibfnamefont {B.~P.}\ \bibnamefont
  {Abbott}} \emph {et~al.} (\bibinfo {collaboration} {LIGO Scientific
  Collaboration and Virgo Collaboration}),\ }\href {\doibase
  10.1103/PhysRevLett.116.241102} {\bibfield  {journal} {\bibinfo  {journal}
  {Phys. Rev. Lett.}\ }\textbf {\bibinfo {volume} {116}},\ \bibinfo {pages}
  {241102} (\bibinfo {year} {2016}{\natexlab{e}})},\ \Eprint
  {http://arxiv.org/abs/1602.03840} {arXiv:1602.03840 [gr-qc]} \BibitemShut
  {NoStop}%
\bibitem [{\citenamefont {Abbott}\ \emph
  {et~al.}(2016{\natexlab{f}})\citenamefont {Abbott} \emph
  {et~al.}}]{TheLIGOScientific:2016src}%
  \BibitemOpen
  \bibfield  {author} {\bibinfo {author} {\bibfnamefont {B.~P.}\ \bibnamefont
  {Abbott}} \emph {et~al.} (\bibinfo {collaboration} {LIGO Scientific
  Collaboration and Virgo Collaboration}),\ }\href {\doibase
  10.1103/PhysRevLett.116.221101} {\bibfield  {journal} {\bibinfo  {journal}
  {Phys. Rev. Lett.}\ }\textbf {\bibinfo {volume} {116}},\ \bibinfo {pages}
  {221101} (\bibinfo {year} {2016}{\natexlab{f}})},\ \Eprint
  {http://arxiv.org/abs/1602.03841} {arXiv:1602.03841 [gr-qc]} \BibitemShut
  {NoStop}%
\bibitem [{\citenamefont {Frederiks}\ \emph {et~al.}(2013)\citenamefont
  {Frederiks} \emph {et~al.}}]{Frederiks:2013cga}%
  \BibitemOpen
  \bibfield  {author} {\bibinfo {author} {\bibfnamefont {D.~D.}\ \bibnamefont
  {Frederiks}} \emph {et~al.},\ }\href {\doibase 10.1088/0004-637X/779/2/151}
  {\bibfield  {journal} {\bibinfo  {journal} {Astrophys. J.}\ }\textbf
  {\bibinfo {volume} {779}},\ \bibinfo {pages} {151} (\bibinfo {year}
  {2013})},\ \Eprint {http://arxiv.org/abs/1311.5734} {arXiv:1311.5734
  [astro-ph.HE]} \BibitemShut {NoStop}%
\bibitem [{\citenamefont {Jim{\'e}nez~Forteza}\ \emph
  {et~al.}(2016)\citenamefont {Jim{\'e}nez~Forteza}, \citenamefont {Keitel},
  \citenamefont {Husa}, \citenamefont {Hannam}, \citenamefont {Khan},
  \citenamefont {London},\ and\ \citenamefont {P{\"u}rrer}}]{T1600018}%
  \BibitemOpen
  \bibfield  {author} {\bibinfo {author} {\bibfnamefont {X.}~\bibnamefont
  {Jim{\'e}nez~Forteza}}, \bibinfo {author} {\bibfnamefont {D.}~\bibnamefont
  {Keitel}}, \bibinfo {author} {\bibfnamefont {S.}~\bibnamefont {Husa}},
  \bibinfo {author} {\bibfnamefont {M.}~\bibnamefont {Hannam}}, \bibinfo
  {author} {\bibfnamefont {S.}~\bibnamefont {Khan}}, \bibinfo {author}
  {\bibfnamefont {L.}~\bibnamefont {London}}, \ and\ \bibinfo {author}
  {\bibfnamefont {M.}~\bibnamefont {P{\"u}rrer}},\ }\href
  {https://dcc.ligo.org/LIGO-T1600018-v4/public} {\emph {\bibinfo {title}
  {{Phenomenological fit of the peak luminosity from non-precessing
  binary-black-hole coalescences}}}},\ \bibinfo {type} {Tech. Rep.}\ \bibinfo
  {number} {LIGO-T1600018-v4}\ (\bibinfo {year} {2016})\BibitemShut {NoStop}%
\bibitem [{\citenamefont {Aasi}\ \emph {et~al.}(2013)\citenamefont {Aasi} \emph
  {et~al.}}]{Aasi:2013jjl}%
  \BibitemOpen
  \bibfield  {author} {\bibinfo {author} {\bibfnamefont {J.}~\bibnamefont
  {Aasi}} \emph {et~al.} (\bibinfo {collaboration} {LIGO Scientific
  Collaboration and Virgo Collaboration}),\ }\href {\doibase
  10.1103/PhysRevD.88.062001} {\bibfield  {journal} {\bibinfo  {journal} {Phys.
  Rev.}\ }\textbf {\bibinfo {volume} {D88}},\ \bibinfo {pages} {062001}
  (\bibinfo {year} {2013})},\ \Eprint {http://arxiv.org/abs/1304.1775}
  {arXiv:1304.1775 [gr-qc]} \BibitemShut {NoStop}%
\bibitem [{\citenamefont {{Veitch}}\ \emph {et~al.}(2015)\citenamefont
  {{Veitch}}, \citenamefont {{Raymond}}, \citenamefont {{Farr}}, \citenamefont
  {{Farr}}, \citenamefont {{Graff}}, \citenamefont {{Vitale}}, \citenamefont
  {{Aylott}}, \citenamefont {{Blackburn}}, \citenamefont {{Christensen}},
  \citenamefont {{Coughlin}}, \citenamefont {{Del Pozzo}}, \citenamefont
  {{Feroz}}, \citenamefont {{Gair}}, \citenamefont {{Haster}}, \citenamefont
  {{Kalogera}}, \citenamefont {{Littenberg}}, \citenamefont {{Mandel}},
  \citenamefont {{O'Shaughnessy}}, \citenamefont {{Pitkin}}, \citenamefont
  {{Rodriguez}}, \citenamefont {{R{\"o}ver}}, \citenamefont {{Sidery}},
  \citenamefont {{Smith}}, \citenamefont {{Van Der Sluys}}, \citenamefont
  {{Vecchio}}, \citenamefont {{Vousden}},\ and\ \citenamefont
  {{Wade}}}]{Veitch:2014wba}%
  \BibitemOpen
  \bibfield  {author} {\bibinfo {author} {\bibfnamefont {J.}~\bibnamefont
  {{Veitch}}}, \bibinfo {author} {\bibfnamefont {V.}~\bibnamefont {{Raymond}}},
  \bibinfo {author} {\bibfnamefont {B.}~\bibnamefont {{Farr}}}, \bibinfo
  {author} {\bibfnamefont {W.}~\bibnamefont {{Farr}}}, \bibinfo {author}
  {\bibfnamefont {P.}~\bibnamefont {{Graff}}}, \bibinfo {author} {\bibfnamefont
  {S.}~\bibnamefont {{Vitale}}}, \bibinfo {author} {\bibfnamefont
  {B.}~\bibnamefont {{Aylott}}}, \bibinfo {author} {\bibfnamefont
  {K.}~\bibnamefont {{Blackburn}}}, \bibinfo {author} {\bibfnamefont
  {N.}~\bibnamefont {{Christensen}}}, \bibinfo {author} {\bibfnamefont
  {M.}~\bibnamefont {{Coughlin}}}, \bibinfo {author} {\bibfnamefont
  {W.}~\bibnamefont {{Del Pozzo}}}, \bibinfo {author} {\bibfnamefont
  {F.}~\bibnamefont {{Feroz}}}, \bibinfo {author} {\bibfnamefont
  {J.}~\bibnamefont {{Gair}}}, \bibinfo {author} {\bibfnamefont {C.-J.}\
  \bibnamefont {{Haster}}}, \bibinfo {author} {\bibfnamefont {V.}~\bibnamefont
  {{Kalogera}}}, \bibinfo {author} {\bibfnamefont {T.}~\bibnamefont
  {{Littenberg}}}, \bibinfo {author} {\bibfnamefont {I.}~\bibnamefont
  {{Mandel}}}, \bibinfo {author} {\bibfnamefont {R.}~\bibnamefont
  {{O'Shaughnessy}}}, \bibinfo {author} {\bibfnamefont {M.}~\bibnamefont
  {{Pitkin}}}, \bibinfo {author} {\bibfnamefont {C.}~\bibnamefont
  {{Rodriguez}}}, \bibinfo {author} {\bibfnamefont {C.}~\bibnamefont
  {{R{\"o}ver}}}, \bibinfo {author} {\bibfnamefont {T.}~\bibnamefont
  {{Sidery}}}, \bibinfo {author} {\bibfnamefont {R.}~\bibnamefont {{Smith}}},
  \bibinfo {author} {\bibfnamefont {M.}~\bibnamefont {{Van Der Sluys}}},
  \bibinfo {author} {\bibfnamefont {A.}~\bibnamefont {{Vecchio}}}, \bibinfo
  {author} {\bibfnamefont {W.}~\bibnamefont {{Vousden}}}, \ and\ \bibinfo
  {author} {\bibfnamefont {L.}~\bibnamefont {{Wade}}},\ }\href {\doibase
  10.1103/PhysRevD.91.042003} {\bibfield  {journal} {\bibinfo  {journal} {Phys.
  Rev.}\ }\textbf {\bibinfo {volume} {D91}},\ \bibinfo {pages} {042003}
  (\bibinfo {year} {2015})},\ \Eprint {http://arxiv.org/abs/1409.7215}
  {arXiv:1409.7215 [gr-qc]} \BibitemShut {NoStop}%
\bibitem [{\citenamefont {Abbott}\ \emph
  {et~al.}(2016{\natexlab{g}})\citenamefont {Abbott} \emph
  {et~al.}}]{Abbott:2016izl}%
  \BibitemOpen
  \bibfield  {author} {\bibinfo {author} {\bibfnamefont {B.}~\bibnamefont
  {Abbott}} \emph {et~al.} (\bibinfo {collaboration} {LIGO Scientific
  Collaboration and Virgo Collaboration}),\ }\href {\doibase
  10.1103/PhysRevX.6.041014} {\bibfield  {journal} {\bibinfo  {journal} {Phys.
  Rev.}\ }\textbf {\bibinfo {volume} {X6}},\ \bibinfo {pages} {041014}
  (\bibinfo {year} {2016}{\natexlab{g}})},\ \Eprint
  {http://arxiv.org/abs/1606.01210} {arXiv:1606.01210 [gr-qc]} \BibitemShut
  {NoStop}%
\bibitem [{\citenamefont {Hannam}\ \emph {et~al.}(2014)\citenamefont {Hannam},
  \citenamefont {Schmidt}, \citenamefont {Boh{\'e}}, \citenamefont {Haegel},
  \citenamefont {Husa}, \citenamefont {Ohme}, \citenamefont {Pratten},\ and\
  \citenamefont {P{\"u}rrer}}]{Hannam:2013oca}%
  \BibitemOpen
  \bibfield  {author} {\bibinfo {author} {\bibfnamefont {M.}~\bibnamefont
  {Hannam}}, \bibinfo {author} {\bibfnamefont {P.}~\bibnamefont {Schmidt}},
  \bibinfo {author} {\bibfnamefont {A.}~\bibnamefont {Boh{\'e}}}, \bibinfo
  {author} {\bibfnamefont {L.}~\bibnamefont {Haegel}}, \bibinfo {author}
  {\bibfnamefont {S.}~\bibnamefont {Husa}}, \bibinfo {author} {\bibfnamefont
  {F.}~\bibnamefont {Ohme}}, \bibinfo {author} {\bibfnamefont {G.}~\bibnamefont
  {Pratten}}, \ and\ \bibinfo {author} {\bibfnamefont {M.}~\bibnamefont
  {P{\"u}rrer}},\ }\href {\doibase 10.1103/PhysRevLett.113.151101} {\bibfield
  {journal} {\bibinfo  {journal} {Phys. Rev. Lett.}\ }\textbf {\bibinfo
  {volume} {113}},\ \bibinfo {pages} {151101} (\bibinfo {year} {2014})},\
  \Eprint {http://arxiv.org/abs/1308.3271} {arXiv:1308.3271 [gr-qc]}
  \BibitemShut {NoStop}%
\bibitem [{\citenamefont {Boh{\'e}}\ \emph {et~al.}(2016)\citenamefont
  {Boh{\'e}}, \citenamefont {Hannam}, \citenamefont {Husa}, \citenamefont
  {Ohme}, \citenamefont {P{\"u}rrer},\ and\ \citenamefont
  {Schmidt}}]{T1500602}%
  \BibitemOpen
  \bibfield  {author} {\bibinfo {author} {\bibfnamefont {A.}~\bibnamefont
  {Boh{\'e}}}, \bibinfo {author} {\bibfnamefont {M.}~\bibnamefont {Hannam}},
  \bibinfo {author} {\bibfnamefont {S.}~\bibnamefont {Husa}}, \bibinfo {author}
  {\bibfnamefont {F.}~\bibnamefont {Ohme}}, \bibinfo {author} {\bibfnamefont
  {M.}~\bibnamefont {P{\"u}rrer}}, \ and\ \bibinfo {author} {\bibfnamefont
  {P.}~\bibnamefont {Schmidt}},\ }\href {https://dcc.ligo.org/T1500602} {\emph
  {\bibinfo {title} {{PhenomPv2 -- technical notes for the LAL
  implementation}}}},\ \bibinfo {type} {Tech. Rep.}\ \bibinfo {number}
  {LIGO-T1500602}\ (\bibinfo {year} {2016})\BibitemShut {NoStop}%
\bibitem [{\citenamefont {Boh{\'e}}\ \emph {et~al.}()\citenamefont {Boh{\'e}},
  \citenamefont {P{\"u}rrer}, \citenamefont {Hannam}, \citenamefont {Husa},
  \citenamefont {Ohme}, \citenamefont {Schmidt} \emph
  {et~al.}}]{PhenomPv2Paper}%
  \BibitemOpen
  \bibfield  {author} {\bibinfo {author} {\bibfnamefont {A.}~\bibnamefont
  {Boh{\'e}}}, \bibinfo {author} {\bibfnamefont {M.}~\bibnamefont
  {P{\"u}rrer}}, \bibinfo {author} {\bibfnamefont {M.}~\bibnamefont {Hannam}},
  \bibinfo {author} {\bibfnamefont {S.}~\bibnamefont {Husa}}, \bibinfo {author}
  {\bibfnamefont {F.}~\bibnamefont {Ohme}}, \bibinfo {author} {\bibfnamefont
  {P.}~\bibnamefont {Schmidt}},  \emph {et~al.},\ }\href@noop {} {\enquote
  {\bibinfo {title} {{IMRPhenomPv2 publication}},}\ }\bibinfo {note} {{in
  preparation}}\BibitemShut {NoStop}%
\bibitem [{\citenamefont {Husa}\ \emph {et~al.}(2016)\citenamefont {Husa},
  \citenamefont {Khan}, \citenamefont {Hannam}, \citenamefont {P{\"u}rrer},
  \citenamefont {Ohme}, \citenamefont {Jim{\'e}nez~Forteza},\ and\
  \citenamefont {Boh{\'e}}}]{Husa:2015iqa}%
  \BibitemOpen
  \bibfield  {author} {\bibinfo {author} {\bibfnamefont {S.}~\bibnamefont
  {Husa}}, \bibinfo {author} {\bibfnamefont {S.}~\bibnamefont {Khan}}, \bibinfo
  {author} {\bibfnamefont {M.}~\bibnamefont {Hannam}}, \bibinfo {author}
  {\bibfnamefont {M.}~\bibnamefont {P{\"u}rrer}}, \bibinfo {author}
  {\bibfnamefont {F.}~\bibnamefont {Ohme}}, \bibinfo {author} {\bibfnamefont
  {X.}~\bibnamefont {Jim{\'e}nez~Forteza}}, \ and\ \bibinfo {author}
  {\bibfnamefont {A.}~\bibnamefont {Boh{\'e}}},\ }\href {\doibase
  10.1103/PhysRevD.93.044006} {\bibfield  {journal} {\bibinfo  {journal} {Phys.
  Rev.}\ }\textbf {\bibinfo {volume} {D93}},\ \bibinfo {pages} {044006}
  (\bibinfo {year} {2016})},\ \Eprint {http://arxiv.org/abs/1508.07250}
  {arXiv:1508.07250 [gr-qc]} \BibitemShut {NoStop}%
\bibitem [{\citenamefont {Khan}\ \emph {et~al.}(2016)\citenamefont {Khan},
  \citenamefont {Husa}, \citenamefont {Hannam}, \citenamefont {Ohme},
  \citenamefont {P{\"u}rrer}, \citenamefont {Jim{\'e}nez~Forteza},\ and\
  \citenamefont {Boh{\'e}}}]{Khan:2015jqa}%
  \BibitemOpen
  \bibfield  {author} {\bibinfo {author} {\bibfnamefont {S.}~\bibnamefont
  {Khan}}, \bibinfo {author} {\bibfnamefont {S.}~\bibnamefont {Husa}}, \bibinfo
  {author} {\bibfnamefont {M.}~\bibnamefont {Hannam}}, \bibinfo {author}
  {\bibfnamefont {F.}~\bibnamefont {Ohme}}, \bibinfo {author} {\bibfnamefont
  {M.}~\bibnamefont {P{\"u}rrer}}, \bibinfo {author} {\bibfnamefont
  {X.}~\bibnamefont {Jim{\'e}nez~Forteza}}, \ and\ \bibinfo {author}
  {\bibfnamefont {A.}~\bibnamefont {Boh{\'e}}},\ }\href {\doibase
  10.1103/PhysRevD.93.044007} {\bibfield  {journal} {\bibinfo  {journal} {Phys.
  Rev.}\ }\textbf {\bibinfo {volume} {D93}},\ \bibinfo {pages} {044007}
  (\bibinfo {year} {2016})},\ \Eprint {http://arxiv.org/abs/1508.07253}
  {arXiv:1508.07253 [gr-qc]} \BibitemShut {NoStop}%
\bibitem [{\citenamefont {Pan}\ \emph {et~al.}(2014)\citenamefont {Pan},
  \citenamefont {Buonanno}, \citenamefont {Taracchini}, \citenamefont {Kidder},
  \citenamefont {Mrou{\'e}}, \citenamefont {Pfeiffer}, \citenamefont {Scheel},\
  and\ \citenamefont {Szil{\'a}gyi}}]{Pan:2013rra}%
  \BibitemOpen
  \bibfield  {author} {\bibinfo {author} {\bibfnamefont {Y.}~\bibnamefont
  {Pan}}, \bibinfo {author} {\bibfnamefont {A.}~\bibnamefont {Buonanno}},
  \bibinfo {author} {\bibfnamefont {A.}~\bibnamefont {Taracchini}}, \bibinfo
  {author} {\bibfnamefont {L.~E.}\ \bibnamefont {Kidder}}, \bibinfo {author}
  {\bibfnamefont {A.~H.}\ \bibnamefont {Mrou{\'e}}}, \bibinfo {author}
  {\bibfnamefont {H.~P.}\ \bibnamefont {Pfeiffer}}, \bibinfo {author}
  {\bibfnamefont {M.~A.}\ \bibnamefont {Scheel}}, \ and\ \bibinfo {author}
  {\bibfnamefont {B.}~\bibnamefont {Szil{\'a}gyi}},\ }\href {\doibase
  10.1103/PhysRevD.89.084006} {\bibfield  {journal} {\bibinfo  {journal} {Phys.
  Rev.}\ }\textbf {\bibinfo {volume} {D89}},\ \bibinfo {pages} {084006}
  (\bibinfo {year} {2014})},\ \Eprint {http://arxiv.org/abs/1307.6232}
  {arXiv:1307.6232 [gr-qc]} \BibitemShut {NoStop}%
\bibitem [{\citenamefont {P{\"u}rrer}(2014)}]{Purrer:2014fza}%
  \BibitemOpen
  \bibfield  {author} {\bibinfo {author} {\bibfnamefont {M.}~\bibnamefont
  {P{\"u}rrer}},\ }\href {\doibase 10.1088/0264-9381/31/19/195010} {\bibfield
  {journal} {\bibinfo  {journal} {Class. Quant. Grav.}\ }\textbf {\bibinfo
  {volume} {31}},\ \bibinfo {pages} {195010} (\bibinfo {year} {2014})},\
  \Eprint {http://arxiv.org/abs/1402.4146} {arXiv:1402.4146 [gr-qc]}
  \BibitemShut {NoStop}%
\bibitem [{\citenamefont {P{\"u}rrer}(2016)}]{Purrer:2015tud}%
  \BibitemOpen
  \bibfield  {author} {\bibinfo {author} {\bibfnamefont {M.}~\bibnamefont
  {P{\"u}rrer}},\ }\href {\doibase 10.1103/PhysRevD.93.064041} {\bibfield
  {journal} {\bibinfo  {journal} {Phys. Rev.}\ }\textbf {\bibinfo {volume}
  {D93}},\ \bibinfo {pages} {064041} (\bibinfo {year} {2016})},\ \Eprint
  {http://arxiv.org/abs/1512.02248} {arXiv:1512.02248 [gr-qc]} \BibitemShut
  {NoStop}%
\bibitem [{\citenamefont {Babak}\ \emph {et~al.}(2017)\citenamefont {Babak},
  \citenamefont {Taracchini},\ and\ \citenamefont {Buonanno}}]{Babak:2016tgq}%
  \BibitemOpen
  \bibfield  {author} {\bibinfo {author} {\bibfnamefont {S.}~\bibnamefont
  {Babak}}, \bibinfo {author} {\bibfnamefont {A.}~\bibnamefont {Taracchini}}, \
  and\ \bibinfo {author} {\bibfnamefont {A.}~\bibnamefont {Buonanno}},\ }\href
  {\doibase 10.1103/PhysRevD.95.024010} {\bibfield  {journal} {\bibinfo
  {journal} {Phys. Rev.}\ }\textbf {\bibinfo {volume} {D95}},\ \bibinfo {pages}
  {024010} (\bibinfo {year} {2017})},\ \Eprint
  {http://arxiv.org/abs/1607.05661} {arXiv:1607.05661 [gr-qc]} \BibitemShut
  {NoStop}%
\bibitem [{\citenamefont {Abbott}\ \emph {et~al.}(2017)\citenamefont {Abbott}
  \emph {et~al.}}]{Abbott:2016wiq}%
  \BibitemOpen
  \bibfield  {author} {\bibinfo {author} {\bibfnamefont {B.~P.}\ \bibnamefont
  {Abbott}} \emph {et~al.} (\bibinfo {collaboration} {LIGO Scientific
  Collaboration and Virgo Collaboration}),\ }\href {\doibase
  10.1088/1361-6382/aa6854} {\bibfield  {journal} {\bibinfo  {journal} {Class.
  Quant. Grav.}\ }\textbf {\bibinfo {volume} {34}},\ \bibinfo {pages} {104002}
  (\bibinfo {year} {2017})},\ \Eprint {http://arxiv.org/abs/1611.07531}
  {arXiv:1611.07531 [gr-qc]} \BibitemShut {NoStop}%
\bibitem [{\citenamefont {Healy}\ \emph {et~al.}(2014)\citenamefont {Healy},
  \citenamefont {Lousto},\ and\ \citenamefont {Zlochower}}]{Healy:2014yta}%
  \BibitemOpen
  \bibfield  {author} {\bibinfo {author} {\bibfnamefont {J.}~\bibnamefont
  {Healy}}, \bibinfo {author} {\bibfnamefont {C.~O.}\ \bibnamefont {Lousto}}, \
  and\ \bibinfo {author} {\bibfnamefont {Y.}~\bibnamefont {Zlochower}},\ }\href
  {\doibase 10.1103/PhysRevD.90.104004} {\bibfield  {journal} {\bibinfo
  {journal} {Phys. Rev.}\ }\textbf {\bibinfo {volume} {D90}},\ \bibinfo {pages}
  {104004} (\bibinfo {year} {2014})},\ \Eprint {http://arxiv.org/abs/1406.7295}
  {arXiv:1406.7295 [gr-qc]} \BibitemShut {NoStop}%
\bibitem [{\citenamefont {Healy}\ and\ \citenamefont
  {Lousto}(2017)}]{Healy:2016lce}%
  \BibitemOpen
  \bibfield  {author} {\bibinfo {author} {\bibfnamefont {J.}~\bibnamefont
  {Healy}}\ and\ \bibinfo {author} {\bibfnamefont {C.~O.}\ \bibnamefont
  {Lousto}},\ }\href {\doibase 10.1103/PhysRevD.95.024037} {\bibfield
  {journal} {\bibinfo  {journal} {Phys. Rev.}\ }\textbf {\bibinfo {volume}
  {D95}},\ \bibinfo {pages} {024037} (\bibinfo {year} {2017})},\ \Eprint
  {http://arxiv.org/abs/1610.09713} {arXiv:1610.09713 [gr-qc]} \BibitemShut
  {NoStop}%
\bibitem [{\citenamefont {Hofmann}\ \emph {et~al.}(2016)\citenamefont
  {Hofmann}, \citenamefont {Barausse},\ and\ \citenamefont
  {Rezzolla}}]{Hofmann:2016yih}%
  \BibitemOpen
  \bibfield  {author} {\bibinfo {author} {\bibfnamefont {F.}~\bibnamefont
  {Hofmann}}, \bibinfo {author} {\bibfnamefont {E.}~\bibnamefont {Barausse}}, \
  and\ \bibinfo {author} {\bibfnamefont {L.}~\bibnamefont {Rezzolla}},\ }\href
  {\doibase 10.3847/2041-8205/825/2/L19} {\bibfield  {journal} {\bibinfo
  {journal} {Astrophys. J.}\ }\textbf {\bibinfo {volume} {825}},\ \bibinfo
  {pages} {L19} (\bibinfo {year} {2016})},\ \Eprint
  {http://arxiv.org/abs/1605.01938} {arXiv:1605.01938 [gr-qc]} \BibitemShut
  {NoStop}%
\bibitem [{\citenamefont {Jim{\'e}nez-Forteza}\ \emph
  {et~al.}(2017)\citenamefont {Jim{\'e}nez-Forteza}, \citenamefont {Keitel},
  \citenamefont {Husa}, \citenamefont {Hannam}, \citenamefont {Khan},\ and\
  \citenamefont {P{\"u}rrer}}]{Jimenez-Forteza:2016oae}%
  \BibitemOpen
  \bibfield  {author} {\bibinfo {author} {\bibfnamefont {X.}~\bibnamefont
  {Jim{\'e}nez-Forteza}}, \bibinfo {author} {\bibfnamefont {D.}~\bibnamefont
  {Keitel}}, \bibinfo {author} {\bibfnamefont {S.}~\bibnamefont {Husa}},
  \bibinfo {author} {\bibfnamefont {M.}~\bibnamefont {Hannam}}, \bibinfo
  {author} {\bibfnamefont {S.}~\bibnamefont {Khan}}, \ and\ \bibinfo {author}
  {\bibfnamefont {M.}~\bibnamefont {P{\"u}rrer}},\ }\href {\doibase
  10.1103/PhysRevD.95.064024} {\bibfield  {journal} {\bibinfo  {journal} {Phys.
  Rev.}\ }\textbf {\bibinfo {volume} {D95}},\ \bibinfo {pages} {064024}
  (\bibinfo {year} {2017})},\ \Eprint {http://arxiv.org/abs/1611.00332}
  {arXiv:1611.00332 [gr-qc]} \BibitemShut {NoStop}%
\bibitem [{\citenamefont {Baker}\ \emph {et~al.}(2008)\citenamefont {Baker},
  \citenamefont {Boggs}, \citenamefont {Centrella}, \citenamefont {Kelly},
  \citenamefont {McWilliams},\ and\ \citenamefont {van Meter}}]{Baker:2008mj}%
  \BibitemOpen
  \bibfield  {author} {\bibinfo {author} {\bibfnamefont {J.~G.}\ \bibnamefont
  {Baker}}, \bibinfo {author} {\bibfnamefont {W.~D.}\ \bibnamefont {Boggs}},
  \bibinfo {author} {\bibfnamefont {J.}~\bibnamefont {Centrella}}, \bibinfo
  {author} {\bibfnamefont {B.~J.}\ \bibnamefont {Kelly}}, \bibinfo {author}
  {\bibfnamefont {S.~T.}\ \bibnamefont {McWilliams}}, \ and\ \bibinfo {author}
  {\bibfnamefont {J.~R.}\ \bibnamefont {van Meter}},\ }\href {\doibase
  10.1103/PhysRevD.78.044046} {\bibfield  {journal} {\bibinfo  {journal} {Phys.
  Rev.}\ }\textbf {\bibinfo {volume} {D78}},\ \bibinfo {pages} {044046}
  (\bibinfo {year} {2008})},\ \Eprint {http://arxiv.org/abs/0805.1428}
  {arXiv:0805.1428 [gr-qc]} \BibitemShut {NoStop}%
\bibitem [{\citenamefont {Nagar}\ \emph {et~al.}(2007)\citenamefont {Nagar},
  \citenamefont {Damour},\ and\ \citenamefont {Tartaglia}}]{Nagar:2006xv}%
  \BibitemOpen
  \bibfield  {author} {\bibinfo {author} {\bibfnamefont {A.}~\bibnamefont
  {Nagar}}, \bibinfo {author} {\bibfnamefont {T.}~\bibnamefont {Damour}}, \
  and\ \bibinfo {author} {\bibfnamefont {A.}~\bibnamefont {Tartaglia}},\ }\href
  {\doibase 10.1088/0264-9381/24/12/S08} {\bibfield  {journal} {\bibinfo
  {journal} {Class. Quant. Grav.}\ }\textbf {\bibinfo {volume} {24}},\ \bibinfo
  {pages} {S109} (\bibinfo {year} {2007})},\ \Eprint
  {http://arxiv.org/abs/gr-qc/0612096} {arXiv:gr-qc/0612096 [gr-qc]}
  \BibitemShut {NoStop}%
\bibitem [{\citenamefont {Bernuzzi}\ \emph
  {et~al.}(2011{\natexlab{a}})\citenamefont {Bernuzzi}, \citenamefont {Nagar},\
  and\ \citenamefont {Zengino{\u g}lu}}]{Bernuzzi:2011aj}%
  \BibitemOpen
  \bibfield  {author} {\bibinfo {author} {\bibfnamefont {S.}~\bibnamefont
  {Bernuzzi}}, \bibinfo {author} {\bibfnamefont {A.}~\bibnamefont {Nagar}}, \
  and\ \bibinfo {author} {\bibfnamefont {A.}~\bibnamefont {Zengino{\u g}lu}},\
  }\href {\doibase 10.1103/PhysRevD.84.084026} {\bibfield  {journal} {\bibinfo
  {journal} {Phys. Rev.}\ }\textbf {\bibinfo {volume} {D84}},\ \bibinfo {pages}
  {084026} (\bibinfo {year} {2011}{\natexlab{a}})},\ \Eprint
  {http://arxiv.org/abs/1107.5402} {arXiv:1107.5402 [gr-qc]} \BibitemShut
  {NoStop}%
\bibitem [{\citenamefont {Harms}\ \emph {et~al.}(2014)\citenamefont {Harms},
  \citenamefont {Bernuzzi}, \citenamefont {Nagar},\ and\ \citenamefont
  {Zengino{\u g}lu}}]{Harms:2014dqa}%
  \BibitemOpen
  \bibfield  {author} {\bibinfo {author} {\bibfnamefont {E.}~\bibnamefont
  {Harms}}, \bibinfo {author} {\bibfnamefont {S.}~\bibnamefont {Bernuzzi}},
  \bibinfo {author} {\bibfnamefont {A.}~\bibnamefont {Nagar}}, \ and\ \bibinfo
  {author} {\bibfnamefont {A.}~\bibnamefont {Zengino{\u g}lu}},\ }\href
  {\doibase 10.1088/0264-9381/31/24/245004} {\bibfield  {journal} {\bibinfo
  {journal} {Class. Quant. Grav.}\ }\textbf {\bibinfo {volume} {31}},\ \bibinfo
  {pages} {245004} (\bibinfo {year} {2014})},\ \Eprint
  {http://arxiv.org/abs/1406.5983} {arXiv:1406.5983 [gr-qc]} \BibitemShut
  {NoStop}%
\bibitem [{\citenamefont {{Amaro Seoane}}\ \emph
  {et~al.}(2013{\natexlab{a}})\citenamefont {{Amaro Seoane}} \emph
  {et~al.}}]{AmaroSeoane:2012km}%
  \BibitemOpen
  \bibfield  {author} {\bibinfo {author} {\bibfnamefont {P.}~\bibnamefont
  {{Amaro Seoane}}} \emph {et~al.},\ }\href@noop {} {\bibfield  {journal}
  {\bibinfo  {journal} {GW Notes}\ }\textbf {\bibinfo {volume} {6}},\ \bibinfo
  {pages} {4} (\bibinfo {year} {2013}{\natexlab{a}})},\ \Eprint
  {http://arxiv.org/abs/1201.3621} {arXiv:1201.3621 [astro-ph.CO]} \BibitemShut
  {NoStop}%
\bibitem [{\citenamefont {{Amaro Seoane}}\ \emph
  {et~al.}(2013{\natexlab{b}})\citenamefont {{Amaro Seoane}} \emph
  {et~al.}}]{Seoane:2013qna}%
  \BibitemOpen
  \bibfield  {author} {\bibinfo {author} {\bibfnamefont {P.}~\bibnamefont
  {{Amaro Seoane}}} \emph {et~al.} (\bibinfo {collaboration} {eLISA
  Consortium}),\ }\href@noop {} {\  (\bibinfo {year} {2013}{\natexlab{b}})},\
  \Eprint {http://arxiv.org/abs/1305.5720} {arXiv:1305.5720 [astro-ph.CO]}
  \BibitemShut {NoStop}%
\bibitem [{\citenamefont {{Klein}}\ \emph {et~al.}(2016)\citenamefont
  {{Klein}}, \citenamefont {{Barausse}}, \citenamefont {{Sesana}},
  \citenamefont {{Petiteau}}, \citenamefont {{Berti}}, \citenamefont {{Babak}},
  \citenamefont {{Gair}}, \citenamefont {{Aoudia}}, \citenamefont {{Hinder}},
  \citenamefont {{Ohme}},\ and\ \citenamefont {{Wardell}}}]{Klein:2015hvg}%
  \BibitemOpen
  \bibfield  {author} {\bibinfo {author} {\bibfnamefont {A.}~\bibnamefont
  {{Klein}}}, \bibinfo {author} {\bibfnamefont {E.}~\bibnamefont {{Barausse}}},
  \bibinfo {author} {\bibfnamefont {A.}~\bibnamefont {{Sesana}}}, \bibinfo
  {author} {\bibfnamefont {A.}~\bibnamefont {{Petiteau}}}, \bibinfo {author}
  {\bibfnamefont {E.}~\bibnamefont {{Berti}}}, \bibinfo {author} {\bibfnamefont
  {S.}~\bibnamefont {{Babak}}}, \bibinfo {author} {\bibfnamefont
  {J.}~\bibnamefont {{Gair}}}, \bibinfo {author} {\bibfnamefont
  {S.}~\bibnamefont {{Aoudia}}}, \bibinfo {author} {\bibfnamefont
  {I.}~\bibnamefont {{Hinder}}}, \bibinfo {author} {\bibfnamefont
  {F.}~\bibnamefont {{Ohme}}}, \ and\ \bibinfo {author} {\bibfnamefont
  {B.}~\bibnamefont {{Wardell}}},\ }\href {\doibase 10.1103/PhysRevD.93.024003}
  {\bibfield  {journal} {\bibinfo  {journal} {Phys. Rev.}\ }\textbf {\bibinfo
  {volume} {D93}},\ \bibinfo {pages} {024003} (\bibinfo {year} {2016})},\
  \Eprint {http://arxiv.org/abs/1511.05581} {arXiv:1511.05581 [gr-qc]}
  \BibitemShut {NoStop}%
\bibitem [{\citenamefont {Rajagopal}\ and\ \citenamefont
  {Romani}(1995)}]{Rajagopal:1994zj}%
  \BibitemOpen
  \bibfield  {author} {\bibinfo {author} {\bibfnamefont {M.}~\bibnamefont
  {Rajagopal}}\ and\ \bibinfo {author} {\bibfnamefont {R.~W.}\ \bibnamefont
  {Romani}},\ }\href {\doibase 10.1086/175813} {\bibfield  {journal} {\bibinfo
  {journal} {Astrophys. J.}\ }\textbf {\bibinfo {volume} {446}},\ \bibinfo
  {pages} {543} (\bibinfo {year} {1995})},\ \Eprint
  {http://arxiv.org/abs/astro-ph/9412038} {arXiv:astro-ph/9412038 [astro-ph]}
  \BibitemShut {NoStop}%
\bibitem [{\citenamefont {Sesana}\ \emph {et~al.}(2009)\citenamefont {Sesana},
  \citenamefont {Vecchio},\ and\ \citenamefont {Volonteri}}]{Sesana:2008xk}%
  \BibitemOpen
  \bibfield  {author} {\bibinfo {author} {\bibfnamefont {A.}~\bibnamefont
  {Sesana}}, \bibinfo {author} {\bibfnamefont {A.}~\bibnamefont {Vecchio}}, \
  and\ \bibinfo {author} {\bibfnamefont {M.}~\bibnamefont {Volonteri}},\ }\href
  {\doibase 10.1111/j.1365-2966.2009.14499.x} {\bibfield  {journal} {\bibinfo
  {journal} {Mon. Not. Roy. Astron. Soc.}\ }\textbf {\bibinfo {volume} {394}},\
  \bibinfo {pages} {2255} (\bibinfo {year} {2009})},\ \Eprint
  {http://arxiv.org/abs/0809.3412} {arXiv:0809.3412 [astro-ph]} \BibitemShut
  {NoStop}%
\bibitem [{\citenamefont {Sesana}\ and\ \citenamefont
  {Vecchio}(2010)}]{Sesana:2010ac}%
  \BibitemOpen
  \bibfield  {author} {\bibinfo {author} {\bibfnamefont {A.}~\bibnamefont
  {Sesana}}\ and\ \bibinfo {author} {\bibfnamefont {A.}~\bibnamefont
  {Vecchio}},\ }\href {\doibase 10.1103/PhysRevD.81.104008} {\bibfield
  {journal} {\bibinfo  {journal} {Phys. Rev.}\ }\textbf {\bibinfo {volume}
  {D81}},\ \bibinfo {pages} {104008} (\bibinfo {year} {2010})},\ \Eprint
  {http://arxiv.org/abs/1003.0677} {arXiv:1003.0677 [astro-ph.CO]} \BibitemShut
  {NoStop}%
\bibitem [{\citenamefont {Schnittman}(2013)}]{Schnittman:2013qxa}%
  \BibitemOpen
  \bibfield  {author} {\bibinfo {author} {\bibfnamefont {J.~D.}\ \bibnamefont
  {Schnittman}},\ }\href {\doibase 10.1088/0264-9381/30/24/244007} {\bibfield
  {journal} {\bibinfo  {journal} {Class. Quant. Grav.}\ }\textbf {\bibinfo
  {volume} {30}},\ \bibinfo {pages} {244007} (\bibinfo {year} {2013})},\
  \Eprint {http://arxiv.org/abs/1307.3542} {arXiv:1307.3542 [gr-qc]}
  \BibitemShut {NoStop}%
\bibitem [{\citenamefont {Kocsis}\ and\ \citenamefont
  {Loeb}(2008)}]{Kocsis:2008aa}%
  \BibitemOpen
  \bibfield  {author} {\bibinfo {author} {\bibfnamefont {B.}~\bibnamefont
  {Kocsis}}\ and\ \bibinfo {author} {\bibfnamefont {A.}~\bibnamefont {Loeb}},\
  }\href {\doibase 10.1103/PhysRevLett.101.041101} {\bibfield  {journal}
  {\bibinfo  {journal} {Phys. Rev. Lett.}\ }\textbf {\bibinfo {volume} {101}},\
  \bibinfo {pages} {041101} (\bibinfo {year} {2008})},\ \Eprint
  {http://arxiv.org/abs/0803.0003} {arXiv:0803.0003 [astro-ph]} \BibitemShut
  {NoStop}%
\bibitem [{\citenamefont {Li}\ \emph {et~al.}(2012)\citenamefont {Li},
  \citenamefont {Kocsis},\ and\ \citenamefont {Loeb}}]{Li:2012dta}%
  \BibitemOpen
  \bibfield  {author} {\bibinfo {author} {\bibfnamefont {G.}~\bibnamefont
  {Li}}, \bibinfo {author} {\bibfnamefont {B.}~\bibnamefont {Kocsis}}, \ and\
  \bibinfo {author} {\bibfnamefont {A.}~\bibnamefont {Loeb}},\ }\href {\doibase
  10.1111/j.1365-2966.2012.21206.x} {\bibfield  {journal} {\bibinfo  {journal}
  {Mon. Not. Roy. Astron. Soc.}\ }\textbf {\bibinfo {volume} {425}},\ \bibinfo
  {pages} {2407} (\bibinfo {year} {2012})},\ \Eprint
  {http://arxiv.org/abs/1203.0317} {arXiv:1203.0317 [astro-ph.HE]} \BibitemShut
  {NoStop}%
\bibitem [{\citenamefont {Connaughton}\ \emph {et~al.}(2016)\citenamefont
  {Connaughton} \emph {et~al.}}]{Connaughton:2016umz}%
  \BibitemOpen
  \bibfield  {author} {\bibinfo {author} {\bibfnamefont {V.}~\bibnamefont
  {Connaughton}} \emph {et~al.},\ }\href {\doibase 10.3847/2041-8205/826/1/L6}
  {\bibfield  {journal} {\bibinfo  {journal} {Astrophys. J.}\ }\textbf
  {\bibinfo {volume} {826}},\ \bibinfo {pages} {L6} (\bibinfo {year} {2016})},\
  \Eprint {http://arxiv.org/abs/1602.03920} {arXiv:1602.03920 [astro-ph.HE]}
  \BibitemShut {NoStop}%
\bibitem [{\citenamefont {Cornish}\ and\ \citenamefont
  {Littenberg}(2015)}]{Cornish:2014kda}%
  \BibitemOpen
  \bibfield  {author} {\bibinfo {author} {\bibfnamefont {N.~J.}\ \bibnamefont
  {Cornish}}\ and\ \bibinfo {author} {\bibfnamefont {T.~B.}\ \bibnamefont
  {Littenberg}},\ }\href {\doibase 10.1088/0264-9381/32/13/135012} {\bibfield
  {journal} {\bibinfo  {journal} {Class. Quant. Grav.}\ }\textbf {\bibinfo
  {volume} {32}},\ \bibinfo {pages} {135012} (\bibinfo {year} {2015})},\
  \Eprint {http://arxiv.org/abs/1410.3835} {arXiv:1410.3835 [gr-qc]}
  \BibitemShut {NoStop}%
\bibitem [{\citenamefont {Klimenko}\ \emph {et~al.}(2016)\citenamefont
  {Klimenko} \emph {et~al.}}]{Klimenko:2015ypf}%
  \BibitemOpen
  \bibfield  {author} {\bibinfo {author} {\bibfnamefont {S.}~\bibnamefont
  {Klimenko}} \emph {et~al.},\ }\href {\doibase 10.1103/PhysRevD.93.042004}
  {\bibfield  {journal} {\bibinfo  {journal} {Phys. Rev.}\ }\textbf {\bibinfo
  {volume} {D93}},\ \bibinfo {pages} {042004} (\bibinfo {year} {2016})},\
  \Eprint {http://arxiv.org/abs/1511.05999} {arXiv:1511.05999 [gr-qc]}
  \BibitemShut {NoStop}%
\bibitem [{\citenamefont {Lynch}\ \emph {et~al.}(2017)\citenamefont {Lynch},
  \citenamefont {Vitale}, \citenamefont {Essick}, \citenamefont
  {Katsavounidis},\ and\ \citenamefont {Robinet}}]{Lynch:2015yin}%
  \BibitemOpen
  \bibfield  {author} {\bibinfo {author} {\bibfnamefont {R.}~\bibnamefont
  {Lynch}}, \bibinfo {author} {\bibfnamefont {S.}~\bibnamefont {Vitale}},
  \bibinfo {author} {\bibfnamefont {R.}~\bibnamefont {Essick}}, \bibinfo
  {author} {\bibfnamefont {E.}~\bibnamefont {Katsavounidis}}, \ and\ \bibinfo
  {author} {\bibfnamefont {F.}~\bibnamefont {Robinet}},\ }\href {\doibase
  10.1103/PhysRevD.95.104046} {\bibfield  {journal} {\bibinfo  {journal} {Phys.
  Rev.}\ }\textbf {\bibinfo {volume} {D95}},\ \bibinfo {pages} {104046}
  (\bibinfo {year} {2017})},\ \Eprint {http://arxiv.org/abs/1511.05955}
  {arXiv:1511.05955 [gr-qc]} \BibitemShut {NoStop}%
\bibitem [{\citenamefont {B{\'e}csy}\ \emph {et~al.}(2017)\citenamefont
  {B{\'e}csy}, \citenamefont {Raffai}, \citenamefont {Cornish}, \citenamefont
  {Essick}, \citenamefont {Kanner}, \citenamefont {Katsavounidis},
  \citenamefont {Littenberg}, \citenamefont {Millhouse},\ and\ \citenamefont
  {Vitale}}]{Becsy:2016ofp}%
  \BibitemOpen
  \bibfield  {author} {\bibinfo {author} {\bibfnamefont {B.}~\bibnamefont
  {B{\'e}csy}}, \bibinfo {author} {\bibfnamefont {P.}~\bibnamefont {Raffai}},
  \bibinfo {author} {\bibfnamefont {N.~J.}\ \bibnamefont {Cornish}}, \bibinfo
  {author} {\bibfnamefont {R.}~\bibnamefont {Essick}}, \bibinfo {author}
  {\bibfnamefont {J.}~\bibnamefont {Kanner}}, \bibinfo {author} {\bibfnamefont
  {E.}~\bibnamefont {Katsavounidis}}, \bibinfo {author} {\bibfnamefont {T.~B.}\
  \bibnamefont {Littenberg}}, \bibinfo {author} {\bibfnamefont
  {M.}~\bibnamefont {Millhouse}}, \ and\ \bibinfo {author} {\bibfnamefont
  {S.}~\bibnamefont {Vitale}},\ }\href {\doibase 10.3847/1538-4357/aa63ef}
  {\bibfield  {journal} {\bibinfo  {journal} {Astrophys. J.}\ }\textbf
  {\bibinfo {volume} {839}},\ \bibinfo {pages} {15} (\bibinfo {year} {2017})},\
  \Eprint {http://arxiv.org/abs/1612.02003} {arXiv:1612.02003 [astro-ph.HE]}
  \BibitemShut {NoStop}%
\bibitem [{\citenamefont {Abbott}\ \emph
  {et~al.}(2016{\natexlab{h}})\citenamefont {Abbott} \emph
  {et~al.}}]{TheLIGOScientific:2016uux}%
  \BibitemOpen
  \bibfield  {author} {\bibinfo {author} {\bibfnamefont {B.~P.}\ \bibnamefont
  {Abbott}} \emph {et~al.} (\bibinfo {collaboration} {LIGO Scientific
  Collaboration and Virgo Collaboration}),\ }\href {\doibase
  10.1103/PhysRevD.94.069903, 10.1103/PhysRevD.93.122004} {\bibfield  {journal}
  {\bibinfo  {journal} {Phys. Rev.}\ }\textbf {\bibinfo {volume} {D93}},\
  \bibinfo {pages} {122004} (\bibinfo {year} {2016}{\natexlab{h}})},\ \bibinfo
  {note} {[Addendum: Phys. Rev.D94,no.6,069903(2016)]},\ \Eprint
  {http://arxiv.org/abs/1602.03843} {arXiv:1602.03843 [gr-qc]} \BibitemShut
  {NoStop}%
\bibitem [{\citenamefont {Bruegmann}\ \emph {et~al.}(2008)\citenamefont
  {Bruegmann}, \citenamefont {Gonzalez}, \citenamefont {Hannam}, \citenamefont
  {Husa}, \citenamefont {Sperhake},\ and\ \citenamefont
  {Tichy}}]{Bruegmann:2006at}%
  \BibitemOpen
  \bibfield  {author} {\bibinfo {author} {\bibfnamefont {B.}~\bibnamefont
  {Bruegmann}}, \bibinfo {author} {\bibfnamefont {J.~A.}\ \bibnamefont
  {Gonzalez}}, \bibinfo {author} {\bibfnamefont {M.}~\bibnamefont {Hannam}},
  \bibinfo {author} {\bibfnamefont {S.}~\bibnamefont {Husa}}, \bibinfo {author}
  {\bibfnamefont {U.}~\bibnamefont {Sperhake}}, \ and\ \bibinfo {author}
  {\bibfnamefont {W.}~\bibnamefont {Tichy}},\ }\href {\doibase
  10.1103/PhysRevD.77.024027} {\bibfield  {journal} {\bibinfo  {journal} {Phys.
  Rev.}\ }\textbf {\bibinfo {volume} {D77}},\ \bibinfo {pages} {024027}
  (\bibinfo {year} {2008})},\ \Eprint {http://arxiv.org/abs/gr-qc/0610128}
  {arXiv:gr-qc/0610128 [gr-qc]} \BibitemShut {NoStop}%
\bibitem [{\citenamefont {Husa}\ \emph {et~al.}(2008)\citenamefont {Husa},
  \citenamefont {Gonzalez}, \citenamefont {Hannam}, \citenamefont {Bruegmann},\
  and\ \citenamefont {Sperhake}}]{Husa:2007hp}%
  \BibitemOpen
  \bibfield  {author} {\bibinfo {author} {\bibfnamefont {S.}~\bibnamefont
  {Husa}}, \bibinfo {author} {\bibfnamefont {J.~A.}\ \bibnamefont {Gonzalez}},
  \bibinfo {author} {\bibfnamefont {M.}~\bibnamefont {Hannam}}, \bibinfo
  {author} {\bibfnamefont {B.}~\bibnamefont {Bruegmann}}, \ and\ \bibinfo
  {author} {\bibfnamefont {U.}~\bibnamefont {Sperhake}},\ }\href {\doibase
  10.1088/0264-9381/25/10/105006} {\bibfield  {journal} {\bibinfo  {journal}
  {Class. Quant. Grav.}\ }\textbf {\bibinfo {volume} {25}},\ \bibinfo {pages}
  {105006} (\bibinfo {year} {2008})},\ \Eprint {http://arxiv.org/abs/0706.0740}
  {arXiv:0706.0740 [gr-qc]} \BibitemShut {NoStop}%
\bibitem [{\citenamefont {{Mrou{\'e}}}\ \emph {et~al.}(2013)\citenamefont
  {{Mrou{\'e}}}, \citenamefont {{Scheel}}, \citenamefont {{Szil{\'a}gyi}},
  \citenamefont {{Pfeiffer}}, \citenamefont {{Boyle}}, \citenamefont
  {{Hemberger}}, \citenamefont {{Kidder}}, \citenamefont {{Lovelace}},
  \citenamefont {{Ossokine}}, \citenamefont {{Taylor}}, \citenamefont
  {{Zengino{\u g}lu}}, \citenamefont {{Buchman}}, \citenamefont {{Chu}},
  \citenamefont {{Foley}}, \citenamefont {{Giesler}}, \citenamefont {{Owen}},\
  and\ \citenamefont {{Teukolsky}}}]{Mroue:2013xna}%
  \BibitemOpen
  \bibfield  {author} {\bibinfo {author} {\bibfnamefont {A.~H.}\ \bibnamefont
  {{Mrou{\'e}}}}, \bibinfo {author} {\bibfnamefont {M.~A.}\ \bibnamefont
  {{Scheel}}}, \bibinfo {author} {\bibfnamefont {B.}~\bibnamefont
  {{Szil{\'a}gyi}}}, \bibinfo {author} {\bibfnamefont {H.~P.}\ \bibnamefont
  {{Pfeiffer}}}, \bibinfo {author} {\bibfnamefont {M.}~\bibnamefont {{Boyle}}},
  \bibinfo {author} {\bibfnamefont {D.~A.}\ \bibnamefont {{Hemberger}}},
  \bibinfo {author} {\bibfnamefont {L.~E.}\ \bibnamefont {{Kidder}}}, \bibinfo
  {author} {\bibfnamefont {G.}~\bibnamefont {{Lovelace}}}, \bibinfo {author}
  {\bibfnamefont {S.}~\bibnamefont {{Ossokine}}}, \bibinfo {author}
  {\bibfnamefont {N.~W.}\ \bibnamefont {{Taylor}}}, \bibinfo {author}
  {\bibfnamefont {A.}~\bibnamefont {{Zengino{\u g}lu}}}, \bibinfo {author}
  {\bibfnamefont {L.~T.}\ \bibnamefont {{Buchman}}}, \bibinfo {author}
  {\bibfnamefont {T.}~\bibnamefont {{Chu}}}, \bibinfo {author} {\bibfnamefont
  {E.}~\bibnamefont {{Foley}}}, \bibinfo {author} {\bibfnamefont
  {M.}~\bibnamefont {{Giesler}}}, \bibinfo {author} {\bibfnamefont
  {R.}~\bibnamefont {{Owen}}}, \ and\ \bibinfo {author} {\bibfnamefont {S.~A.}\
  \bibnamefont {{Teukolsky}}},\ }\href {\doibase
  10.1103/PhysRevLett.111.241104} {\bibfield  {journal} {\bibinfo  {journal}
  {Phys. Rev. Lett.}\ }\textbf {\bibinfo {volume} {111}},\ \bibinfo {pages}
  {241104} (\bibinfo {year} {2013})},\ \Eprint {http://arxiv.org/abs/1304.6077}
  {arXiv:1304.6077 [gr-qc]} \BibitemShut {NoStop}%
\bibitem [{\citenamefont {{SXS
  Collaboration}}(2016{\natexlab{a}})}]{SXScatalog}%
  \BibitemOpen
  \bibfield  {author} {\bibinfo {author} {\bibnamefont {{SXS Collaboration}}},\
  }\href {http://www.black-holes.org/waveforms/} {\enquote {\bibinfo {title}
  {{SXS Gravitational Waveform Database}},}\ } (\bibinfo {year}
  {2016}{\natexlab{a}})\BibitemShut {NoStop}%
\bibitem [{\citenamefont {{SXS Collaboration}}(2016{\natexlab{b}})}]{SpEC}%
  \BibitemOpen
  \bibfield  {author} {\bibinfo {author} {\bibnamefont {{SXS Collaboration}}},\
  }\href {http://www.black-holes.org/SpEC.html} {\enquote {\bibinfo {title}
  {{SpEC: Spectral Einstein Code}},}\ } (\bibinfo {year}
  {2016}{\natexlab{b}})\BibitemShut {NoStop}%
\bibitem [{\citenamefont {Jani}\ \emph
  {et~al.}(2016{\natexlab{a}})\citenamefont {Jani}, \citenamefont {Healy},
  \citenamefont {Clark}, \citenamefont {London}, \citenamefont {Laguna},\ and\
  \citenamefont {Shoemaker}}]{Jani:2016wkt}%
  \BibitemOpen
  \bibfield  {author} {\bibinfo {author} {\bibfnamefont {K.}~\bibnamefont
  {Jani}}, \bibinfo {author} {\bibfnamefont {J.}~\bibnamefont {Healy}},
  \bibinfo {author} {\bibfnamefont {J.~A.}\ \bibnamefont {Clark}}, \bibinfo
  {author} {\bibfnamefont {L.}~\bibnamefont {London}}, \bibinfo {author}
  {\bibfnamefont {P.}~\bibnamefont {Laguna}}, \ and\ \bibinfo {author}
  {\bibfnamefont {D.}~\bibnamefont {Shoemaker}},\ }\href {\doibase
  10.1088/0264-9381/33/20/204001} {\bibfield  {journal} {\bibinfo  {journal}
  {Class. Quant. Grav.}\ }\textbf {\bibinfo {volume} {33}},\ \bibinfo {pages}
  {204001} (\bibinfo {year} {2016}{\natexlab{a}})},\ \Eprint
  {http://arxiv.org/abs/1605.03204} {arXiv:1605.03204 [gr-qc]} \BibitemShut
  {NoStop}%
\bibitem [{\citenamefont {Jani}\ \emph
  {et~al.}(2016{\natexlab{b}})\citenamefont {Jani}, \citenamefont {Healy},
  \citenamefont {Clark}, \citenamefont {London}, \citenamefont {Laguna},\ and\
  \citenamefont {Shoemaker}}]{GaTechcatalog}%
  \BibitemOpen
  \bibfield  {author} {\bibinfo {author} {\bibfnamefont {K.}~\bibnamefont
  {Jani}}, \bibinfo {author} {\bibfnamefont {J.}~\bibnamefont {Healy}},
  \bibinfo {author} {\bibfnamefont {J.~A.}\ \bibnamefont {Clark}}, \bibinfo
  {author} {\bibfnamefont {L.}~\bibnamefont {London}}, \bibinfo {author}
  {\bibfnamefont {P.}~\bibnamefont {Laguna}}, \ and\ \bibinfo {author}
  {\bibfnamefont {D.}~\bibnamefont {Shoemaker}},\ }\href
  {http://www.einstein.gatech.edu/catalog/} {\enquote {\bibinfo {title}
  {{Georgia Tech} catalog of binary black hole simulations},}\ } (\bibinfo
  {year} {2016}{\natexlab{b}})\BibitemShut {NoStop}%
\bibitem [{\citenamefont {Herrmann}\ \emph {et~al.}(2007)\citenamefont
  {Herrmann}, \citenamefont {Hinder}, \citenamefont {Shoemaker},\ and\
  \citenamefont {Laguna}}]{Herrmann:2006ks}%
  \BibitemOpen
  \bibfield  {author} {\bibinfo {author} {\bibfnamefont {F.}~\bibnamefont
  {Herrmann}}, \bibinfo {author} {\bibfnamefont {I.}~\bibnamefont {Hinder}},
  \bibinfo {author} {\bibfnamefont {D.}~\bibnamefont {Shoemaker}}, \ and\
  \bibinfo {author} {\bibfnamefont {P.}~\bibnamefont {Laguna}},\ }\href
  {\doibase 10.1088/0264-9381/24/12/S04} {\bibfield  {journal} {\bibinfo
  {journal} {Class. Quant. Grav.}\ }\textbf {\bibinfo {volume} {24}},\ \bibinfo
  {pages} {S33} (\bibinfo {year} {2007})},\ \Eprint
  {http://arxiv.org/abs/gr-qc/0601026} {arXiv:gr-qc/0601026 [gr-qc]}
  \BibitemShut {NoStop}%
\bibitem [{\citenamefont {Vaishnav}\ \emph {et~al.}(2007)\citenamefont
  {Vaishnav}, \citenamefont {Hinder}, \citenamefont {Herrmann},\ and\
  \citenamefont {Shoemaker}}]{Vaishnav:2007nm}%
  \BibitemOpen
  \bibfield  {author} {\bibinfo {author} {\bibfnamefont {B.}~\bibnamefont
  {Vaishnav}}, \bibinfo {author} {\bibfnamefont {I.}~\bibnamefont {Hinder}},
  \bibinfo {author} {\bibfnamefont {F.}~\bibnamefont {Herrmann}}, \ and\
  \bibinfo {author} {\bibfnamefont {D.}~\bibnamefont {Shoemaker}},\ }\href
  {\doibase 10.1103/PhysRevD.76.084020} {\bibfield  {journal} {\bibinfo
  {journal} {Phys. Rev.}\ }\textbf {\bibinfo {volume} {D76}},\ \bibinfo {pages}
  {084020} (\bibinfo {year} {2007})},\ \Eprint {http://arxiv.org/abs/0705.3829}
  {arXiv:0705.3829 [gr-qc]} \BibitemShut {NoStop}%
\bibitem [{\citenamefont {Healy}\ \emph {et~al.}(2009)\citenamefont {Healy},
  \citenamefont {Levin},\ and\ \citenamefont {Shoemaker}}]{Healy:2009zm}%
  \BibitemOpen
  \bibfield  {author} {\bibinfo {author} {\bibfnamefont {J.}~\bibnamefont
  {Healy}}, \bibinfo {author} {\bibfnamefont {J.}~\bibnamefont {Levin}}, \ and\
  \bibinfo {author} {\bibfnamefont {D.}~\bibnamefont {Shoemaker}},\ }\href
  {\doibase 10.1103/PhysRevLett.103.131101} {\bibfield  {journal} {\bibinfo
  {journal} {Phys. Rev. Lett.}\ }\textbf {\bibinfo {volume} {103}},\ \bibinfo
  {pages} {131101} (\bibinfo {year} {2009})},\ \Eprint
  {http://arxiv.org/abs/0907.0671} {arXiv:0907.0671 [gr-qc]} \BibitemShut
  {NoStop}%
\bibitem [{\citenamefont {Pekowsky}\ \emph
  {et~al.}(2013{\natexlab{a}})\citenamefont {Pekowsky}, \citenamefont
  {O'Shaughnessy}, \citenamefont {Healy},\ and\ \citenamefont
  {Shoemaker}}]{Pekowsky:2013ska}%
  \BibitemOpen
  \bibfield  {author} {\bibinfo {author} {\bibfnamefont {L.}~\bibnamefont
  {Pekowsky}}, \bibinfo {author} {\bibfnamefont {R.}~\bibnamefont
  {O'Shaughnessy}}, \bibinfo {author} {\bibfnamefont {J.}~\bibnamefont
  {Healy}}, \ and\ \bibinfo {author} {\bibfnamefont {D.}~\bibnamefont
  {Shoemaker}},\ }\href {\doibase 10.1103/PhysRevD.88.024040} {\bibfield
  {journal} {\bibinfo  {journal} {Phys. Rev.}\ }\textbf {\bibinfo {volume}
  {D88}},\ \bibinfo {pages} {024040} (\bibinfo {year} {2013}{\natexlab{a}})},\
  \Eprint {http://arxiv.org/abs/1304.3176} {arXiv:1304.3176 [gr-qc]}
  \BibitemShut {NoStop}%
\bibitem [{\citenamefont {Healy}\ \emph {et~al.}(2017)\citenamefont {Healy},
  \citenamefont {Lousto}, \citenamefont {Zlochower},\ and\ \citenamefont
  {Campanelli}}]{Healy:2017psd}%
  \BibitemOpen
  \bibfield  {author} {\bibinfo {author} {\bibfnamefont {J.}~\bibnamefont
  {Healy}}, \bibinfo {author} {\bibfnamefont {C.~O.}\ \bibnamefont {Lousto}},
  \bibinfo {author} {\bibfnamefont {Y.}~\bibnamefont {Zlochower}}, \ and\
  \bibinfo {author} {\bibfnamefont {M.}~\bibnamefont {Campanelli}},\
  }\href@noop {} {\  (\bibinfo {year} {2017})},\ \Eprint
  {http://arxiv.org/abs/1703.03423} {arXiv:1703.03423 [gr-qc]} \BibitemShut
  {NoStop}%
\bibitem [{\citenamefont {Campanelli}\ \emph {et~al.}(2016)\citenamefont
  {Campanelli}, \citenamefont {Healy}, \citenamefont {Lousto},\ and\
  \citenamefont {Zlochower}}]{RITcatalog}%
  \BibitemOpen
  \bibfield  {author} {\bibinfo {author} {\bibfnamefont {M.}~\bibnamefont
  {Campanelli}}, \bibinfo {author} {\bibfnamefont {J.}~\bibnamefont {Healy}},
  \bibinfo {author} {\bibfnamefont {C.}~\bibnamefont {Lousto}}, \ and\ \bibinfo
  {author} {\bibfnamefont {Y.}~\bibnamefont {Zlochower}},\ }\href
  {http://ccrg.rit.edu/~RITCatalog/} {\enquote {\bibinfo {title} {{CCRG@RIT
  Catalog of Numerical Simulations}},}\ } (\bibinfo {year} {2016})\BibitemShut
  {NoStop}%
\bibitem [{\citenamefont {Zlochower}\ \emph {et~al.}(2005)\citenamefont
  {Zlochower}, \citenamefont {Baker}, \citenamefont {Campanelli},\ and\
  \citenamefont {Lousto}}]{Zlochower:2005bj}%
  \BibitemOpen
  \bibfield  {author} {\bibinfo {author} {\bibfnamefont {Y.}~\bibnamefont
  {Zlochower}}, \bibinfo {author} {\bibfnamefont {J.~G.}\ \bibnamefont
  {Baker}}, \bibinfo {author} {\bibfnamefont {M.}~\bibnamefont {Campanelli}}, \
  and\ \bibinfo {author} {\bibfnamefont {C.~O.}\ \bibnamefont {Lousto}},\
  }\href {\doibase 10.1103/PhysRevD.72.024021} {\bibfield  {journal} {\bibinfo
  {journal} {Phys. Rev.}\ }\textbf {\bibinfo {volume} {D72}},\ \bibinfo {pages}
  {024021} (\bibinfo {year} {2005})},\ \Eprint
  {http://arxiv.org/abs/gr-qc/0505055} {arXiv:gr-qc/0505055 [gr-qc]}
  \BibitemShut {NoStop}%
\bibitem [{\citenamefont {Baumgarte}\ and\ \citenamefont
  {Shapiro}(1998)}]{Baumgarte:1998te}%
  \BibitemOpen
  \bibfield  {author} {\bibinfo {author} {\bibfnamefont {T.~W.}\ \bibnamefont
  {Baumgarte}}\ and\ \bibinfo {author} {\bibfnamefont {S.~L.}\ \bibnamefont
  {Shapiro}},\ }\href {\doibase 10.1103/PhysRevD.59.024007} {\bibfield
  {journal} {\bibinfo  {journal} {Phys. Rev.}\ }\textbf {\bibinfo {volume}
  {D59}},\ \bibinfo {pages} {024007} (\bibinfo {year} {1998})},\ \Eprint
  {http://arxiv.org/abs/gr-qc/9810065} {arXiv:gr-qc/9810065 [gr-qc]}
  \BibitemShut {NoStop}%
\bibitem [{\citenamefont {Campanelli}\ \emph {et~al.}(2006)\citenamefont
  {Campanelli}, \citenamefont {Lousto}, \citenamefont {Marronetti},\ and\
  \citenamefont {Zlochower}}]{Campanelli:2005dd}%
  \BibitemOpen
  \bibfield  {author} {\bibinfo {author} {\bibfnamefont {M.}~\bibnamefont
  {Campanelli}}, \bibinfo {author} {\bibfnamefont {C.~O.}\ \bibnamefont
  {Lousto}}, \bibinfo {author} {\bibfnamefont {P.}~\bibnamefont {Marronetti}},
  \ and\ \bibinfo {author} {\bibfnamefont {Y.}~\bibnamefont {Zlochower}},\
  }\href {\doibase 10.1103/PhysRevLett.96.111101} {\bibfield  {journal}
  {\bibinfo  {journal} {Phys. Rev. Lett.}\ }\textbf {\bibinfo {volume} {96}},\
  \bibinfo {pages} {111101} (\bibinfo {year} {2006})},\ \Eprint
  {http://arxiv.org/abs/gr-qc/0511048} {arXiv:gr-qc/0511048 [gr-qc]}
  \BibitemShut {NoStop}%
\bibitem [{\citenamefont {Baker}\ \emph {et~al.}(2006)\citenamefont {Baker},
  \citenamefont {Centrella}, \citenamefont {Choi}, \citenamefont {Koppitz},\
  and\ \citenamefont {van Meter}}]{Baker:2005vv}%
  \BibitemOpen
  \bibfield  {author} {\bibinfo {author} {\bibfnamefont {J.~G.}\ \bibnamefont
  {Baker}}, \bibinfo {author} {\bibfnamefont {J.}~\bibnamefont {Centrella}},
  \bibinfo {author} {\bibfnamefont {D.-I.}\ \bibnamefont {Choi}}, \bibinfo
  {author} {\bibfnamefont {M.}~\bibnamefont {Koppitz}}, \ and\ \bibinfo
  {author} {\bibfnamefont {J.}~\bibnamefont {van Meter}},\ }\href {\doibase
  10.1103/PhysRevLett.96.111102} {\bibfield  {journal} {\bibinfo  {journal}
  {Phys. Rev. Lett.}\ }\textbf {\bibinfo {volume} {96}},\ \bibinfo {pages}
  {111102} (\bibinfo {year} {2006})},\ \Eprint
  {http://arxiv.org/abs/gr-qc/0511103} {arXiv:gr-qc/0511103 [gr-qc]}
  \BibitemShut {NoStop}%
\bibitem [{\citenamefont {Boyle}(2013)}]{Boyle:2013nka}%
  \BibitemOpen
  \bibfield  {author} {\bibinfo {author} {\bibfnamefont {M.}~\bibnamefont
  {Boyle}},\ }\href {\doibase 10.1103/PhysRevD.87.104006} {\bibfield  {journal}
  {\bibinfo  {journal} {Phys. Rev.}\ }\textbf {\bibinfo {volume} {D87}},\
  \bibinfo {pages} {104006} (\bibinfo {year} {2013})},\ \Eprint
  {http://arxiv.org/abs/1302.2919} {arXiv:1302.2919 [gr-qc]} \BibitemShut
  {NoStop}%
\bibitem [{\citenamefont {Boyle}\ \emph {et~al.}(2014)\citenamefont {Boyle},
  \citenamefont {Kidder}, \citenamefont {Ossokine},\ and\ \citenamefont
  {Pfeiffer}}]{Boyle:2014ioa}%
  \BibitemOpen
  \bibfield  {author} {\bibinfo {author} {\bibfnamefont {M.}~\bibnamefont
  {Boyle}}, \bibinfo {author} {\bibfnamefont {L.~E.}\ \bibnamefont {Kidder}},
  \bibinfo {author} {\bibfnamefont {S.}~\bibnamefont {Ossokine}}, \ and\
  \bibinfo {author} {\bibfnamefont {H.~P.}\ \bibnamefont {Pfeiffer}},\
  }\href@noop {} {\  (\bibinfo {year} {2014})},\ \Eprint
  {http://arxiv.org/abs/1409.4431} {arXiv:1409.4431 [gr-qc]} \BibitemShut
  {NoStop}%
\bibitem [{\citenamefont {Boyle}(2016)}]{Boyle:2015nqa}%
  \BibitemOpen
  \bibfield  {author} {\bibinfo {author} {\bibfnamefont {M.}~\bibnamefont
  {Boyle}},\ }\href {\doibase 10.1103/PhysRevD.93.084031} {\bibfield  {journal}
  {\bibinfo  {journal} {Phys. Rev.}\ }\textbf {\bibinfo {volume} {D93}},\
  \bibinfo {pages} {084031} (\bibinfo {year} {2016})},\ \Eprint
  {http://arxiv.org/abs/1509.00862} {arXiv:1509.00862 [gr-qc]} \BibitemShut
  {NoStop}%
\bibitem [{\citenamefont {Reisswig}\ and\ \citenamefont
  {Pollney}(2011)}]{Reisswig:2010di}%
  \BibitemOpen
  \bibfield  {author} {\bibinfo {author} {\bibfnamefont {C.}~\bibnamefont
  {Reisswig}}\ and\ \bibinfo {author} {\bibfnamefont {D.}~\bibnamefont
  {Pollney}},\ }\href {\doibase 10.1088/0264-9381/28/19/195015} {\bibfield
  {journal} {\bibinfo  {journal} {Class. Quant. Grav.}\ }\textbf {\bibinfo
  {volume} {28}},\ \bibinfo {pages} {195015} (\bibinfo {year} {2011})},\
  \Eprint {http://arxiv.org/abs/1006.1632} {arXiv:1006.1632 [gr-qc]}
  \BibitemShut {NoStop}%
\bibitem [{\citenamefont {Bardeen}\ \emph {et~al.}(1972)\citenamefont
  {Bardeen}, \citenamefont {Press},\ and\ \citenamefont
  {Teukolsky}}]{Bardeen:1972fi}%
  \BibitemOpen
  \bibfield  {author} {\bibinfo {author} {\bibfnamefont {J.~M.}\ \bibnamefont
  {Bardeen}}, \bibinfo {author} {\bibfnamefont {W.~H.}\ \bibnamefont {Press}},
  \ and\ \bibinfo {author} {\bibfnamefont {S.~A.}\ \bibnamefont {Teukolsky}},\
  }\href {\doibase 10.1086/151796} {\bibfield  {journal} {\bibinfo  {journal}
  {Astrophys. J.}\ }\textbf {\bibinfo {volume} {178}},\ \bibinfo {pages} {347}
  (\bibinfo {year} {1972})}\BibitemShut {NoStop}%
\bibitem [{\citenamefont {Teukolsky}\ and\ \citenamefont
  {Press}(1974)}]{Teukolsky:1974yv}%
  \BibitemOpen
  \bibfield  {author} {\bibinfo {author} {\bibfnamefont {S.~A.}\ \bibnamefont
  {Teukolsky}}\ and\ \bibinfo {author} {\bibfnamefont {W.~H.}\ \bibnamefont
  {Press}},\ }\href {\doibase 10.1086/153180} {\bibfield  {journal} {\bibinfo
  {journal} {Astrophys. J.}\ }\textbf {\bibinfo {volume} {193}},\ \bibinfo
  {pages} {443} (\bibinfo {year} {1974})}\BibitemShut {NoStop}%
\bibitem [{\citenamefont {Fujita}(2015)}]{Fujita:2014eta}%
  \BibitemOpen
  \bibfield  {author} {\bibinfo {author} {\bibfnamefont {R.}~\bibnamefont
  {Fujita}},\ }\href {\doibase 10.1093/ptep/ptv012} {\bibfield  {journal}
  {\bibinfo  {journal} {PTEP}\ }\textbf {\bibinfo {volume} {2015}},\ \bibinfo
  {pages} {033E01} (\bibinfo {year} {2015})},\ \Eprint
  {http://arxiv.org/abs/1412.5689} {arXiv:1412.5689 [gr-qc]} \BibitemShut
  {NoStop}%
\bibitem [{\citenamefont {Bernuzzi}\ and\ \citenamefont
  {Nagar}(2010)}]{Bernuzzi:2010ty}%
  \BibitemOpen
  \bibfield  {author} {\bibinfo {author} {\bibfnamefont {S.}~\bibnamefont
  {Bernuzzi}}\ and\ \bibinfo {author} {\bibfnamefont {A.}~\bibnamefont
  {Nagar}},\ }\href {\doibase 10.1103/PhysRevD.81.084056} {\bibfield  {journal}
  {\bibinfo  {journal} {Phys. Rev.}\ }\textbf {\bibinfo {volume} {D81}},\
  \bibinfo {pages} {084056} (\bibinfo {year} {2010})},\ \Eprint
  {http://arxiv.org/abs/1003.0597} {arXiv:1003.0597 [gr-qc]} \BibitemShut
  {NoStop}%
\bibitem [{\citenamefont {Damour}\ and\ \citenamefont
  {Nagar}(2007)}]{Damour:2007xr}%
  \BibitemOpen
  \bibfield  {author} {\bibinfo {author} {\bibfnamefont {T.}~\bibnamefont
  {Damour}}\ and\ \bibinfo {author} {\bibfnamefont {A.}~\bibnamefont {Nagar}},\
  }\href {\doibase 10.1103/PhysRevD.76.064028} {\bibfield  {journal} {\bibinfo
  {journal} {Phys. Rev.}\ }\textbf {\bibinfo {volume} {D76}},\ \bibinfo {pages}
  {064028} (\bibinfo {year} {2007})},\ \Eprint {http://arxiv.org/abs/0705.2519}
  {arXiv:0705.2519 [gr-qc]} \BibitemShut {NoStop}%
\bibitem [{\citenamefont {Damour}\ \emph {et~al.}(2009)\citenamefont {Damour},
  \citenamefont {Iyer},\ and\ \citenamefont {Nagar}}]{Damour:2008gu}%
  \BibitemOpen
  \bibfield  {author} {\bibinfo {author} {\bibfnamefont {T.}~\bibnamefont
  {Damour}}, \bibinfo {author} {\bibfnamefont {B.~R.}\ \bibnamefont {Iyer}}, \
  and\ \bibinfo {author} {\bibfnamefont {A.}~\bibnamefont {Nagar}},\ }\href
  {\doibase 10.1103/PhysRevD.79.064004} {\bibfield  {journal} {\bibinfo
  {journal} {Phys. Rev.}\ }\textbf {\bibinfo {volume} {D79}},\ \bibinfo {pages}
  {064004} (\bibinfo {year} {2009})},\ \Eprint {http://arxiv.org/abs/0811.2069}
  {arXiv:0811.2069 [gr-qc]} \BibitemShut {NoStop}%
\bibitem [{\citenamefont {Pan}\ \emph {et~al.}(2011)\citenamefont {Pan},
  \citenamefont {Buonanno}, \citenamefont {Fujita}, \citenamefont {Racine},\
  and\ \citenamefont {Tagoshi}}]{Pan:2010hz}%
  \BibitemOpen
  \bibfield  {author} {\bibinfo {author} {\bibfnamefont {Y.}~\bibnamefont
  {Pan}}, \bibinfo {author} {\bibfnamefont {A.}~\bibnamefont {Buonanno}},
  \bibinfo {author} {\bibfnamefont {R.}~\bibnamefont {Fujita}}, \bibinfo
  {author} {\bibfnamefont {E.}~\bibnamefont {Racine}}, \ and\ \bibinfo {author}
  {\bibfnamefont {H.}~\bibnamefont {Tagoshi}},\ }\href {\doibase
  10.1103/PhysRevD.83.064003, 10.1103/PhysRevD.87.109901} {\bibfield  {journal}
  {\bibinfo  {journal} {Phys. Rev.}\ }\textbf {\bibinfo {volume} {D83}},\
  \bibinfo {pages} {064003} (\bibinfo {year} {2011})},\ \bibinfo {note}
  {[Erratum: Phys. Rev.D87,no.10,109901(2013)]},\ \Eprint
  {http://arxiv.org/abs/1006.0431} {arXiv:1006.0431 [gr-qc]} \BibitemShut
  {NoStop}%
\bibitem [{\citenamefont {Fujita}(2012)}]{Fujita:2012cm}%
  \BibitemOpen
  \bibfield  {author} {\bibinfo {author} {\bibfnamefont {R.}~\bibnamefont
  {Fujita}},\ }\href {\doibase 10.1143/PTP.128.971} {\bibfield  {journal}
  {\bibinfo  {journal} {Prog. Theor. Phys.}\ }\textbf {\bibinfo {volume}
  {128}},\ \bibinfo {pages} {971} (\bibinfo {year} {2012})},\ \Eprint
  {http://arxiv.org/abs/1211.5535} {arXiv:1211.5535 [gr-qc]} \BibitemShut
  {NoStop}%
\bibitem [{\citenamefont {Nagar}\ and\ \citenamefont
  {Shah}(2016)}]{Nagar:2016ayt}%
  \BibitemOpen
  \bibfield  {author} {\bibinfo {author} {\bibfnamefont {A.}~\bibnamefont
  {Nagar}}\ and\ \bibinfo {author} {\bibfnamefont {A.}~\bibnamefont {Shah}},\
  }\href {\doibase 10.1103/PhysRevD.94.104017} {\bibfield  {journal} {\bibinfo
  {journal} {Phys. Rev.}\ }\textbf {\bibinfo {volume} {D94}},\ \bibinfo {pages}
  {104017} (\bibinfo {year} {2016})},\ \Eprint
  {http://arxiv.org/abs/1606.00207} {arXiv:1606.00207 [gr-qc]} \BibitemShut
  {NoStop}%
\bibitem [{\citenamefont {Bernuzzi}\ \emph
  {et~al.}(2011{\natexlab{b}})\citenamefont {Bernuzzi}, \citenamefont {Nagar},\
  and\ \citenamefont {Zengino{\u g}lu}}]{Bernuzzi:2010xj}%
  \BibitemOpen
  \bibfield  {author} {\bibinfo {author} {\bibfnamefont {S.}~\bibnamefont
  {Bernuzzi}}, \bibinfo {author} {\bibfnamefont {A.}~\bibnamefont {Nagar}}, \
  and\ \bibinfo {author} {\bibfnamefont {A.}~\bibnamefont {Zengino{\u g}lu}},\
  }\href {\doibase 10.1103/PhysRevD.83.064010} {\bibfield  {journal} {\bibinfo
  {journal} {Phys. Rev.}\ }\textbf {\bibinfo {volume} {D83}},\ \bibinfo {pages}
  {064010} (\bibinfo {year} {2011}{\natexlab{b}})},\ \Eprint
  {http://arxiv.org/abs/1012.2456} {arXiv:1012.2456 [gr-qc]} \BibitemShut
  {NoStop}%
\bibitem [{\citenamefont {Damour}\ \emph {et~al.}(2013)\citenamefont {Damour},
  \citenamefont {Nagar},\ and\ \citenamefont {Bernuzzi}}]{Damour:2012ky}%
  \BibitemOpen
  \bibfield  {author} {\bibinfo {author} {\bibfnamefont {T.}~\bibnamefont
  {Damour}}, \bibinfo {author} {\bibfnamefont {A.}~\bibnamefont {Nagar}}, \
  and\ \bibinfo {author} {\bibfnamefont {S.}~\bibnamefont {Bernuzzi}},\ }\href
  {\doibase 10.1103/PhysRevD.87.084035} {\bibfield  {journal} {\bibinfo
  {journal} {Phys. Rev.}\ }\textbf {\bibinfo {volume} {D87}},\ \bibinfo {pages}
  {084035} (\bibinfo {year} {2013})},\ \Eprint {http://arxiv.org/abs/1212.4357}
  {arXiv:1212.4357 [gr-qc]} \BibitemShut {NoStop}%
\bibitem [{\citenamefont {Nagar}(2013)}]{Nagar:2013sga}%
  \BibitemOpen
  \bibfield  {author} {\bibinfo {author} {\bibfnamefont {A.}~\bibnamefont
  {Nagar}},\ }\href {\doibase 10.1103/PhysRevD.88.121501} {\bibfield  {journal}
  {\bibinfo  {journal} {Phys. Rev.}\ }\textbf {\bibinfo {volume} {D88}},\
  \bibinfo {pages} {121501} (\bibinfo {year} {2013})},\ \Eprint
  {http://arxiv.org/abs/1306.6299} {arXiv:1306.6299 [gr-qc]} \BibitemShut
  {NoStop}%
\bibitem [{\citenamefont {Nagar}\ \emph {et~al.}(2014)\citenamefont {Nagar},
  \citenamefont {Harms}, \citenamefont {Bernuzzi},\ and\ \citenamefont
  {Zengino{\u g}lu}}]{Nagar:2014kha}%
  \BibitemOpen
  \bibfield  {author} {\bibinfo {author} {\bibfnamefont {A.}~\bibnamefont
  {Nagar}}, \bibinfo {author} {\bibfnamefont {E.}~\bibnamefont {Harms}},
  \bibinfo {author} {\bibfnamefont {S.}~\bibnamefont {Bernuzzi}}, \ and\
  \bibinfo {author} {\bibfnamefont {A.}~\bibnamefont {Zengino{\u g}lu}},\
  }\href {\doibase 10.1103/PhysRevD.90.124086} {\bibfield  {journal} {\bibinfo
  {journal} {Phys. Rev.}\ }\textbf {\bibinfo {volume} {D90}},\ \bibinfo {pages}
  {124086} (\bibinfo {year} {2014})},\ \Eprint {http://arxiv.org/abs/1407.5033}
  {arXiv:1407.5033 [gr-qc]} \BibitemShut {NoStop}%
\bibitem [{\citenamefont {Akaike}(1974)}]{Akaike:1974}%
  \BibitemOpen
  \bibfield  {author} {\bibinfo {author} {\bibfnamefont {H.}~\bibnamefont
  {Akaike}},\ }\href {\doibase 10.1109/TAC.1974.1100705} {\bibfield  {journal}
  {\bibinfo  {journal} {IEEE Trans. Autom. Control}\ }\textbf {\bibinfo
  {volume} {19}},\ \bibinfo {pages} {716} (\bibinfo {year} {1974})}\BibitemShut
  {NoStop}%
\bibitem [{\citenamefont {Schwarz}(1978)}]{Schwarz:1978}%
  \BibitemOpen
  \bibfield  {author} {\bibinfo {author} {\bibfnamefont {G.~E.}\ \bibnamefont
  {Schwarz}},\ }\href {\doibase 10.1214/aos/1176344136} {\bibfield  {journal}
  {\bibinfo  {journal} {Annals of Statistics}\ }\textbf {\bibinfo {volume}
  {6}},\ \bibinfo {pages} {461} (\bibinfo {year} {1978})}\BibitemShut {NoStop}%
\bibitem [{\citenamefont {Liddle}(2007)}]{Liddle:2007fy}%
  \BibitemOpen
  \bibfield  {author} {\bibinfo {author} {\bibfnamefont {A.~R.}\ \bibnamefont
  {Liddle}},\ }\href {\doibase 10.1111/j.1745-3933.2007.00306.x} {\bibfield
  {journal} {\bibinfo  {journal} {Mon. Not. Roy. Astron. Soc.}\ }\textbf
  {\bibinfo {volume} {377}},\ \bibinfo {pages} {L74} (\bibinfo {year}
  {2007})},\ \Eprint {http://arxiv.org/abs/astro-ph/0701113}
  {arXiv:astro-ph/0701113 [astro-ph]} \BibitemShut {NoStop}%
\bibitem [{\citenamefont {Berti}\ \emph {et~al.}(2007)\citenamefont {Berti},
  \citenamefont {Cardoso}, \citenamefont {Gonzalez}, \citenamefont {Sperhake},
  \citenamefont {Hannam}, \citenamefont {Husa},\ and\ \citenamefont
  {Bruegmann}}]{Berti:2007fi}%
  \BibitemOpen
  \bibfield  {author} {\bibinfo {author} {\bibfnamefont {E.}~\bibnamefont
  {Berti}}, \bibinfo {author} {\bibfnamefont {V.}~\bibnamefont {Cardoso}},
  \bibinfo {author} {\bibfnamefont {J.~A.}\ \bibnamefont {Gonzalez}}, \bibinfo
  {author} {\bibfnamefont {U.}~\bibnamefont {Sperhake}}, \bibinfo {author}
  {\bibfnamefont {M.}~\bibnamefont {Hannam}}, \bibinfo {author} {\bibfnamefont
  {S.}~\bibnamefont {Husa}}, \ and\ \bibinfo {author} {\bibfnamefont
  {B.}~\bibnamefont {Bruegmann}},\ }\href {\doibase 10.1103/PhysRevD.76.064034}
  {\bibfield  {journal} {\bibinfo  {journal} {Phys. Rev.}\ }\textbf {\bibinfo
  {volume} {D76}},\ \bibinfo {pages} {064034} (\bibinfo {year} {2007})},\
  \Eprint {http://arxiv.org/abs/gr-qc/0703053} {arXiv:gr-qc/0703053 [GR-QC]}
  \BibitemShut {NoStop}%
\bibitem [{\citenamefont {Kelly}\ \emph {et~al.}(2011)\citenamefont {Kelly},
  \citenamefont {Baker}, \citenamefont {Boggs}, \citenamefont {McWilliams},\
  and\ \citenamefont {Centrella}}]{Kelly:2011bp}%
  \BibitemOpen
  \bibfield  {author} {\bibinfo {author} {\bibfnamefont {B.~J.}\ \bibnamefont
  {Kelly}}, \bibinfo {author} {\bibfnamefont {J.~G.}\ \bibnamefont {Baker}},
  \bibinfo {author} {\bibfnamefont {W.~D.}\ \bibnamefont {Boggs}}, \bibinfo
  {author} {\bibfnamefont {S.~T.}\ \bibnamefont {McWilliams}}, \ and\ \bibinfo
  {author} {\bibfnamefont {J.}~\bibnamefont {Centrella}},\ }\href {\doibase
  10.1103/PhysRevD.84.084009} {\bibfield  {journal} {\bibinfo  {journal} {Phys.
  Rev.}\ }\textbf {\bibinfo {volume} {D84}},\ \bibinfo {pages} {084009}
  (\bibinfo {year} {2011})},\ \Eprint {http://arxiv.org/abs/1107.1181}
  {arXiv:1107.1181 [gr-qc]} \BibitemShut {NoStop}%
\bibitem [{\citenamefont {Pekowsky}\ \emph
  {et~al.}(2013{\natexlab{b}})\citenamefont {Pekowsky}, \citenamefont {Healy},
  \citenamefont {Shoemaker},\ and\ \citenamefont {Laguna}}]{Pekowsky:2012sr}%
  \BibitemOpen
  \bibfield  {author} {\bibinfo {author} {\bibfnamefont {L.}~\bibnamefont
  {Pekowsky}}, \bibinfo {author} {\bibfnamefont {J.}~\bibnamefont {Healy}},
  \bibinfo {author} {\bibfnamefont {D.}~\bibnamefont {Shoemaker}}, \ and\
  \bibinfo {author} {\bibfnamefont {P.}~\bibnamefont {Laguna}},\ }\href
  {\doibase 10.1103/PhysRevD.87.084008} {\bibfield  {journal} {\bibinfo
  {journal} {Phys. Rev.}\ }\textbf {\bibinfo {volume} {D87}},\ \bibinfo {pages}
  {084008} (\bibinfo {year} {2013}{\natexlab{b}})},\ \Eprint
  {http://arxiv.org/abs/1210.1891} {arXiv:1210.1891 [gr-qc]} \BibitemShut
  {NoStop}%
\bibitem [{\citenamefont {Calder{\'o}n~Bustillo}\ \emph
  {et~al.}(2016)\citenamefont {Calder{\'o}n~Bustillo}, \citenamefont {Husa},
  \citenamefont {Sintes},\ and\ \citenamefont {P{\"u}rrer}}]{Bustillo:2015qty}%
  \BibitemOpen
  \bibfield  {author} {\bibinfo {author} {\bibfnamefont {J.}~\bibnamefont
  {Calder{\'o}n~Bustillo}}, \bibinfo {author} {\bibfnamefont {S.}~\bibnamefont
  {Husa}}, \bibinfo {author} {\bibfnamefont {A.~M.}\ \bibnamefont {Sintes}}, \
  and\ \bibinfo {author} {\bibfnamefont {M.}~\bibnamefont {P{\"u}rrer}},\
  }\href {\doibase 10.1103/PhysRevD.93.084019} {\bibfield  {journal} {\bibinfo
  {journal} {Phys. Rev.}\ }\textbf {\bibinfo {volume} {D93}},\ \bibinfo {pages}
  {084019} (\bibinfo {year} {2016})},\ \Eprint
  {http://arxiv.org/abs/1511.02060} {arXiv:1511.02060 [gr-qc]} \BibitemShut
  {NoStop}%
\bibitem [{\citenamefont {Boh{\'e}}\ \emph {et~al.}(2013)\citenamefont
  {Boh{\'e}}, \citenamefont {Marsat},\ and\ \citenamefont
  {Blanchet}}]{Bohe:2013cla}%
  \BibitemOpen
  \bibfield  {author} {\bibinfo {author} {\bibfnamefont {A.}~\bibnamefont
  {Boh{\'e}}}, \bibinfo {author} {\bibfnamefont {S.}~\bibnamefont {Marsat}}, \
  and\ \bibinfo {author} {\bibfnamefont {L.}~\bibnamefont {Blanchet}},\ }\href
  {\doibase 10.1088/0264-9381/30/13/135009} {\bibfield  {journal} {\bibinfo
  {journal} {Class. Quant. Grav.}\ }\textbf {\bibinfo {volume} {30}},\ \bibinfo
  {pages} {135009} (\bibinfo {year} {2013})},\ \Eprint
  {http://arxiv.org/abs/1303.7412} {arXiv:1303.7412 [gr-qc]} \BibitemShut
  {NoStop}%
\bibitem [{\citenamefont {Marsat}\ \emph {et~al.}(2014)\citenamefont {Marsat},
  \citenamefont {Boh{\'e}}, \citenamefont {Blanchet},\ and\ \citenamefont
  {Buonanno}}]{Marsat:2013caa}%
  \BibitemOpen
  \bibfield  {author} {\bibinfo {author} {\bibfnamefont {S.}~\bibnamefont
  {Marsat}}, \bibinfo {author} {\bibfnamefont {A.}~\bibnamefont {Boh{\'e}}},
  \bibinfo {author} {\bibfnamefont {L.}~\bibnamefont {Blanchet}}, \ and\
  \bibinfo {author} {\bibfnamefont {A.}~\bibnamefont {Buonanno}},\ }\href
  {\doibase 10.1088/0264-9381/31/2/025023} {\bibfield  {journal} {\bibinfo
  {journal} {Class. Quant. Grav.}\ }\textbf {\bibinfo {volume} {31}},\ \bibinfo
  {pages} {025023} (\bibinfo {year} {2014})},\ \Eprint
  {http://arxiv.org/abs/1307.6793} {arXiv:1307.6793 [gr-qc]} \BibitemShut
  {NoStop}%
\bibitem [{\citenamefont {Creighton}\ and\ \citenamefont
  {Anderson}(2011)}]{Creighton:2011zz}%
  \BibitemOpen
  \bibfield  {author} {\bibinfo {author} {\bibfnamefont {J.~D.~E.}\
  \bibnamefont {Creighton}}\ and\ \bibinfo {author} {\bibfnamefont {W.~G.}\
  \bibnamefont {Anderson}},\ }\href@noop {} {\emph {\bibinfo {title}
  {{Gravitational-wave physics and astronomy: An introduction to theory,
  experiment and data analysis}}}}\ (\bibinfo  {publisher} {Wiley-VCH},\
  \bibinfo {address} {Weinheim, Germany},\ \bibinfo {year} {2011})\BibitemShut
  {NoStop}%
\bibitem [{\citenamefont {Keitel}\ \emph
  {et~al.}(2017{\natexlab{a}})\citenamefont {Keitel}, \citenamefont
  {Jim{\'e}nez-Forteza}, \citenamefont {Husa}, \citenamefont {London} \emph
  {et~al.}}]{Keitel:2016krm-suppl}%
  \BibitemOpen
  \bibfield  {author} {\bibinfo {author} {\bibfnamefont {D.}~\bibnamefont
  {Keitel}}, \bibinfo {author} {\bibfnamefont {X.}~\bibnamefont
  {Jim{\'e}nez-Forteza}}, \bibinfo {author} {\bibfnamefont {S.}~\bibnamefont
  {Husa}}, \bibinfo {author} {\bibfnamefont {L.}~\bibnamefont {London}},  \emph
  {et~al.},\ }\href
  {http://link.aps.org/supplemental/10.1103/PhysRevD.96.024006} {\enquote
  {\bibinfo {title} {{Phys. Rev. D96, 024006 Supplemental Material}},}\ }
  (\bibinfo {year} {2017}{\natexlab{a}})\BibitemShut {NoStop}%
\bibitem [{\citenamefont {Keitel}\ \emph
  {et~al.}(2017{\natexlab{b}})\citenamefont {Keitel}, \citenamefont
  {Jim{\'e}nez-Forteza}, \citenamefont {Husa}, \citenamefont {London} \emph
  {et~al.}}]{Keitel:2016krm-anc}%
  \BibitemOpen
  \bibfield  {author} {\bibinfo {author} {\bibfnamefont {D.}~\bibnamefont
  {Keitel}}, \bibinfo {author} {\bibfnamefont {X.}~\bibnamefont
  {Jim{\'e}nez-Forteza}}, \bibinfo {author} {\bibfnamefont {S.}~\bibnamefont
  {Husa}}, \bibinfo {author} {\bibfnamefont {L.}~\bibnamefont {London}},  \emph
  {et~al.},\ }\href {https://arxiv.org/src/1612.09566v2/anc} {\enquote
  {\bibinfo {title} {{arXiv:1612.09566v2 ancillary files}},}\ } (\bibinfo
  {year} {2017}{\natexlab{b}})\BibitemShut {NoStop}%
\bibitem [{\citenamefont {{LIGO Scientific Collaboration}}()}]{lalsuite}%
  \BibitemOpen
  \bibfield  {author} {\bibinfo {author} {\bibnamefont {{LIGO Scientific
  Collaboration}}},\ }\href {https://wiki.ligo.org/DASWG/LALSuite} {\enquote
  {\bibinfo {title} {{LSC} {A}lgorithm {L}ibrary - {LALS}uite},}\ }\bibinfo
  {howpublished} {free software}\BibitemShut {NoStop}%
\bibitem [{\citenamefont {Acernese}\ \emph {et~al.}(2015)\citenamefont
  {Acernese} \emph {et~al.}}]{TheVirgo:2014hva}%
  \BibitemOpen
  \bibfield  {author} {\bibinfo {author} {\bibfnamefont {F.}~\bibnamefont
  {Acernese}} \emph {et~al.} (\bibinfo {collaboration} {Virgo Collaboration}),\
  }\href {\doibase 10.1088/0264-9381/32/2/024001} {\bibfield  {journal}
  {\bibinfo  {journal} {Class. Quant. Grav.}\ }\textbf {\bibinfo {volume}
  {32}},\ \bibinfo {pages} {024001} (\bibinfo {year} {2015})},\ \Eprint
  {http://arxiv.org/abs/1408.3978} {arXiv:1408.3978 [gr-qc]} \BibitemShut
  {NoStop}%
\bibitem [{\citenamefont {Somiya}(2012)}]{Somiya:2011np}%
  \BibitemOpen
  \bibfield  {author} {\bibinfo {author} {\bibfnamefont {K.}~\bibnamefont
  {Somiya}} (\bibinfo {collaboration} {KAGRA Collaboration}),\ }\href {\doibase
  10.1088/0264-9381/29/12/124007} {\bibfield  {journal} {\bibinfo  {journal}
  {Class. Quant. Grav.}\ }\textbf {\bibinfo {volume} {29}},\ \bibinfo {pages}
  {124007} (\bibinfo {year} {2012})},\ \Eprint {http://arxiv.org/abs/1111.7185}
  {arXiv:1111.7185 [gr-qc]} \BibitemShut {NoStop}%
\bibitem [{\citenamefont {Aso}\ \emph {et~al.}(2013)\citenamefont {Aso},
  \citenamefont {Michimura}, \citenamefont {Somiya}, \citenamefont {Ando},
  \citenamefont {Miyakawa}, \citenamefont {Sekiguchi}, \citenamefont
  {Tatsumi},\ and\ \citenamefont {Yamamoto}}]{Aso:2013eba}%
  \BibitemOpen
  \bibfield  {author} {\bibinfo {author} {\bibfnamefont {Y.}~\bibnamefont
  {Aso}}, \bibinfo {author} {\bibfnamefont {Y.}~\bibnamefont {Michimura}},
  \bibinfo {author} {\bibfnamefont {K.}~\bibnamefont {Somiya}}, \bibinfo
  {author} {\bibfnamefont {M.}~\bibnamefont {Ando}}, \bibinfo {author}
  {\bibfnamefont {O.}~\bibnamefont {Miyakawa}}, \bibinfo {author}
  {\bibfnamefont {T.}~\bibnamefont {Sekiguchi}}, \bibinfo {author}
  {\bibfnamefont {D.}~\bibnamefont {Tatsumi}}, \ and\ \bibinfo {author}
  {\bibfnamefont {H.}~\bibnamefont {Yamamoto}} (\bibinfo {collaboration} {KAGRA
  Collaboration}),\ }\href {\doibase 10.1103/PhysRevD.88.043007} {\bibfield
  {journal} {\bibinfo  {journal} {Phys. Rev.}\ }\textbf {\bibinfo {volume}
  {D88}},\ \bibinfo {pages} {043007} (\bibinfo {year} {2013})},\ \Eprint
  {http://arxiv.org/abs/1306.6747} {arXiv:1306.6747 [gr-qc]} \BibitemShut
  {NoStop}%
\bibitem [{\citenamefont {Unnikrishnan}(2013)}]{Unnikrishnan:2013qwa}%
  \BibitemOpen
  \bibfield  {author} {\bibinfo {author} {\bibfnamefont {C.~S.}\ \bibnamefont
  {Unnikrishnan}},\ }\href {\doibase 10.1142/S0218271813410101} {\bibfield
  {journal} {\bibinfo  {journal} {Int. J. Mod. Phys.}\ }\textbf {\bibinfo
  {volume} {D22}},\ \bibinfo {pages} {1341010} (\bibinfo {year} {2013})},\
  \Eprint {http://arxiv.org/abs/1510.06059} {arXiv:1510.06059
  [physics.ins-det]} \BibitemShut {NoStop}%
\bibitem [{\citenamefont {Aasi}\ \emph {et~al.}(2016)\citenamefont {Aasi} \emph
  {et~al.}}]{Aasi:2013wya}%
  \BibitemOpen
  \bibfield  {author} {\bibinfo {author} {\bibfnamefont {J.}~\bibnamefont
  {Aasi}} \emph {et~al.} (\bibinfo {collaboration} {LIGO Scientific
  Collaboration and Virgo Collaboration}),\ }\href {\doibase
  10.1007/lrr-2016-1} {\bibfield  {journal} {\bibinfo  {journal} {Living Rev.
  Rel.}\ }\textbf {\bibinfo {volume} {19}},\ \bibinfo {pages} {1} (\bibinfo
  {year} {2016})},\ \Eprint {http://arxiv.org/abs/1304.0670} {arXiv:1304.0670
  [gr-qc]} \BibitemShut {NoStop}%
\bibitem [{\citenamefont {Boyle}\ \emph {et~al.}(2007)\citenamefont {Boyle},
  \citenamefont {Brown}, \citenamefont {Kidder}, \citenamefont {Mroue},
  \citenamefont {Pfeiffer}, \citenamefont {Scheel}, \citenamefont {Cook},\ and\
  \citenamefont {Teukolsky}}]{Boyle:2007ft}%
  \BibitemOpen
  \bibfield  {author} {\bibinfo {author} {\bibfnamefont {M.}~\bibnamefont
  {Boyle}}, \bibinfo {author} {\bibfnamefont {D.~A.}\ \bibnamefont {Brown}},
  \bibinfo {author} {\bibfnamefont {L.~E.}\ \bibnamefont {Kidder}}, \bibinfo
  {author} {\bibfnamefont {A.~H.}\ \bibnamefont {Mroue}}, \bibinfo {author}
  {\bibfnamefont {H.~P.}\ \bibnamefont {Pfeiffer}}, \bibinfo {author}
  {\bibfnamefont {M.~A.}\ \bibnamefont {Scheel}}, \bibinfo {author}
  {\bibfnamefont {G.~B.}\ \bibnamefont {Cook}}, \ and\ \bibinfo {author}
  {\bibfnamefont {S.~A.}\ \bibnamefont {Teukolsky}},\ }\href {\doibase
  10.1103/PhysRevD.76.124038} {\bibfield  {journal} {\bibinfo  {journal} {Phys.
  Rev.}\ }\textbf {\bibinfo {volume} {D76}},\ \bibinfo {pages} {124038}
  (\bibinfo {year} {2007})},\ \Eprint {http://arxiv.org/abs/0710.0158}
  {arXiv:0710.0158 [gr-qc]} \BibitemShut {NoStop}%
\bibitem [{\citenamefont {Calder{\'o}n~Bustillo}\ \emph
  {et~al.}(2015)\citenamefont {Calder{\'o}n~Bustillo}, \citenamefont
  {Boh{\'e}}, \citenamefont {Husa}, \citenamefont {Sintes}, \citenamefont
  {Hannam},\ and\ \citenamefont {P{\"u}rrer}}]{Bustillo:2015ova}%
  \BibitemOpen
  \bibfield  {author} {\bibinfo {author} {\bibfnamefont {J.}~\bibnamefont
  {Calder{\'o}n~Bustillo}}, \bibinfo {author} {\bibfnamefont {A.}~\bibnamefont
  {Boh{\'e}}}, \bibinfo {author} {\bibfnamefont {S.}~\bibnamefont {Husa}},
  \bibinfo {author} {\bibfnamefont {A.~M.}\ \bibnamefont {Sintes}}, \bibinfo
  {author} {\bibfnamefont {M.}~\bibnamefont {Hannam}}, \ and\ \bibinfo {author}
  {\bibfnamefont {M.}~\bibnamefont {P{\"u}rrer}},\ }\href@noop {} {\  (\bibinfo
  {year} {2015})},\ \Eprint {http://arxiv.org/abs/1501.00918} {arXiv:1501.00918
  [gr-qc]} \BibitemShut {NoStop}%
\bibitem [{\citenamefont {Regge}\ and\ \citenamefont
  {Wheeler}(1957)}]{Regge:1957td}%
  \BibitemOpen
  \bibfield  {author} {\bibinfo {author} {\bibfnamefont {T.}~\bibnamefont
  {Regge}}\ and\ \bibinfo {author} {\bibfnamefont {J.~A.}\ \bibnamefont
  {Wheeler}},\ }\href {\doibase 10.1103/PhysRev.108.1063} {\bibfield  {journal}
  {\bibinfo  {journal} {Phys. Rev.}\ }\textbf {\bibinfo {volume} {108}},\
  \bibinfo {pages} {1063} (\bibinfo {year} {1957})}\BibitemShut {NoStop}%
\bibitem [{\citenamefont {Zerilli}(1970)}]{Zerilli:1970se}%
  \BibitemOpen
  \bibfield  {author} {\bibinfo {author} {\bibfnamefont {F.~J.}\ \bibnamefont
  {Zerilli}},\ }\href {\doibase 10.1103/PhysRevLett.24.737} {\bibfield
  {journal} {\bibinfo  {journal} {Phys. Rev. Lett.}\ }\textbf {\bibinfo
  {volume} {24}},\ \bibinfo {pages} {737} (\bibinfo {year} {1970})}\BibitemShut
  {NoStop}%
\bibitem [{\citenamefont {Sarbach}\ and\ \citenamefont
  {Tiglio}(2001)}]{Sarbach:2001qq}%
  \BibitemOpen
  \bibfield  {author} {\bibinfo {author} {\bibfnamefont {O.}~\bibnamefont
  {Sarbach}}\ and\ \bibinfo {author} {\bibfnamefont {M.}~\bibnamefont
  {Tiglio}},\ }\href {\doibase 10.1103/PhysRevD.64.084016} {\bibfield
  {journal} {\bibinfo  {journal} {Phys. Rev.}\ }\textbf {\bibinfo {volume}
  {D64}},\ \bibinfo {pages} {084016} (\bibinfo {year} {2001})},\ \Eprint
  {http://arxiv.org/abs/gr-qc/0104061} {arXiv:gr-qc/0104061 [gr-qc]}
  \BibitemShut {NoStop}%
\bibitem [{\citenamefont {Rinne}\ \emph {et~al.}(2009)\citenamefont {Rinne},
  \citenamefont {Buchman}, \citenamefont {Scheel},\ and\ \citenamefont
  {Pfeiffer}}]{Rinne:2008vn}%
  \BibitemOpen
  \bibfield  {author} {\bibinfo {author} {\bibfnamefont {O.}~\bibnamefont
  {Rinne}}, \bibinfo {author} {\bibfnamefont {L.~T.}\ \bibnamefont {Buchman}},
  \bibinfo {author} {\bibfnamefont {M.~A.}\ \bibnamefont {Scheel}}, \ and\
  \bibinfo {author} {\bibfnamefont {H.~P.}\ \bibnamefont {Pfeiffer}},\ }\href
  {\doibase 10.1088/0264-9381/26/7/075009} {\bibfield  {journal} {\bibinfo
  {journal} {Class. Quant. Grav.}\ }\textbf {\bibinfo {volume} {26}},\ \bibinfo
  {pages} {075009} (\bibinfo {year} {2009})},\ \Eprint
  {http://arxiv.org/abs/0811.3593} {arXiv:0811.3593 [gr-qc]} \BibitemShut
  {NoStop}%
\bibitem [{\citenamefont {Nakano}\ \emph {et~al.}(2015)\citenamefont {Nakano},
  \citenamefont {Healy}, \citenamefont {Lousto},\ and\ \citenamefont
  {Zlochower}}]{Nakano:2015pta}%
  \BibitemOpen
  \bibfield  {author} {\bibinfo {author} {\bibfnamefont {H.}~\bibnamefont
  {Nakano}}, \bibinfo {author} {\bibfnamefont {J.}~\bibnamefont {Healy}},
  \bibinfo {author} {\bibfnamefont {C.~O.}\ \bibnamefont {Lousto}}, \ and\
  \bibinfo {author} {\bibfnamefont {Y.}~\bibnamefont {Zlochower}},\ }\href
  {\doibase 10.1103/PhysRevD.91.104022} {\bibfield  {journal} {\bibinfo
  {journal} {Phys. Rev.}\ }\textbf {\bibinfo {volume} {D91}},\ \bibinfo {pages}
  {104022} (\bibinfo {year} {2015})},\ \Eprint
  {http://arxiv.org/abs/1503.00718} {arXiv:1503.00718 [gr-qc]} \BibitemShut
  {NoStop}%
\bibitem [{\citenamefont {Lovelace}\ \emph {et~al.}(2016)\citenamefont
  {Lovelace} \emph {et~al.}}]{Lovelace:2016uwp}%
  \BibitemOpen
  \bibfield  {author} {\bibinfo {author} {\bibfnamefont {G.}~\bibnamefont
  {Lovelace}} \emph {et~al.},\ }\href {\doibase 10.1088/0264-9381/33/24/244002}
  {\bibfield  {journal} {\bibinfo  {journal} {Class. Quant. Grav.}\ }\textbf
  {\bibinfo {volume} {33}},\ \bibinfo {pages} {244002} (\bibinfo {year}
  {2016})},\ \Eprint {http://arxiv.org/abs/1607.05377} {arXiv:1607.05377
  [gr-qc]} \BibitemShut {NoStop}%
\end{thebibliography}%

\end{document}